\documentclass[aps,reprint,showpacs]{revtex4-1}
\usepackage{amsmath}
\usepackage{amssymb}
\usepackage{graphicx}
\usepackage{epstopdf}
\usepackage{color}
\usepackage{hyperref}
\newcommand{\eq}[1]{Eq.~(\ref{#1})}

\begin{document}

\title{Dielectric function and plasmons in graphene: \\
A self-consistent-field calculation within a Markovian master equation formalism}

\author{F. Karimi}\email{karimi2@wisc.edu}

\author{A. H. Davoody}

\author{I. Knezevic}\email{iknezevic@wisc.edu}

\affiliation{Department of Electrical and Computer Engineering, University of Wisconsin-Madison, Madison, Wisconsin 53706, USA}

\date{\today}

\begin{abstract}
We introduce a method for calculating the dielectric function of nanostructures with an arbitrary band dispersion and Bloch wave functions. The linear response of a dissipative electronic system to an external electromagnetic field is calculated by a self-consistent-field approach within a Markovian master equation formalism (SCF-MMEF) coupled with full-wave electromagnetic equations. The SCF-MMEF accurately accounts for several concurrent scattering mechanisms. The method captures interband electron-hole-pair generation, as well as the interband and intraband electron scattering with phonons and impurities. We employ the SCF-MMEF to calculate the dielectric function, complex conductivity, and loss function for supported graphene. From the loss-function maximum, we obtain plasmon dispersion and propagation length for different substrate types [nonpolar diamondlike carbon (DLC) and polar  SiO$_2$ and hBN], impurity densities, carrier densities, and temperatures. Plasmons on the two polar substrates are suppressed below the highest surface phonon energy, while the spectrum is broad on the nonpolar DLC. Plasmon propagation lengths are comparable on polar and nonpolar substrates and are on the order of tens of nanometers, considerably shorter than previously reported. They improve with fewer impurities, at lower temperatures, and at higher carrier densities.
\end{abstract}


\maketitle

\section{Introduction}
Surface plasmon-polaritons, often referred to as plasmons for brevity, are hybrid excitations that can arise from the interaction of the electron plasma in good conductors with external electromagnetic fields \cite{maier2001plasmonics,schuller2010plasmonics}. Plasmonics, the field of study of plasmon dynamics, has attracted considerable interest in recent years as a promising path towards the miniaturization of nanophotonics \cite{maier2005plasmonics,maier2001plasmonics,karalis2005surface,karalis2005surface}. Conventionally, noble metals are considered as plasmonic materials, with diverse applications such as integrated photonic systems \cite{gramotnev2010plasmonics,novotny2011antennas,schuller2010plasmonics,schuller2010plasmonics}, magneto-photonic
structures \cite{Yu2008One,Khanikav2008One,abbasi2015one}, metamaterials and cloaking \cite{shalaev2007optical,luk2010fano,kawata2009plasmonics,alu2005achieving}, biosensing \cite{anker2008biosensing,kabashin2009plasmonic} and photovoltaic devices \cite{atwater2010plasmonics} in the visible to near-infrared spectrum \cite{low2014graphene}. However, at THz to mid-infrared frequencies, which have various applications \cite{tonouchi2007cutting,ferguson2002materials,soref2010mid} in information and communication, biology, chemical and biological sensing, homeland security, and spectroscopy, metal plasmonic materials are quite lossy. Therefore, there is considerable interest in low-loss plasmonic materials at these frequencies \cite{west2010searching}.

Graphene, the two-dimensional allotrope of carbon \cite{novoselov2004electric,geim2007rise,neto2009electronic,sarma2011electronic,sarma2011electronic,mayorov2011micrometer}, is a promising plasmonic material \cite{stauber2014plasmonics,christensen2014classical,kravets2014graphene}. Its resonance typically falls in the THz to mid-infrared range \cite{grigorenko2012graphene,koppens2011graphene,hwang2007dielectric,jablan2009plasmonics} and it shows significantly different screening properties and collective excitations than the quasi-2D systems with parabolic electron dispersions \cite{ando1982electronic}. Moreoever, graphene's carrier density is easily tunable by an external gate voltage \cite{neto2009electronic}, which enables electrostatic control of its electronic and optical properties.

Plasmons in graphene have been experimentally excited and visualized by several methods, such as the electron energy-loss spectroscopy (EELS) \cite{liou2015pi,eberlein2008plasmon,liu2010plasmon,kramberger2008linear}, Fourier-transform infrared spectroscopy (FTIR)  \cite{brar2014hybrid,yan2013damping,yan2013damping}, and scanning near-filed optical microscopy (SNOM) \cite{woessner2014highly,fei2012gate}. Theoretical studies have also been performed: the momentum-independent graphene conductivity  was calculated within the local phase approximation used along with finite-element \cite{brar2014hybrid} and transfer matrix methods  \cite{woessner2014highly} in order to solve the electromagnetic equations for the plasmon modes. However, plasmons are in the nonretarded regime at mid-infrared frequencies, so the momentum dependence of the optical response should be considered. The Lindhard dielectric function  $\varepsilon(\mathbf{q},\omega)$ \cite{hwang2007dielectric} has a dependence on both frequency ($\omega$) and wave vector ($\mathbf{q}$). It captures interband transitions due to the electromagnetic field and is based on the random-phase approximation (RPA). The RPA can also explain the coupling between graphene plasmons and surface optical (SO) phonons on polar substrates    \cite{lu2009plasmon,hwang2010plasmon,hwang2013surface,ahn2014inelastic,jablan2011unconventional,ong2012theory}, a phenomenon that has been captured experimentally \cite{brar2014hybrid,woessner2014highly,yan2013damping}.

Graphene plasmon modes have much shorter wave lengths than light with the same frequencies, and their propagation length is very sensitive to the damping pathways, such as intrinsic phonons \cite{perebeinos2010inelastic,sule2012phonon,li2010surface}, ionized impurities \cite{stauber2007electronic,hwang2007carrier,sule2014clustered}, and SO phonons on polar substrates \cite{fratini2008substrate,schiefele2012temperature,konar2010effect}. The damping of graphene plasmons has been calculated based on the Mermin-Lindhard (ML) dielectric function \cite{jablan2009plasmonics} and its simplified version, the Drude dielectric function \cite{yan2013damping}. The ML dielectric function stems from Mermin's derivation \cite{mermin1970lindhard} of the Lindhard expression based on a master equation and within the relaxation-time approximation (RTA), with a single effective relaxation time accounting for all scattering. However, the ML dielectric function was derived for electronic systems with a parabolic band dispersion and is also only valid when intraband dissipation mechanisms dominate. In the case of nanostructures with nonparabolic dispersions,  densely spaced energy subbands, or in the presence of efficient interband dissipation mechanisms (e.g., optical phonons), the ML dielectric function provides an incomplete picture.

With multiple concurrent scattering mechanisms, it is common to extract an averaged relaxation time corresponding to each mechanism separately and then calculate the total relaxation time based on Mattheissen's rule \cite{jablan2009plasmonics,yan2013damping}. However, Matthiessen's rule technically holds only when all mechanisms have the same relaxation time vs energy dependence \cite{lundstrom2009fundamentals}. Also, employing a single energy-independent relaxation time is generally not a good approximation for systems with pronounced non-Coulomb scattering mechanisms \cite{YagerJAP36,HillDissadoJPC85,BeardPRB2000,willis2013generalized}, such as phonons in nonpolar materials; indeed, the use of a single relaxation time has been shown not to accurately capture the loss mechanisms in suspended and, to a lower degree, supported graphene \cite{sule2014terahertz}. What is needed is an accurate (and, ideally, computationally inexpensive) theoretical approach that can treat  the interaction of light with charge carries in graphene (and in related nanomaterials) in the presence of both interband and intraband transitions due to multiple competing scattering mechanisms, where the transition rates can have pronounced and widely differing dependencies on both carrier energy and momentum.

In this paper, we present a method for calculating the dielectric function of a dissipative electronic system with an arbitrary band dispersion and Bloch wave functions by a self-consistent-field approach within an open-system Markovian master-equation formalism (SCF-MMEF) coupled with full-wave electromagnetic equations (Sec. \ref{sect2}). We derive a generalized Markovian master equation \cite{breuer2002theory,knezevic2013time} of the Lindblad form (i.e., conserving the positivity of the density matrix), which includes the interaction of the electronic system with an external electromagnetic field (to first order) and with a dissipative environment (to second order). We solve for the time evolution of coherences and calculate the induced charge density as a function of the self-consistent field within linear response. Based on the electrodynamic relation between the induced charge density and the induced potential, we obtain the expressions for the polarization, dielectric function, and conductivity (Sec. \ref{sect2}). Numerical implementation is achieved with the aid of Brillouin-zone discretization, and the resulting matrix equations are readily solved using modern linear solvers (Sec. \ref{numerics}).

We use the SCF-MMEF to calculate the graphene dielectric function, complex conductivity, and loss function (Sec. \ref{sect3}). The complex conductivity agrees well with experiment and its DC limit shows a well-known sublinearity at high carrier densities. At low carrier densities, the screening length depends on the impurity density, which is a phenomenon that the ML approach cannot capture (Sec. \ref{sect3a}). The plasmon modes and their propagation length are obtained from the loss-function maximum (Sec. \ref{sect3c}). We did our calculations for three substrates: diamondlike carbon (DLC) \cite{yan2013damping,wu2012state} as a non-polar substrate, and $\text{SiO}_2$ \cite{chen2008intrinsic,dorgan2010mobility,fei2011infrared} and hBN \cite{principi2014plasmon,wang2013one,dean2010boron} as polar substrates, with the effect of SO phonons captured via a weak Fr{\"o}hlich term. We investigate the effects of substrate type, substrate impurity density, carrier density, and temperature on the dielectric function and plasmon characteristics. Plasmon dispersion is fairly insensitive to varying impurity density or temperature, but is pushed towards shorter wave vectors with increasing carrier density. Plasmons are strongly suppressed on polar substrates below the substrate optical-phonon energies. Plasmon damping -- inversely proportional to the number of wave lengths that a plasmon propagates before dying out -- worsens with more pronounced scattering, such as when the impurity density or temperature are increased. However, damping drops with increasing carrier density, benefiting plasmon propagation. Overall, the plasmon propagation length in absolute units improves on substrates with fewer impurities, and is better at low temperatures and at higher carrier densities. Plasmon propagation lengths on polar and nonpolar substrates are comparable and in the tens of nanometers, an order of magnitude lower than previously reported \cite{jablan2009plasmonics,yan2013damping}. We conclude with Sec. \ref{sect4}.

\section{\label{sect2}Theory}

In this section, we present a detailed derivation of the SCF-MMEF in order to calculate the dielectric function and the conductivity of graphene, a two-dimensional material. We note that the SCF-MMEF is quite general and can be applied to electronic systems of arbitrary dimensionality (see Appendix \ref{app} for quasi-one-dimensional materials).

{\color{black} Single-layer graphene has a density-independent Wigner-Seitz radius $r_s<1$ ($r_s\approx 0.5$ for graphene on SiO$_2$ \cite{hwang2007dielectric, sarma2011electronic}), owing to linear electron dispersion. Therefore, it can be considered a weakly interacting system in which the random-phase approximation (RPA, the equivalent of the SCF approximation we use here) is valid. Intuitively, the question is whether the carriers in graphene are effective at screening. For graphene on a substrate, the affirmative answer is supported by experimental observations of Drudelike behavior of the electronic system in frequency-dependent conductivity measurements \cite{Ren_NanoLett2012,Rouhi2012,ju2011graphene,sule2014terahertz}. This holds even at low carrier densities, as electrons and holes form puddles, so the carrier density locally exceeds the impurity density  \cite{Rossi_PRL2008,sule2014clustered}. (We note that, in bilayer graphene and  at low carrier densities, the SCF/RPA approximation becomes more difficult to justify \cite{Abergel_PRB2013,Kharitonov_PRB2008}.)}

\subsection{The self-consistent-field approximation}

The dielectric function describes how the electron system screens a perturbing potential $\mathbb{V}_{\text{ext}}$. The system response results in an induced potential ${\mathbb {V}_{\text{ind}}}$ that stems from electron-electron interactions. The two give rise to a combined self-consistent field, $\mathbb {V}_{\text{SCF}}= {\mathbb {V}_{\text{ind}}}+{\mathbb{V}_{\text{ext}}}$.

We assume the graphene sheet is in the $xy$-plane ($z=0$). The induced charge density $n$ will be of the form

\begin{equation}
    n(\mathbf{r},z,t)=n_s(\mathbf{r},t)\delta(z).
\end{equation}
$\delta(\cdot)$ denotes the Dirac delta function, $\mathbf{r}$ is a vector in the $xy$-plane, and $n_s$ is the sheet density. Then, by taking a temporal Fourier transform ($\omega$ is the angular frequency) and a two-dimensional spatial Fourier transform over $\mathbf{r}$ ($\mathbf{q}$ is the in-plane wave vector), the inhomogeneous wave equation for the induced potential is given by
\begin{equation}\label{fullwave}
\left[\frac{\partial}{\partial^2 z} +(iQ)^2\right] V_{\text{ind}}(\mathbf{q},z,\omega) = -\frac{-e^2}{\varepsilon_b(\omega)\varepsilon_0}n_s(\mathbf{q},\omega)\delta(z),
\end{equation}
where $\varepsilon_b(\omega)\equiv\frac{1+\varepsilon_s(\omega)}{2}$ represents the background lattice dielectric function, while $\varepsilon_s(\omega)$ is the complex dielectric function of the substrate (Appendix \ref{app2}) and  $(iQ)^2=\frac{\varepsilon_b(\omega) \omega^2}{c^2}-\mathbf{q\cdot q}$. In general, the Fourier transform of an arbitrary function $f(\mathbf{r},z,t)$ is defined as $\mathfrak{F}(\mathbf{q},z,\omega)=A^{-1}\int f(\mathbf{r},z,t) e^{-i\mathbf{q \cdot r}+i\omega t} d^2\mathbf{r}~dt$, where $A$ is the sample area.

By employing the Green's function analysis, the solution of \eq{fullwave} becomes
\begin{equation}\label{Vind}
\begin{split}
 V_{\text{ind}}(\mathbf{q},z=0,\omega)=&   \frac{-e^2}{\varepsilon_b(\omega)\varepsilon_0}\frac{n_s(\mathbf{q},\omega)}{2Q}.
 \end{split}
\end{equation}
Here, we assumed that the potential does not vary significantly across the graphene sheet. To simplify the notation, from now on we drop the $z$ argument, and remember that all the wave vectors are in the plane of graphene.

The dielectric function of the graphene sheet is defined as  $\varepsilon (\mathbf{q},\omega)= V_{\text{ext}}(\mathbf{q},\omega) /V_{\text{SCF}}(\mathbf{q},\omega)$. Now, by assuming $Q\approx|\mathbf{q}|=q$ as a valid approximation in the non-retarded regime of interest, and using \eq{Vind},  the dielectric function may be written as
\begin{equation}\label{eps}
\begin{split}
\varepsilon (\mathbf{q},\omega)=&1+\frac{e^2}{\varepsilon_b(\omega)\varepsilon_0}\frac{1}{2q}\frac{n_s(\mathbf{q},\omega)}{V_{\text{SCF}}(\mathbf{q},\omega)}\\
=&1+\frac{e}{\varepsilon_b(\omega)\varepsilon_0}\frac{1}{2q}P_s(\mathbf{q},\omega),
\end{split}
\end{equation}
where we have introduced the surface polarization, $P_s(\mathbf{q},\omega)\equiv \frac{ e n_s(\mathbf{q},\omega)}{V_{\text{SCF}}(\mathbf{q},\omega)}$. Also, we can derive an expression for the conductivity of the graphene sheet. The continuity equation in the frequency and momentum domain reads
\begin{equation}\label{continuity}
\omega e n_s(\mathbf{q},\omega)=\mathbf{q}\cdot \mathbf{J}_s(\mathbf{q},\omega),
\end{equation}
where $\mathbf{J}_s(\mathbf{q},\omega)$ is the surface current density. $\mathbf{J}_s(\mathbf{q},\omega)=\sigma_s(\mathbf{q},\omega)~ \mathbf{E}_{\text{SCF}}(\mathbf{q},z=0,\omega) $ , $\sigma_s$ being the sheet conductivity and $-e \mathbf{E}_{\text{SCF}} = -i\mathbf{q} V_{\text{SCF}}$. Now, the graphene sheet conductivity is written as
\begin{equation}\label{sigma}
\sigma_s(\mathbf{q},\omega) = \frac{ -ie\omega}{q^2 }P_s(\mathbf{q},\omega).
\end{equation}
The surface polarization $P_s(\mathbf{q},\omega)$, needed to obtain  $\varepsilon$ (\ref{eps}) and $\sigma_s$ (\ref{sigma}), carries the information about electron interactions in the material and is calculated next, based on a Markovian master-equation formalism.

\subsection{Markovian master-equation formalism}

We consider a quantum-mechanical electronic system that interacts with an environment. Assuming $\mathbb {H}_{\text{e}}$ to be the unperturbed Hamiltonian of free electrons in a  lattice, its eigenkets and eigenenergies are represented by $|\mathbf{k} l \rangle$ and $\epsilon_{\mathbf{k} l} $, respectively. $\mathbf{k}$ is the in-plane electron wave vector and $l$ denotes the band index. The spatial representation of the single-particle Bloch wave functions (BWF) corresponding to $|\mathbf{k} l \rangle$ has the form of
\begin{equation}
\langle \mathbf{r},z |\mathbf{k} l \rangle ={\frac{1}{\sqrt{N_{\mathrm{uc}}}}} e^ {i \mathbf{k} \cdot\mathbf{r}} u_{\mathbf{k} l}(\mathbf{r},z).
\end{equation}
$N_{\mathrm{uc}}$ is the number of unit cells in a finite-sized graphene sample. The induced charge density in the second quantization form is given by
\begin{equation}
n(\mathbf{r},\omega) = -\frac{1}{N_{\mathrm{uc}}}\sum_{\mathbf{k,q},l',l} u^*_{\mathbf{k+q}l'}u_{\mathbf{k}l} e^{-i\mathbf{q \cdot r}} \langle c^\dagger_{\mathbf{k+q}l'} c_{\mathbf{k}l}\rangle.
\end{equation}
$c^\dagger$ and $c$ are the electron creation and destruction operators, respectively. $\langle \mathbb{O} \rangle \equiv \text{tr}_\text{e}\left\{\mathbb{O}\rho\right\}$ is the expectation value of an electronic operator $\mathbb{O}$ and $\rho$ is the many-body statistical operator (also referred to as the many-body density matrix). Taking the Fourier transform in the $xy$-plane, and integrating over the $z$ direction yield the induced surface charge density
\begin{equation}\label{n}
\begin{split}
n_{s}(\mathbf{q},\omega) &= -\frac{1}{A}  \sum_{\mathbf{k},l',l} \langle c^\dagger_{\mathbf{k}l} c_{\mathbf{k+q}l'}\rangle (\mathbf{k}l|\mathbf{k+q}l')\,\\
\end{split}
\end{equation}
where $A$ is the area of the graphene sheet. We defined the following overlap integrals over a unit cell as
\begin{equation}
\begin{split}
(\mathbf{k+q}l'|\mathbf{k}l) \equiv \int_{\text{uc}} d^3r' u^*_{\mathbf{k+q}l'}(\mathbf{r'})u_{\mathbf{k}l}(\mathbf{r'})\, .
\end{split}
\end{equation}

\noindent Equation (\ref{n}) shows that the expectation value of coherences, $\langle c^\dagger_{\mathbf{k}l} c_{\mathbf{k+q}l'}\rangle$, is required to calculate the induced charge density. The time dependence of the expectation value of coherences is calculated through a quantum master equation.

The total Hamiltonian of an open electronic system within the SCF approximation is
\begin{equation}
\mathbb {H}(t)={\mathbb {H}_{\text{e}}} + \overbrace{\mathbb {V}_{\text{SCF}}(t)+\mathbb {H}_{\text{col}}}^{\displaystyle {\mathbb {H}}_{\text{int}}(t)}+\mathbb{H}_{\text{ph}}\, .
\end{equation}
$\mathbb{H}_{\text{ph}}$ denotes the free Hamiltonian of the phonon bath. $\mathbb {H}_{\text{col}}$, where subscript ``col'' stands for collisions, is the sum of the interaction Hamiltonian of electrons with phonons ($\mathbb {H}_{\text{e-ph}}$) and the interaction Hamiltonian of electrons with ionized impurities ($\mathbb {H}_{\text{e-ii}}$). ${\mathbb {V}_{\text{SCF}}(t)}$, ${\mathbb {H}_{\text{e-ph}}}$, and ${\mathbb {H}_{\text{e-ii}}}$ have the forms
\begin{equation}
 \begin{split}
 &{\mathbb {V}_{\text{SCF}}(t)} =\sum_{\mathbf{k},l'l} \langle\mathbf{k+q}l'|\mathbb {V}_{\text{SCF}}(t)|\mathbf{k}l\rangle c^\dagger_{\mathbf{k+q}l'} c_{\mathbf{k}l},\\
 &{\mathbb {H}_{\text{e-ph}}}=\sum_{\mathbf{kq},l'l} \mathcal{M}_{\text{ph}}(\mathbf{q})(\mathbf{k+q}l'|\mathbf{k}l)c^\dagger_{\mathbf{k+q}l'} c_{\mathbf{k}l} (b_{\mathbf{q}}+b^\dagger_{-\mathbf{q}}),\\
 &{\mathbb {H}_{\text{e-ii}}}=\sum_{\mathbf{kq},l'l} \mathcal{M}_{\text{ii}}(\mathbf{q})(\mathbf{k+q}l'|\mathbf{k}l)c^\dagger_{\mathbf{k+q}l'} c_{\mathbf{k}l}\,.\\
\end{split}
\end{equation}
$b$ and $b^\dagger$ are the phonon creation and destruction operators, respectively. The interaction Hamiltonian can be written as
\begin{equation}\label{Hcol}
 \begin{split}
{\mathbb {H}_{\text{col}}}\equiv\sum_\alpha \mathbb{A}_\alpha \otimes \mathbb{B}_\alpha \, ,
\end{split}
\end{equation}
where $\alpha$ is the set of variables $\{\mathbf{kq},l'lg\}$, and $g$ determines the scattering mechanism. $\mathbb{A}$ operates on the electron and $\mathbb{B}$ on the phonon subspace. For ionized-impurity scattering $\mathbb{B}$ is equal to unity operator.

The equation of motion for the statistical operator in the interaction picture (in units of $\hbar=1$) is
\begin{equation}\label{eom0}
\begin{split}
\frac{d}{dt} \tilde{\rho}(t)=&-i[\mathbb {\tilde{H}}_\text{int}(t),\tilde{\rho}(t)],\\
\tilde{\rho}(t)=&\tilde{\rho}(0)-i\int_0^t[\mathbb {\tilde{H}}_\text{int}(t'),\tilde{\rho}(t')]dt'.\\
\end{split}
\end{equation}
The tilde denotes that the operators are in the interaction picture, i.e.,  $\mathbb{\tilde{O}}(t)=e^{i(\mathbb{H}_e+\mathbb{H}_{ph})t}\mathbb{O}e^{-i(\mathbb{H}_e+\mathbb{H}_{ph})t}$. The collision Hamiltonian in the interaction picture takes the form
\begin{equation}
 \begin{split}\label{Hcolt}
 {\mathbb {\tilde{H}}_{\text{col}}}(t)=&\sum_\alpha e^{-i\Delta_\alpha t} \mathbb{A}_\alpha  \otimes \mathbb{B}_\alpha (t),\\
\end{split}
\end{equation}
where $\mathbb{B}_\alpha(t)\equiv e^{i\mathbb{H}_\text{ph}t} \mathbb{B}_\alpha e^{-i\mathbb{H}_\text{ph}t}$ and $\Delta_\alpha=\epsilon_{\mathbf{k_\alpha}l_\alpha}-\epsilon_{\mathbf{k_\alpha+q_\alpha}l'_\alpha}$ are used.
We assume an uncorrelated initial state of the form $\tilde{\rho}(0)=\tilde{\rho}_e(0)\otimes \tilde{\rho}_\text{ph}$. Within the Born approximation, the interaction is assumed weak so the environment is negligibly affected by it and no considerable system-environment correlations arise due to it on the timescales relevant to the open system, so the the total density matrix can be written as a tensor product $ \tilde{\rho}(t)=\tilde{\rho}_e(t)\otimes \tilde{\rho}_\text{ph}$ at all times \cite{breuer2002theory,knezevic2013time}. Within the Markov approximation, the system is considered memoryless so the evolution of the density matrix does not depend on its past, but only on its present state. Putting the integral form  in \eq{eom0} into the right-hand side (RHS) of the differential form, applying the Born and Markov approximations, and taking the trace over the phonon reservoir yield

\begin{equation}\label{eom1}
\begin{split}
\hspace{-0.3cm}\frac{d}{dt}& \tilde{\rho}_e(t)=-i\text{tr}_\text{ph}\left\{[\mathbb {\tilde{H}}_\text{int}(t),\tilde{\rho}_e(0)\otimes \tilde{\rho}_\text{ph}]\right\}\\
&\hspace{0.2cm}-\int_0^tds~\text{tr}_\text{ph}\left\{[\mathbb {\tilde{H}}_\text{int}(t),[\mathbb {\tilde{H}}_\text{int}(s),\tilde{\rho}_e(t)\otimes \tilde{\rho}_\text{ph}]]\right\}.
\end{split}
\end{equation}
The above equation is the Redfield equation. We can substitute $s$ by $t-s$; the new $s$ denotes the time difference from $t$ and, because we expect the integrand to be negligible for large values of $s$, we can let the upper limit of the integral go to infinity:
\begin{equation}\label{eom2}
\begin{split}
\hspace{-0.3cm}\frac{d}{dt}&\tilde{\rho}_e(t)=-i\text{tr}_\text{ph}\left\{[\mathbb {\tilde{H}}_\text{int}(t),\tilde{\rho}_e(0)\otimes \tilde{\rho}_\text{ph}]\right\}\\
&-\int_0^\infty \hspace{-.1 cm}ds~\text{tr}_\text{ph}\left\{[\mathbb {\tilde{H}}_\text{int}(t),[\mathbb {\tilde{H}}_\text{int}(t-s),\tilde{\rho}_e(t)\otimes \tilde{\rho}_\text{ph}]]\right\}.
\end{split}
\end{equation}
Now, to remove the temporal dependence of interaction Hamiltonians, we switch back to the Schr\"{o}dinger picture
\begin{equation}\label{eom3}
\begin{split}
&\hspace{-0.35cm}\frac{d\rho_e(t)}{dt}=-i[\mathbb {H}_{\text{e}}, \rho_e(t) ]\hspace{-0.1cm}-\hspace{-0.1cm}i\text{tr}_\text{ph}\left\{[ \mathbb {{H}}_\text{int}(t), \rho_e(0) \otimes {\rho}_\text{ph}]\right\}\\
&-\int_0^\infty ds~ \text{tr}_\text{ph}\left\{[\mathbb {{H}}_\text{int},[\mathbb {\tilde{H}}_\text{int}(-s),{\rho}_e(t)\otimes {\rho}_\text{ph}]]\right\}.\\
\end{split}
\end{equation}
In order to obtain the time evolution of $\langle c^\dagger_{\mathbf{k}l} c_{\mathbf{k+q}l'}\rangle$, which we need to calculate the charge density in \eq{n}, we multiply \eq{eom3} by $c^\dagger_{\mathbf{k}l} c_{\mathbf{k+q}l'}$, and take a trace over the electron subspace. Also, we replace $\mathbb {{H}}_\text{int}(t)$ by $\mathbb {{V}}_\text{SCF}(t)+\mathbb {{H}}_\text{e-ii}(t)+\mathbb{{H}}_\text{e-ph}(t)$. Because we are seeking linear response, we keep the linear terms of $\mathbb {V}_{\text{SCF}}(t)$, and neglect the higher-order terms. As a result, we can approximate $\mathbb {{V}}_{\text{SCF}}(t){\rho}_e(t)\approx \mathbb {{V}}_{\text{SCF}}(t){\rho}_e(0)$. ${\rho}_e(0)$ denotes the initial density matrix which is the density matrix of the unperturbed system. Thus, \eq{eom3} can be rewritten as
\begin{widetext}
\begin{equation}\label{eom4}
\begin{split}
\frac{d \langle c^\dagger_{\mathbf{k}l} c_{\mathbf{k+q}l'}\rangle }{dt}=&-i\text{tr}_\text{e}\left\{[ \mathbb {H}_{\text{e}}, \rho_e(t) ]c^\dagger_{\mathbf{k}l} c_{\mathbf{k+q}l'} \right\}-i \text{tr}_\text{e}\left\{\text{tr}_\text{ph}\left\{[\mathbb {{H}}_\text{int}(t), \rho_e(0) \otimes {\rho}_\text{ph}]\right\}c^\dagger_{\mathbf{k}l} c_{\mathbf{k+q}l'}\right\}\\
&\hspace{-0.8cm}-\int_0^\infty ds~ \text{tr}_\text{e}\left\{\text{tr}_\text{ph}\left\{
[\mathbb {{H}}_\text{e-ph},[\mathbb {\tilde{H}}_\text{e-ph}(-s),{\rho}_e(t)\otimes {\rho}_\text{ph}]]+[\mathbb {{H}}_\text{e-ii},[\mathbb {\tilde{H}}_\text{e-ii}(-s),{\rho}_e(t)\otimes {\rho}_\text{ph}]]
\right\}c^\dagger_{\mathbf{k}l} c_{\mathbf{k+q}l'} \right\}\\
&\hspace{-0.8cm}-\int_0^\infty ds~ \text{tr}_\text{e}\left\{\text{tr}_\text{ph}\left\{
[\mathbb {{H}}_\text{e-ph},[\mathbb {\tilde{H}}_\text{e-ii}(-s),{\rho}_e(t)\otimes {\rho}_\text{ph}]]+[\mathbb {{H}}_\text{e-ii},[\mathbb {\tilde{H}}_\text{e-ph}(-s),{\rho}_e(t)\otimes {\rho}_\text{ph}]]
\right\}c^\dagger_{\mathbf{k}l} c_{\mathbf{k+q}l'} \right\}\\
&\hspace{-0.8cm}-\int_0^\infty ds~ \text{tr}_\text{e}\left\{\text{tr}_\text{ph}\left\{
[\mathbb {{V}}_\text{SCF},[\mathbb {\tilde{H}}_\text{e-ph}(-s),{\rho}_e(0)\otimes {\rho}_\text{ph}]]+[\mathbb {{H}}_\text{e-ph},[\mathbb {\tilde{V}}_\text{SCF}(-s),{\rho}_e(0)\otimes {\rho}_\text{ph}]]
\right\}c^\dagger_{\mathbf{k}l} c_{\mathbf{k+q}l'} \right\}\\
&\hspace{-0.8cm}-\int_0^\infty ds~ \text{tr}_\text{e}\left\{\text{tr}_\text{ph}\left\{
[\mathbb {{V}}_\text{SCF},[\mathbb {\tilde{H}}_\text{e-ii}(-s),{\rho}_e(0)\otimes {\rho}_\text{ph}]]+[\mathbb {{H}}_\text{e-ii},[\mathbb {\tilde{V}}_\text{SCF}(-s),{\rho}_e(0)\otimes {\rho}_\text{ph}]]
\right\}c^\dagger_{\mathbf{k}l} c_{\mathbf{k+q}l'} \right\}.\\
\end{split}
\end{equation}
\end{widetext}
As $\text{tr}_\text{ph}\{\mathbb{H}_\text{e-ph}(t)\rho_\text{ph}\}=0$, the second integral on the RHS of \eq{eom4} vanishes. Also, \eq{eom4} contains expectation values of products of four or six creation and destruction operators, e.g., $\langle c^\dagger_{1} c_{2} c^\dagger_{3} c_{4} c^\dagger_{5} c_{6} \rangle = \text{tr}_\text{e} \left\{ c^\dagger_{1} c_{2} c^\dagger_{3} c_{4} c^\dagger_{5} c_{6} \rho_e(t)\right\}$. By~applying Wick's theorem and the mean-field approximation \cite{fetter2003quantum}, these terms are reduced to the expectation values of pairs, e.g.,
\begin{subequations}
\begin{equation}\label{Wick4}
\begin{split}
\langle c^\dagger_1 c_2 c^\dagger_3 c_4 \rangle = \langle c^\dagger_1 c_2 \rangle \langle c^\dagger_3 c_4 \rangle+\langle c^\dagger_1 c_4 \rangle \langle c_2 c^\dagger_3 \rangle
\end{split}
\end{equation}
\begin{equation}\label{Wick6}
\begin{split}
\langle c^\dagger_{1} c_{2} c^\dagger_{3} c_{4} c^\dagger_{5} c_{6} \rangle &\approx
\langle c^\dagger_{1} c_{2} \rangle \langle c^\dagger_{3} c_{4} \rangle \langle c^\dagger_{5} c_{6} \rangle+
\langle c^\dagger_{1} c_{2} \rangle \langle c^\dagger_{3} c_{6} \rangle \langle c_{4} c^\dagger_{5} \rangle\\
&+\langle c^\dagger_{1} c_{4} \rangle \langle c_{2} c^\dagger_{3} \rangle \langle c^\dagger_{5} c_{6} \rangle -
\langle c^\dagger_{1} c_{4} \rangle \langle c_{2} c^\dagger_{5} \rangle \langle c^\dagger_{3} c_{6} \rangle\\
&+\langle c^\dagger_{1} c_{6} \rangle \langle c_{2} c^\dagger_{3} \rangle \langle c_{4} c^\dagger_{5} \rangle+\langle c^\dagger_{1} c_{6} \rangle \langle c_{2} c^\dagger_{5} \rangle \langle c^\dagger_{3} c_{4} \rangle
\end{split}
\end{equation}
\end{subequations}
In the two last integrals in \eq{eom4}, the expectation values should be taken with respect to the unperturbed equilibrium density matrix, i.e.,  $\text{tr}_\text{e} \left\{c^\dagger_{\mathbf{k}l} c_{\mathbf{k+q}l'} \rho_e(0)\right\}=f_{\mathbf{k}l}\delta_{{\mathbf{k}l} ,{\mathbf{k+q}l'}}$,  where $f_{\mathbf{k}l}$ is the Fermi-Dirac distribution function. By using this assumption alongside \eq{Wick6}, it can be shown that the last two integrals in \eq{eom4} also vanish.

Now, we use the electron-phonon interaction and electron-impurity interaction Hamiltonians of the forms given by Eqs. (\ref{Hcol}) and (\ref{Hcolt}). It is useful to introduce a one-sided Fourier transform of the phonon reservoir correlation functions, $\text{tr}_\text{ph}\left\{\mathbb{B}^\dagger_\alpha \mathbb{B}_\beta(-s) \rho_\text{ph}\right\}$, as
\begin{equation}
\Gamma_{\alpha\beta}(\Delta)\equiv \int_0^\infty ds e^{i\Delta s}\text{tr}_\text{ph}\left\{\mathbb{B}^\dagger_\alpha \mathbb{B}_\beta(-s) \rho_\text{ph}\right\},
\end{equation}
where $\beta$, like $\alpha$, is the set of variables $\{\mathbf{kq},l'lg\}$. With the use of Wick's theorem and the mean-field \eq{Wick4}, \eq{eom4} can be written in terms of $\Gamma(\cdot)$ as
\begin{equation}
\begin{split}\label{eom55}
&\frac{d \langle c^\dagger_{\mathbf{k}l} c_{\mathbf{k+q}l'}\rangle}{dt}=-i(\epsilon_{\mathbf{k+q}l'}-\epsilon_{\mathbf{k}l})\langle c^\dagger_{\mathbf{k}l} c_{\mathbf{k+q}l'}\rangle\\
&-i\langle \mathbf{k+q}l'| \mathbb {V}_{\text{SCF}} | \mathbf{k}l\rangle (f_{\mathbf{k}l}-f_{\mathbf{k+q}l'})\\
&+\sum_{\alpha,\beta} \Gamma_{\alpha\beta}(\Delta_\beta)
\text{tr}_e \{\left ( \mathbb{A}_\beta\rho_e(t) \mathbb{A}^\dagger_\alpha - \mathbb{A}^\dagger_\alpha\mathbb{A}_\beta\rho_e(t)\right) c^\dagger_{\mathbf{k}l} c_{\mathbf{k+q}l'}\}\\
&+\sum_{\alpha,\beta}\Gamma^*_{\beta\alpha}(\Delta_\alpha)
\text{tr}_e \{\left(\mathbb{A}_\beta\rho_e(t)\mathbb{A}^\dagger_\alpha-\rho_e(t)\mathbb{A}^\dagger_\alpha\mathbb{A}_\beta\right) c^\dagger_{\mathbf{k}l} c_{\mathbf{k+q}l'}\}\\
\end{split}
\end{equation}
For electron-phonon interaction, $\Gamma$ is
\begin{equation}\label{Gammaph}
\begin{split}
\Gamma_{\alpha\beta}(\Delta)&\approx \delta_{\mathbf{q}_\alpha g_\alpha,\mathbf{q}_\beta g_\beta} \pi [\overbrace{\delta(\Delta + \omega_{\mathbf{q}_\alpha g_\alpha}) N_{\mathbf{q}_\alpha g_\alpha}}^{absorption}]\\
&+\delta_{\mathbf{q}_\alpha g_\alpha,\mathbf{q}_\beta g_\beta} \pi [\overbrace{\delta(\Delta - \omega_{-\mathbf{q}_\alpha g_\alpha}) (N_{-\mathbf{q}_\alpha g_\alpha}+1)}^{emission}]\, .\\
\end{split}
\end{equation}\\
$N_{\mathbf{q}g}$ is the phonon occupation number for the mode in branch $g^{\text{th}}$ and with momentum $ \mathbf{q}$, which has energy $\hbar\omega_{\mathbf{q} g}$. To derive \eq{Gammaph}, the following formula has been used
\begin{equation}
\int_0^\infty e^{-i\epsilon s}ds=\pi\delta(\epsilon)-i\text{P}\frac{1}{\epsilon}=\pi\delta(\epsilon)-i\text{P}\frac{1}{\epsilon}\, ,
\end{equation}
where $\text{P}$ denotes the Cauchy principal value. Here, it leads to a negligible correction to the band structure due to scattering, called the Lamb shift \cite{breuer2002theory}.

$\Gamma$ for electron-ion interaction is
\begin{equation}\label{Gammaii}
\Gamma_{\alpha\beta}(\Delta)\approx \delta_{\mathbf{q}_\alpha,\mathbf{q}_\beta} \pi \delta(\Delta ),
\end{equation}
in which ionized-impurity scattering is assumed to be an elastic process. For the sake of unified notation, the fictitious frequency corresponding to ionized-impurity scattering is defined as $\omega_{\text{ii}}\equiv0$. The electron-ion interaction does not necessarily force  $\mathbf{q}_\alpha=\mathbf{q}_\beta$; however, the corresponding terms are not excited by the external field directly unless $\mathbf{q}_\alpha=\mathbf{q}_\beta$, so we only keep such terms. A similar result can be obtained upon spatial averaging over the random positions of impurity atoms  \cite{fischetti1999master,kohn1957quantum}.

By incorporating Eqs. (\ref{Gammaph}) and (\ref{Gammaii}) for $\Gamma_{\alpha\beta}(\cdot)$ into \eq{eom55}, we obtain

\begin{widetext}
\begin{equation}
\begin{split}\label{eom5}
&\frac{d \langle c^\dagger_{\mathbf{k}l} c_{\mathbf{k+q}l'}\rangle}{dt}=-i(\epsilon_{\mathbf{k+q}l'}-\epsilon_{\mathbf{k}l})\langle c^\dagger_{\mathbf{k}l} c_{\mathbf{k+q}l'}\rangle-i\langle \mathbf{k+q}l'| \mathbb {V}_{\text{SCF}}(t) | \mathbf{k}l\rangle (f_{\mathbf{k}l}-f_{\mathbf{k+q}l'})\\
&+\sum_{\alpha,\beta,\pm} \delta_{\mathbf{q}_\alpha g_\alpha,\mathbf{q}_\beta g_\beta} \pi {\delta(\Delta_\beta \pm \omega_{\mathbf{q}_\beta g_\beta}) (N_{\mathbf{q}_\beta g_\beta}+\frac{1}{2}\mp\frac{1}{2})}\mathcal{M}_{\text{ph},g_\beta}(\mathbf{q}_\beta)\mathcal{M}^*_{\text{ph},g_\alpha}(\mathbf{q_\alpha})(\mathbf{k_\beta+q_\beta}l'_\beta|\mathbf{k_\beta}l_\beta)(\mathbf{k_\alpha}l_\alpha|\mathbf{k_\alpha+q_\alpha}l'_\alpha)
\\
&\hspace{3.3 cm} \times \text{tr}_e \left\{\left ( c^\dagger_{\mathbf{k_\beta}l_\beta} c_{\mathbf{k_\beta+q_\beta}l'_\beta}\rho_e(t) c^\dagger_{\mathbf{k_\alpha+q_\alpha}l'_\alpha} c_{\mathbf{k_\alpha}l_\alpha} - c^\dagger_{\mathbf{k_\alpha+q_\alpha}l'_\alpha} c_{\mathbf{k_\alpha}l_\alpha} c^\dagger_{\mathbf{k_\beta}l_\beta} c_{\mathbf{k_\beta+q_\beta}l'_\beta}\rho_e(t)\right) c^\dagger_{\mathbf{k}l} c_{\mathbf{k+q}l'}\right\}\\
&+\sum_{\alpha,\beta,\pm} \delta_{\mathbf{q}_\alpha g_\alpha,\mathbf{q}_\beta g_\beta} \pi {\delta(\Delta_\alpha \pm \omega_{\mathbf{q}_\alpha g_\alpha}) (N_{\mathbf{q}_\alpha g_\alpha}+\frac{1}{2}\mp\frac{1}{2})}\mathcal{M}_{\text{ph},g_\beta}(\mathbf{q}_\beta)\mathcal{M}^*_{\text{ph},g_\alpha}(\mathbf{q_\alpha})(\mathbf{k_\beta+q_\beta}l'_\beta|\mathbf{k_\beta}l_\beta)(\mathbf{k_\alpha}l_\alpha|\mathbf{k_\alpha+q_\alpha}l'_\alpha)
\\
&\hspace{3.3 cm} \times \text{tr}_e \left\{\left ( c^\dagger_{\mathbf{k_\beta}l_\beta} c_{\mathbf{k_\beta+q_\beta}l'_\beta}\rho_e(t) c^\dagger_{\mathbf{k_\alpha+q_\alpha}l'_\alpha} c_{\mathbf{k_\alpha}l_\alpha} - \rho_e(t)c^\dagger_{\mathbf{k_\alpha+q_\alpha}l'_\alpha} c_{\mathbf{k_\alpha}l_\alpha} c^\dagger_{\mathbf{k_\beta}l_\beta} c_{\mathbf{k_\beta+q_\beta}l'_\beta}\right) c^\dagger_{\mathbf{k}l} c_{\mathbf{k+q}l'}\right\}\\
&+\sum_{\alpha,\beta} \delta_{\mathbf{q}_\alpha ,\mathbf{q}_\beta } \pi {\delta(\Delta_\beta) }\mathcal{M}_{\text{ii}}(\mathbf{q}_\beta)\mathcal{M}^*_{\text{ii}}(\mathbf{q_\alpha})(\mathbf{k_\beta+q_\beta}l'_\beta|\mathbf{k_\beta}l_\beta)(\mathbf{k_\alpha}l_\alpha|\mathbf{k_\alpha+q_\alpha}l'_\alpha)
\\
&\hspace{3.3 cm} \times \text{tr}_e \left\{\left ( c^\dagger_{\mathbf{k_\beta}l_\beta} c_{\mathbf{k_\beta+q_\beta}l'_\beta}\rho_e(t) c^\dagger_{\mathbf{k_\alpha+q_\alpha}l'_\alpha} c_{\mathbf{k_\alpha}l_\alpha} - c^\dagger_{\mathbf{k_\alpha+q_\alpha}l'_\alpha} c_{\mathbf{k_\alpha}l_\alpha} c^\dagger_{\mathbf{k_\beta}l_\beta} c_{\mathbf{k_\beta+q_\beta}l'_\beta}\rho_e(t)\right) c^\dagger_{\mathbf{k}l} c_{\mathbf{k+q}l'}\right\}\\
&+\sum_{\alpha,\beta} \delta_{\mathbf{q}_\alpha ,\mathbf{q}_\beta } \pi {\delta(\Delta_\alpha) }\mathcal{M}_{\text{ii}}(\mathbf{q}_\beta)\mathcal{M}^*_{\text{ii}}(\mathbf{q_\alpha})(\mathbf{k_\beta+q_\beta}l'_\beta|\mathbf{k_\beta}l_\beta)(\mathbf{k_\alpha}l_\alpha|\mathbf{k_\alpha+q_\alpha}l'_\alpha)
\\
&\hspace{3.3 cm} \times \text{tr}_e \left\{\left ( c^\dagger_{\mathbf{k_\beta}l_\beta} c_{\mathbf{k_\beta+q_\beta}l'_\beta}\rho_e(t) c^\dagger_{\mathbf{k_\alpha+q_\alpha}l'_\alpha} c_{\mathbf{k_\alpha}l_\alpha} - \rho_e(t)c^\dagger_{\mathbf{k_\alpha+q_\alpha}l'_\alpha} c_{\mathbf{k_\alpha}l_\alpha} c^\dagger_{\mathbf{k_\beta}l_\beta} c_{\mathbf{k_\beta+q_\beta}l'_\beta}\right) c^\dagger_{\mathbf{k}l} c_{\mathbf{k+q}l'}\right\}.\\
\end{split}
\end{equation}
\end{widetext}
It is worth remembering that $\Delta_\alpha=\epsilon_{\mathbf{k_\alpha}l_\alpha}-\epsilon_{\mathbf{k_\alpha+q_\alpha}l'_\alpha}$. The Markovian master equation (\ref{eom5}) is of the Lindblad form and preserves the positivity of the density matrix \cite{breuer2002theory}; removing terms would violate this important constraint \cite{fischetti1999master,buecking2007theory}. By employing \eq{Wick6}, changing variables, and seeking a time-harmonic solution $\left(\frac{\partial}{\partial t}\rightarrow -i\omega\right)$, the equation of motion for $\langle c^\dagger_{\mathbf{k}l} c_{\mathbf{k+q}l'}\rangle$ becomes
\begin{widetext}
\begin{equation}\label{eom6}
\begin{split}
&-i(\epsilon_{\mathbf{k}l}-\epsilon_{\mathbf{k+q}l'}+\omega) \langle c^\dagger_{\mathbf{k}l} c_{\mathbf{k+q}l'}\rangle\\
=&-i(f_{\mathbf{k}l}-f_{\mathbf{k+q}l'})\langle\mathbf{k+q}l'|\mathbb {V}_{\text{SCF}}(\omega)|\mathbf{k}l\rangle\\
&+\hspace{-0.3 cm}\sum_{\mathbf{ k'} m m' g,\pm}\hspace{-0.1 cm}\pi \delta(\epsilon_{\mathbf{k'+q}m'}-\epsilon_{\mathbf{k'}m}\pm  \omega_g) \left(\mp\Delta\mathfrak{W}_{\mathbf{q},v} \right)
(f_{\mathbf{k+q}l'}-f_{\mathbf{k}l}) (\mathbf{k'} m|\mathbf{k'+q} m' )(\mathbf{k+q}l'|\mathbf{k}l) \langle c^\dagger_{\mathbf{k'}m} c_{\mathbf{k'+q}m'} \rangle\\
&+\hspace{-0.3 cm}\sum_{\mathbf{ k'} m m' g,\pm}\hspace{-0.1 cm}\pi \delta(\epsilon_{\mathbf{k'}m}-\epsilon_{\mathbf{k}l}\pm  \omega_g) \left(\mathfrak{W}^\pm_{\mathbf{k'-k},v} \pm \Delta\mathfrak{W}_{\mathbf{k'-k},v}f_{\mathbf{k}l}\right)
(\mathbf{k+q}l'|\mathbf{k'+q}m') (\mathbf{k'}m|\mathbf{k}l)
\langle c^\dagger_{\mathbf{k'}m} c_{\mathbf{k'+q}m'} \rangle\\
&+\hspace{-0.3 cm}\sum_{\mathbf{ k'} m m' g,\pm}\hspace{-0.1 cm}\pi \delta(\epsilon_{\mathbf{k'+q}m'}-\epsilon_{\mathbf{k+q}l'}\pm  \omega_g) \left(\mathfrak{W}^\pm_{\mathbf{k'-k},v}\pm\Delta\mathfrak{W}_{\mathbf{k'-k},v}f_{\mathbf{k+q}l'}\right)
(\mathbf{k+q}l'|\mathbf{k'+q}m')(\mathbf{k'}m|\mathbf{k}l)
\langle c^\dagger_{{\mathbf{k'}m}} c_{\mathbf{k'+q}m'} \rangle\\
&-\hspace{-0.3 cm}\sum_{\mathbf{ k'} m m' g,\pm}\hspace{-0.1 cm}\pi\delta(\epsilon_{\mathbf{k'}m'}-\epsilon_{\mathbf{k}m}\pm  \omega_g) \left(\mathfrak{W}^\mp_{\mathbf{k'-k},v}\mp \Delta\mathfrak{W}_{\mathbf{k'-k},v}f_{\mathbf{k'}m'}\right)
(\mathbf{k}m|\mathbf{k'}m')(\mathbf{k'}m'|\mathbf{k}l)
\langle c^\dagger_{\mathbf{k}m}  c_{\mathbf{k+q}l'}\rangle\\
&-\hspace{-0.3 cm}\sum_{\mathbf{ k'} m m' g,\pm}\hspace{-0.1 cm}\pi\delta(\epsilon_{\mathbf{k'+q}m}-\epsilon_{\mathbf{k+q}m'}\pm  \omega_g) \left(\mathfrak{W}^\mp_{\mathbf{k'-k},v}\mp\Delta\mathfrak{W}_{\mathbf{k'-k},v}f_{\mathbf{k'+q}m}\right)
 (\mathbf{k+q}l'|\mathbf{k'+q}m)(\mathbf{k'+q}m|\mathbf{k+q}m')
  \langle c^\dagger_{\mathbf{k}l} c_{\mathbf{k+q}m'} \rangle.\\
\end{split}
\end{equation}
\end{widetext}

To simplify the notation, we defined the scattering weights $\mathfrak{W^+}$ and $\mathfrak{W^-}$, which correspond to the absorption and emission processes, respectively. Also, $\Delta \mathfrak{W}_{\mathbf{k-k'},v}\equiv\mathfrak{W}^{-}_{\mathbf{k-k'},v}-\mathfrak{W}^{+}_{\mathbf{k-k'},v}$. Details for each scattering mechanism are provided in Appendix \ref{app3}. It should be mentioned that the first sum is negligible, because the Dirac delta function in it implicitly forces a momentum conservation and, only insignificant number of transitions may satisfy both the energy conservation
and the momentum conservation.

Equation (\ref{eom6}) is the central equation of this paper. It captures the temporal variation of the coherences due to a harmonic field, and carries information about dissipation mechanisms, as well as the band structure and the Bloch-wave  overlaps.

\subsection{Numerical implementation of the SCF-MMEF for graphene}\label{numerics}

We now solve \eq{eom6} for graphene. For simplicity, we use the first-nearest-neighbor tight binding with carbon $p_z$ orbitals to obtain the band structure and Bloch waves \cite{sarma2011electronic}. The overlap integrals are $(\mathbf{k}l |\mathbf{k'}l')\approx\frac{ 1}{2}(1+l'l e^{i\xi_{\mathbf{k',k}}})$, where $\xi_{\mathbf{k',k}}=\arg\left[(k_x'+ik_y')(k_x-ik_y)\right]$. In evaluating the energy terms in \eq{eom6}, the energy dispersion near the Dirac point is approximated as linear and isotropic $\epsilon_{\mathbf{k}l}=l v_F |\mathbf{k}| $, where $l=\pm1$ denotes the valence (-1) and conduction (+1) bands.

The Brillouin zone is gridded up in the polar coordinates, with $N_k$ points in the radial direction and $N_\theta$ in the azimuthal direction. With the aid of this discretization, Eq. (\ref{eom6}) is written in the matrix form and solved numerically for $\mathcal{X}$:

\begin{equation}\label{matrix_eq}
\mathcal{E X=F}+i\mathcal{(R-R'-R'')X}.
\end{equation}

Each pair $(\mathbf{q},\omega)$ results in its own Eq. (\ref{matrix_eq}). The set of variables $\{\mathbf{k}l'l\}$ corresponds to a position in the column $\mathcal{X}$. The matrices and vectors in \eq{matrix_eq} are defined as

\begin{subequations}\label{matrices}
\begin{equation}
\begin{split}
\mathcal{E} _{\{\mathbf{k}l'l\} \{\mathbf{k'}m'm\}}=&\delta_{\{\mathbf{k}l'l\} \{\mathbf{k'}m'm\}}  (\epsilon_{\mathbf{k} l}-\epsilon_{\mathbf{k+q}l'}+\omega),\\
\end{split}
\end{equation}
\begin{equation}
\begin{split}
\mathcal{X} _{\{\mathbf{k}l'l\}}=&\langle c^\dagger_{\mathbf{k} l} c_{\mathbf{k +q }l' }\rangle,\\
\end{split}
\end{equation}
\begin{equation}
\begin{split}
\mathcal{F} _{\{\mathbf{k}l'l\}}=&(f_{\mathbf{k}l}-f_{\mathbf{k+q }l'})(\mathbf{k+q }l'|\mathbf{k}l),\\
\end{split}
\end{equation}
\begin{widetext}
\begin{equation}
\begin{split}
\hspace{-4 cm} \mathcal{R} _{\{\mathbf{k}l'l\} \{\mathbf{k'}m'm\}}=
& \hspace{-0.6 cm} \sum_{\begin{array}{c}{\scriptstyle g}\vspace{-0.15cm}\\ {\scriptstyle\epsilon_{\mathbf{k'+q}m'}=\epsilon_{\mathbf{k+q}l'} \mp \omega_{g}}\end{array}}\hspace{-0.7 cm}
\frac{A}{2 N_\theta}\frac{|\mathbf{k'+q}|}{v_F}
[\mathfrak{W}^{\pm}_{\mathbf{k-k'},g}\pm\Delta \mathfrak{W}_{\mathbf{k-k'},g}f_{\mathbf{k+q}l'}]
(\mathbf{k+q}l'|\mathbf{k'+q}m')(\mathbf{k'}m|\mathbf{k}l)+\\
&\hspace{-0.2 cm} \sum_{\begin{array}{c}{\scriptstyle g}\vspace{-0.15cm}\\ {\scriptstyle\epsilon_{\mathbf{k'}m}=\epsilon_{\mathbf{k}l} \mp \omega_{g}}\end{array}}\hspace{-0.3 cm}
\frac{A}{2 N_\theta}\frac{|\mathbf{k'}|}{v_F}
[\mathfrak{W}^{\pm}_{\mathbf{k-k'},g}\pm\Delta \mathfrak{W}_{\mathbf{k-k'},g}f_{\mathbf{k}l}]
(\mathbf{k+q}l'|\mathbf{k'+q}m')(\mathbf{k'}m|\mathbf{k}l),\\
\end{split}
\end{equation}
\begin{equation}
\begin{split}
\mathcal{R'} _{\{\mathbf{k}l'l\} \{\mathbf{k}m'l\}}=&\hspace{-0.6 cm} \sum_{\begin{array}{c}{\scriptstyle \mathbf{k'}m g}\vspace{-0.15cm}\\ {\scriptstyle\epsilon_{\mathbf{k'+q}m} =\epsilon_{\mathbf{k+q}m'} \pm \omega_{g}}\end{array}}\hspace{-0.7 cm}
\frac{A}{2N_\theta}\frac{|\mathbf{k'+q}|}{v_F}
[\mathfrak{W}^{\pm}_{\mathbf{k-k'},g}\pm\Delta \mathfrak{W}_{\mathbf{k-k'},g}f_{\mathbf{k'+q}m}]
(\mathbf{k+q}l'|\mathbf{k'+q}m)(\mathbf{k'+q}m|\mathbf{k+q}m'),\\
\end{split}
\end{equation}
\begin{equation}
\begin{split}
\hspace{-4.5 cm} \mathcal{R''} _{\{\mathbf{k}l'l\} \{\mathbf{k}l'm\}}=&\hspace{-0.2 cm} \sum_{\begin{array}{c}{\scriptstyle \mathbf{k'}m'g}\vspace{-0.15cm}\\ {\scriptstyle\epsilon_{\mathbf{k'}m'}=\epsilon_{\mathbf{k}m}\pm \omega_{g}}\end{array}} \hspace{-0.5 cm}
\frac{A}{2 N_\theta}\frac{|\mathbf{k'}|}{v_F}
[\mathfrak{W}^{\pm}_{\mathbf{k-k'},g}\pm\Delta \mathfrak{W}_{\mathbf{k-k'},g}f_{\mathbf{k'}m'}]
(\mathbf{k}m|\mathbf{k'}m')(\mathbf{k'}m'|\mathbf{k}l).\\
\end{split}
\end{equation}
\end{widetext}
\end{subequations}

\noindent Equation (\ref{matrix_eq}) is solved for $\mathcal{X}$ for every $(\mathbf{q},\omega)$. With the typical values we used ($N_\theta=12$ or 14 and  $N_k$ between 100 and 200) the square matrices in \eq{matrix_eq} are between 4800$\times$4800 to 11200$\times$ 11200 in size.

We introduce a vector $\mathcal{C}$ as $\mathcal{C}_{\{\mathbf{k}l'l\}}=  (\mathbf{k}l |\mathbf{k+q}l')$, which helps calculate surface polarization as
\begin{equation}\label{polarization}
P_s(\mathbf{q},\omega)= \frac{-e}{A} \mathcal{C}^T\mathcal{X}.
\end{equation}
$P_s$ is then used in Eqs.(\ref{eps}) and (\ref{sigma}) to calculate the dielectric function and the surface conductivity, respectively. For a quasi-one dimensional system, \eq{eom6} can be written in a matrix representation similar to  \eq{matrix_eq}; see Appendix \ref{app} for details.

\section{Results}\label{sect3}

\subsection{\label{sect3a}Complex conductivity of graphene}

\begin{figure}
\includegraphics[width=\columnwidth]{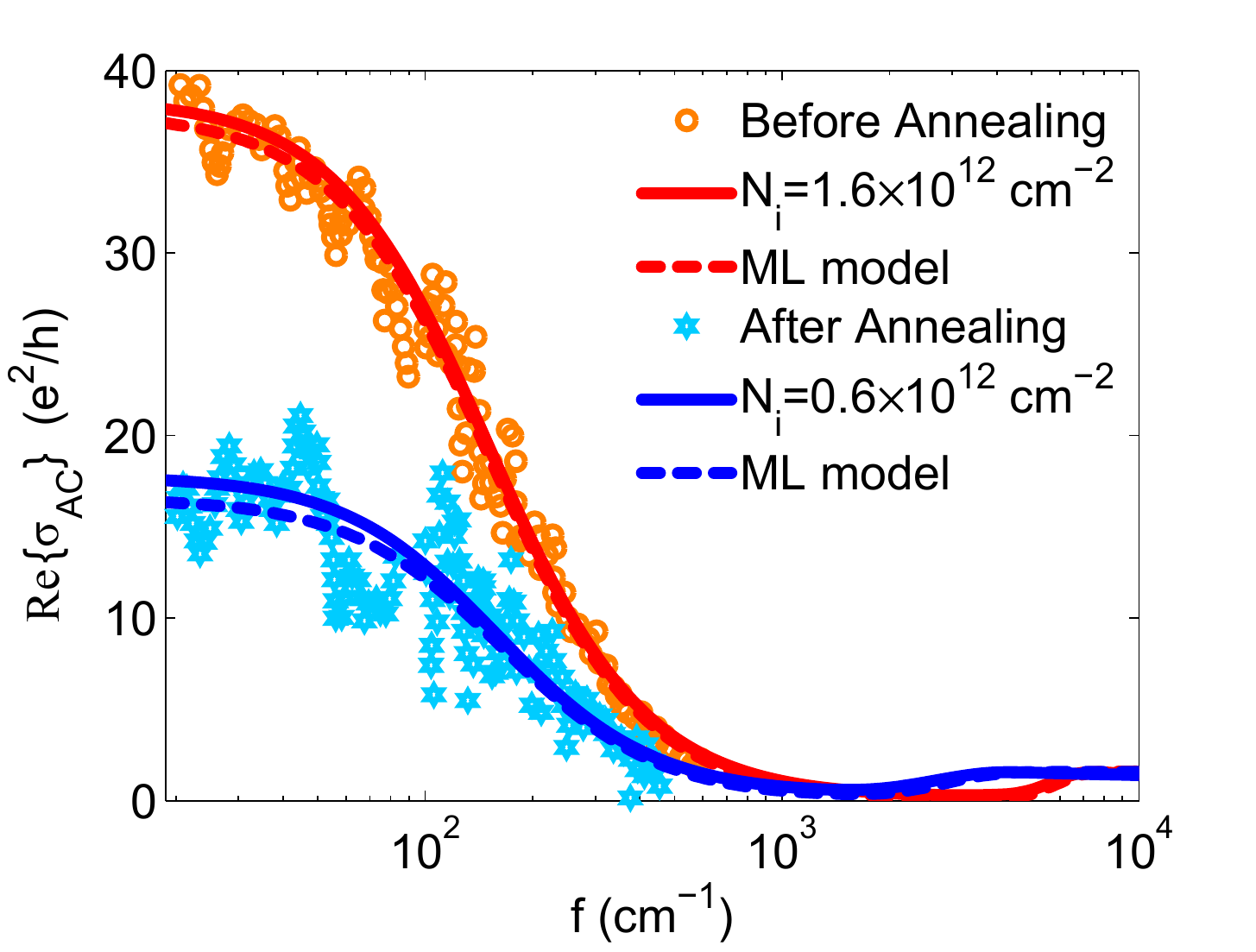}
\caption{\label{fig:sigmaAC}  Real part of $\sigma(\omega)$ as a function of frequency. Experimental results from \cite{ren2012terahertz} are presented with symbols. Theoretical calculations are based on the SCF-MMEF (solid lines) and the ML approach (dashed lines). The two distinct sets of data correspond to before annealing ($n=9.35\times 10^{12} \mathrm{cm}^{-2}$ obtained in experiment and also used in calculations; $N_\text{i}=1.6\times 10^{12}$cm$^{-2}$ used in the calculation to obtain the best fit) and after annealing ($n=2.28\times 10^{12} \mathrm{cm}^{-2}$, $N_\text{i}=0.6\times 10^{12}$cm$^{-2}$). }
\end{figure}

From the polarization, \eq{polarization}, the electrical and optical properties of the graphene sheet are calculated. In Fig. \ref{fig:sigmaAC}, we show the frequency dependence of the real part of the graphene AC conductivity,   $\sigma(\omega)$, as calculated via the SCF-MMEF. The calculation shows excellent agreement with experimental results \cite{ren2012terahertz}. The symbols are the experimental results for graphene on SiO$_2$, for two samples: (circles) before annealing ($n=9.35\times 10^{12} \mathrm{cm}^{-2}$, i.e., $\epsilon_F=354 $ meV) and (stars) after annealing ($n=2.28\times 10^{12} \mathrm{cm}^{-2}$, i.e., $\epsilon_F=170 $ meV). In order to reproduce the measurements, the only variable parameter is the sheet impurity density. We assume that impurities are spread uniformly across a sheet placed 4\AA~ below the interface of graphene and the substrate. The sheet impurity densities of $N_\text{i}=1.6\times 10^{12}$cm$^{-2}$ (before annealing) and $N_\text{i}=0.6\times 10^{12}$cm$^{-2}$ (after annealing) yield the best agreement with experimental data. The obtained impurity density for the after-annealing case is close to the one calculated by the EMC/FDTD/MD method with clustered impurities \cite{sule2014terahertz}. At high enough frequencies (almost twice the Fermi frequency), it can be seen in Fig. \ref{fig:sigmaAC} that the interband conductivity emerges, as expected.

We also calculated the mobility and the corresponding relaxation time for each case, based on the SCF-MMEF, and used them in the Mermin-Lindhard (ML) model \cite{jablan2009plasmonics}. The results are plotted in Fig. \ref{fig:sigmaAC} (dashed lines) for comparison. Even with an appropriate relaxation time, the ML model slightly underestimates the conductivity at lower frequencies.

\begin{figure}
\includegraphics[width=\columnwidth]{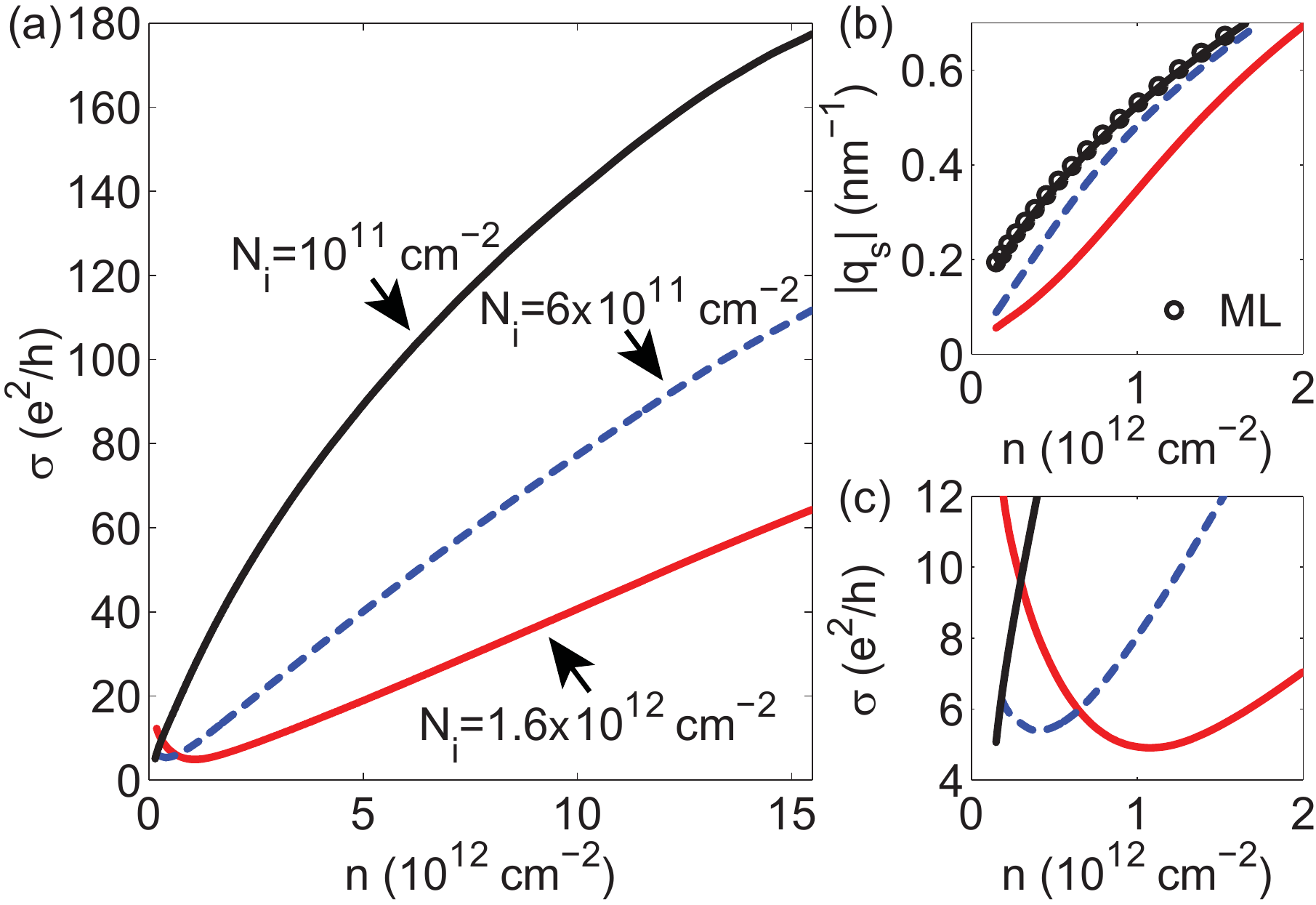}
\caption{\label{fig:sigmaDC}(a) Real part of $\sigma_{DC}$ as a function of the sheet carrier density. For low impurity density, the sublinearity of DC conductivity at high carrier densities can be seen. (b) The absolute value of the screening wave vector as a function of carrier density. The screening wave numbers calculated via the SCF-MMEF are compared with Mermin-Lindhard results. (c)  Real part of $\sigma_{DC}$ at low carrier densities.}
\end{figure}

In Fig.~\ref{fig:sigmaDC}(a),  the real part of the DC conductivity of graphene, $\sigma_{DC}$, is plotted as a function of carrier density for three different values of the impurity density. As expected from experiments, for low impurity density and high carrier density, the sublinear relation between  $\sigma_{DC}$ and carrier density becomes pronounced. The RTA provides a qualitative explanation for this behavior. Within the RTA, $ \sigma_{DC} \approx \frac{e^2 \epsilon_F}{\pi \hbar^2}\tau$ \cite{sarma2011electronic}, and the relaxation time due to intrinsic phonon scattering is inversely proportional to the Fermi energy \cite{sule2012phonon}, therefore at high-enough carrier densities (equivalently at high Fermi levels) phonon scattering dominates over ionized-impurity scattering and $\sigma_{DC}$  gradually becomes less dependent on carrier density. Also, at low carrier densities [Fig. \ref{fig:sigmaDC}(c)], the conductivity does not drop below $4e^2/h$ \cite{sarma2011electronic}. However, these results are obtained for the case of uniformly distributed impurities, whereas it is known that the impurity distribution in realistic samples is nonuniform \cite{sule2014clustered}. The nonuniformity causes the formation of electron-hole puddles at low carrier densities, which significantly affects carrier transport but is not captured in our model. A realistic model for nonuniform distribution of impurities is provided in Ref. \cite{sule2014clustered}.

We also calculate the screening wave number, defined as $q_s=\lim_{q\rightarrow0}~q\left[\varepsilon(\omega=0,\mathbf{q})-1\right]$. Figure \ref{fig:sigmaDC}(b) shows that, for carrier densities comparable to or lower than the impurity density, the screening wave number falls off. However, the screening wave number calculated via the ML model is unable to show this phenomenon and is instead independent of the impurity density.

\subsection{\label{sect3b}Dielectric function of graphene}

In order to study plasmons in graphene, the dielectric function $\varepsilon (\mathbf{q},\omega)$, \eq{eps}, should be calculated. In the case of graphene on a polar substrate, it has been shown that plasmons and SO phonons couple with each other  \cite{lu2009plasmon,hwang2010plasmon,hwang2013surface,ahn2014inelastic,jablan2011unconventional,ong2012theory,brar2014hybrid,woessner2014highly,yan2013damping}. In order to take this phenomenon into account, \eq{eps} should be modified for polar substrates as \cite{hwang2010plasmon}
\begin{equation}
\begin{split}
\varepsilon (\mathbf{q},\omega)=1+&\frac{e}{\varepsilon_b(\omega)\varepsilon_0}\frac{P(\mathbf{q},\omega)}{2q}\\
-&\frac{\sum_j \frac{\tilde{\varepsilon}}{2} e^{-qd} \omega_{\text{SO},j} \mathcal{D}^0_j(\omega)}{1+\sum_j \frac{\tilde{\varepsilon}}{2} e^{-qd} \omega_{\text{SO},j} \mathcal{D}^0_j(\omega)}\, ,\\
\end{split}
\end{equation}
where $\omega_{\mathrm{SO},j}$ is the $j^{\mathrm{th}}$ SO phonon mode, and $\tilde{\varepsilon}=\varepsilon_{s}^\infty\left(\frac{1}{\varepsilon_{s}^\infty+1}-\frac{1}{\varepsilon_{s}^0+1}\right)$, $\varepsilon_{ox}^\infty$ and $\varepsilon_{s}^0$ are the high-frequency and low-frequency permittivity of the substrate, respectively. $d$ is the distance between the graphene sheet and the substrate, which in our calculations is assumed to be 4\AA. In the above equation, $\mathcal{D}^0_j(\omega)$ is the free phonon Green's function
\begin{equation}
\mathcal{D}^0_j(\omega) = \frac{2\omega_{SO,j}}{(\omega+i \tau^{-1}_j)^2-\omega_{SO,j}^2},
\end{equation}
where $\tau_j$ is the relaxation time corresponding to the $j^{\mathrm{th}}$ SO phonon mode. Further information on the surface phonon modes is provided in Appendix \ref{app2}.

Figure~\ref{fig:eps} shows the real and imaginary parts of the graphene dielectric function with the carrier density of $n=3 \times 10^{12} $ cm$^{-2}$ ($\epsilon_F=197$ meV) on SiO$_2$ and DLC. The impurity density of graphene on SiO$_2$ is $N_\text{i}$=$4\times10^{11}$ cm$^{-2}$, which corresponds to the electron mobility of 2500 $ \frac{\text{cm}^2}{\text{Vs}}$, as measured in \cite{gannett2011boron}. The impurity density of graphene on DLC is chosen to be $N_\text{i}$=$4.15\times10^{11}$ cm$^{-2}$, which yields the electron mobility of 3000 $ \frac{\text{cm}^2}{\text{Vs}}$, as reported in \cite{wu2012state}. The effect of SO phonons can be easily seen by comparing the dielectric function of graphene on DLC and SiO$_2$ in Fig. \ref{fig:eps}.


\begin{figure}
\includegraphics[width=\columnwidth]{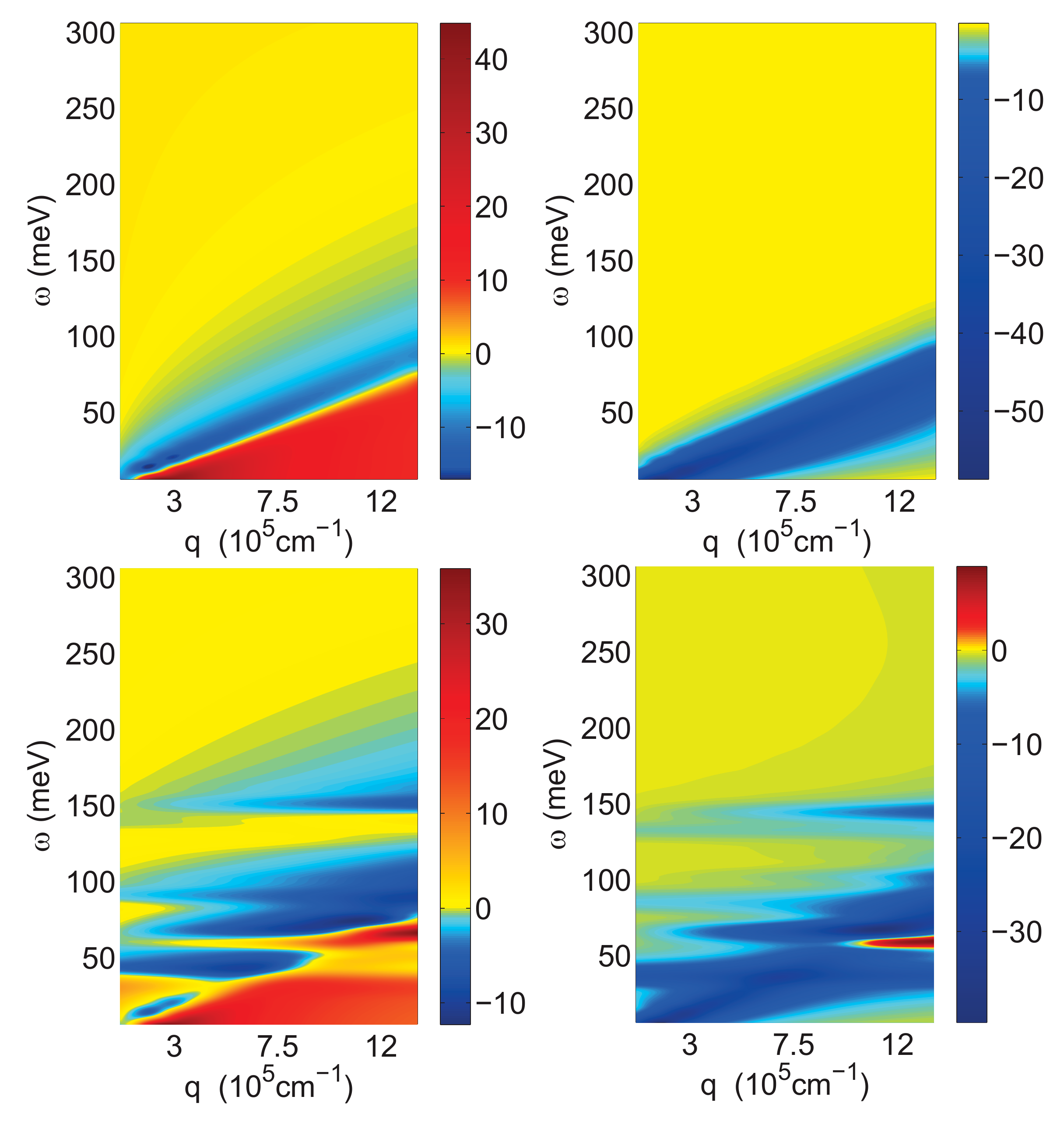}
\caption{\label{fig:eps} Real part (left) and imaginary part (right) of the dielectric function of graphene on DLC (top row) and SiO$_2$ (bottom row). The electron density is $n=3\times 10^{12}$cm$^{-2}$ in both cases. Graphene on DLC has a substrate impurity density of $N_\text{i}$=$4.15\times10^{11}$cm$^{-2}$, equivalent to the electron mobility of 3000$\frac{\text{cm}^2}{\text{Vs}}$ \cite{wu2012state}. Graphene on SiO$_2$  has a substrate impurity density of $N_\text{i}$=$4\times10^{11}$cm$^{-2}$, equivalent to the electron mobility of 2500$\frac{\text{cm}^2}{\text{Vs}}$ \cite{gannett2011boron}.}
\end{figure}

\subsection{\label{sect3c}Plasmons in graphene}

The zeros of the real part of the dielectric function give  an estimate of the plasmon dispersion [Fig.~\ref{fig:eps}], but in the case of graphene on a polar  substrate this approach becomes impractical. A more accurate method is to seek the maxima of the loss function, which is proportional to  $-\Im\{\frac{1}{\varepsilon (\mathbf{q},\omega)}\}$ and can be directly measured. \textcolor{black}{As scattering strongly affects the imaginary part of the dielectric function (without scattering, $\Im\{{\varepsilon (\mathbf{q},\omega)}\}=0$), the loss function is also a sensitive probe  for the role of dissipation. In an ideal, dissipation-free electronic system, the loss function would peak to infinity at plasmon resonances ($\Re\{{\varepsilon (\mathbf{q},\omega)}\}=0$); in realistic dissipative systems, the loss function has finite peaks at plasmon resonances, which higher peaks corresponding to lower plasmon damping.}


\begin{figure*}
\includegraphics[width=\textwidth]{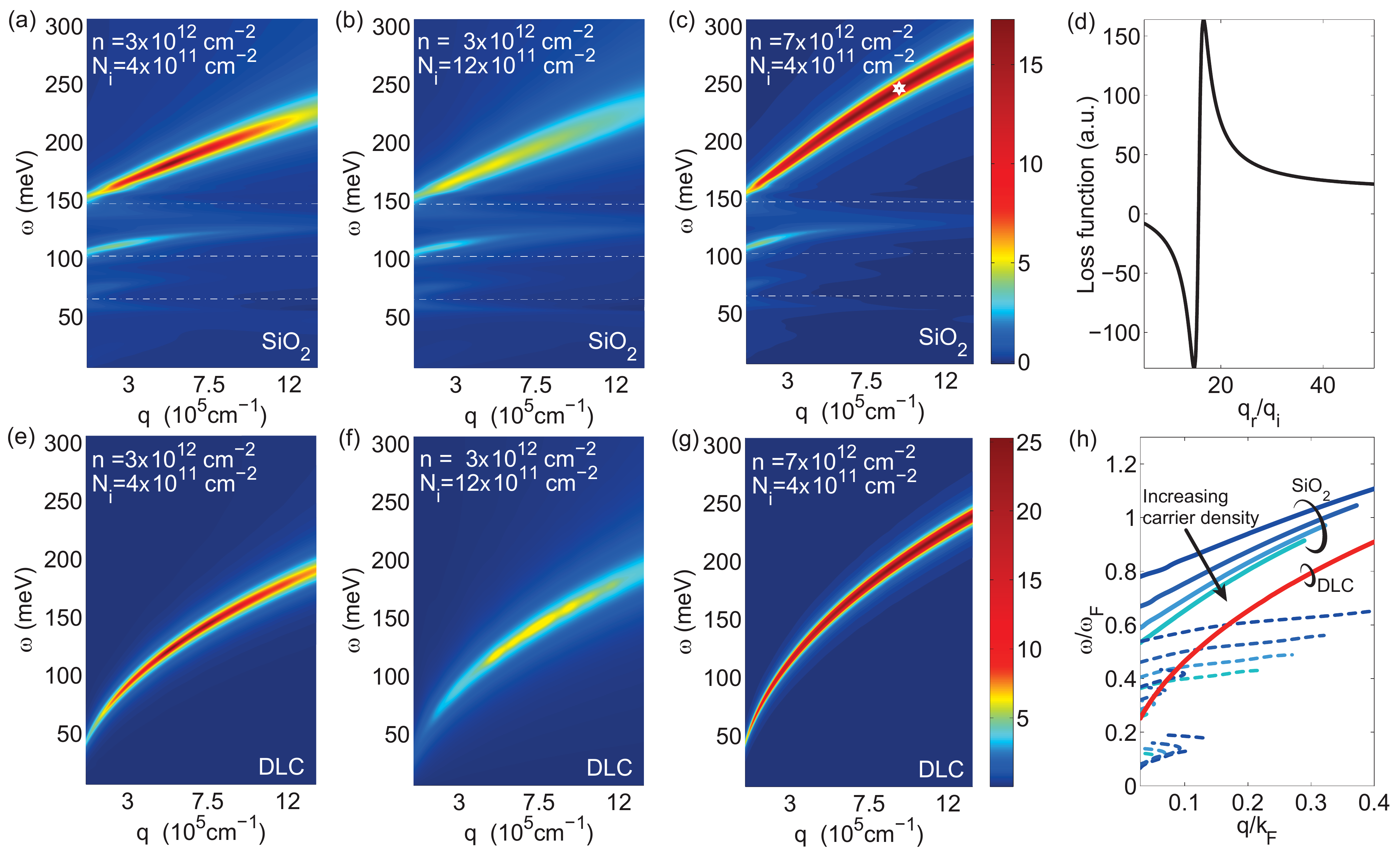}
\caption{\label{fig:3loss_func} (a)--(c) The loss function of graphene on SiO$_2$ (represented via color in arbitrary units; all colorbars have the same scale) as a function of frequency and wave vector for (a) carrier density $n=3 \times 10^{12} $ cm$^{-2}$ and impurity density $N_\text{i}$=$4\times10^{11}$ cm$^{-2}$; (b) $n=3 \times10^{12} $ cm$^{-2}$ and $N_\text{i}$=$1.2\times10^{12}$cm$^{-2}$; (c) $n=7 \times 10^{12} $ cm$^{-2}$ and $N_\text{i}$=$4\times10^{11}$cm$^{-2}$. The dash-dot lines depict the  energies of the SO phonons, which equal 65 meV, 102 meV, and 147 meV. (d) The loss function, in arbitrary units, as a function of $q_r/q_i$ in the vicinity of the star-marked point in panel (c). \color{black}(e)--(g) The loss function of graphene on DLC (represented via color in arbitrary units; all color bars have the same scale) as a function of frequency and wave vector for (e) carrier density $n=3 \times 10^{12} $ cm$^{-2}$ and impurity density $N_\text{i}$=$4\times10^{11}$ cm$^{-2}$; (f) $n=3 \times10^{12} $ cm$^{-2}$ and $N_\text{i}$=$1.2\times10^{12}$cm$^{-2}$; (g) $n=7 \times 10^{12} $ cm$^{-2}$ and $N_\text{i}$=$4\times10^{11}$cm$^{-2}$. (h) The plasmon dispersions for graphene on SiO$_2$ (blue) and DLC (red) for different carrier densities in terms of the scaled wave vector ($q/q_F$) and frequency ($\omega/\omega_F$). On DLC, the plasmon dispersions for different carrier densities coincide (the red curve).}
\end{figure*}

Figures \ref{fig:3loss_func}(a)-(c) show the loss function of graphene on SiO$_2$ with different carrier densities and different impurity densities. Plasmon modes below the highest SO-phonon mode of the substrate (147 meV) are suppressed. Increasing the impurity density (or, equivalently, decreasing the electron mobility) does not change the plasmon dispersion and only enhances  plasmon damping [compare Figs. \ref{fig:3loss_func}(a) and \ref{fig:3loss_func}(b)]. However, decreasing the carrier density (or, equivalently, increasing the Fermi level) not only raises plasmon damping, but also pushes the plasmon dispersion towards higher wave vectors, and both phenomena result in a decreased plasmon propagation length [compare Figs. \ref{fig:3loss_func}(a) and \ref{fig:3loss_func}(c)].

\textcolor{black}{Overall, for graphene on DLC, all dispersions for different carrier and impurity densities, when presented in scaled quantities $q/q_F$ and $\omega/\omega_F$, reduce to a single curve [the red curve in Fig. \ref{fig:3loss_func}(h)]. This dispersion curve is the same one that would be obtained with scattering-free RPA approaches \cite{hwang2007dielectric,barlas1,barlas2,barlas3,barlas4,barlasPhD,Wunsch_NJP06}. For graphene on the polar SiO$_2$ [Fig. \ref{fig:3loss_func}, panels (a)--(c)], we see the effect of SO phonons on plasmon dispersions. There are four dispersion branches associated with interface plasmon-phonon (IPP) modes \cite{ong2012theory}, the hybrid modes that stem from the coupling of plasmons with SO phonons. The three nearly dispersionless low-energy curves are SO-like [also seen in Fig. \ref{fig:3loss_func}(h)], while the highest-energy curve is a plasmonlike IPP branch, in line with Refs. \cite{lu2009plasmon,hwang2010plasmon,hwang2013surface,ahn2014inelastic,jablan2011unconventional,ong2012theory}.}

\begin{figure}
\includegraphics[width=\columnwidth]{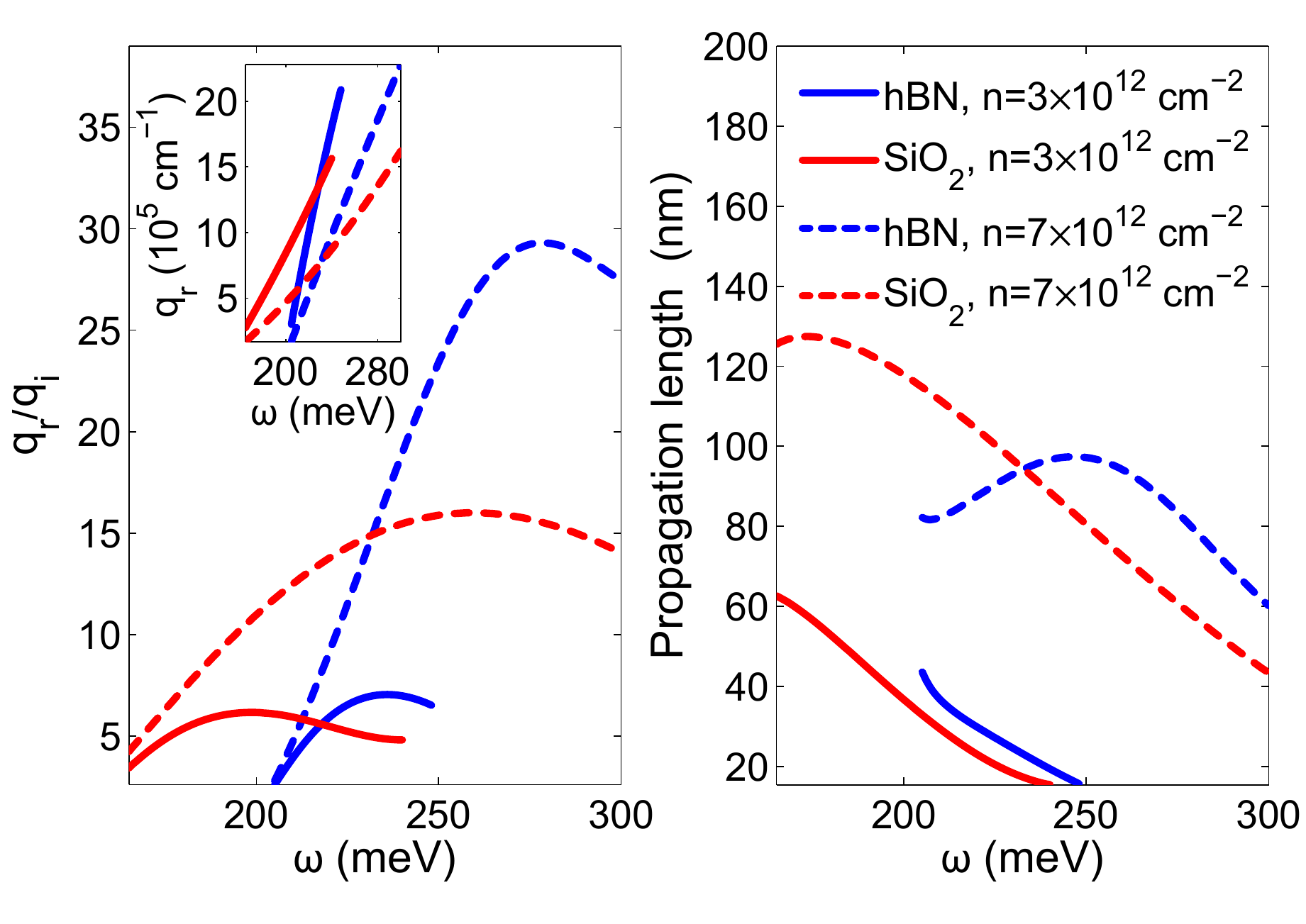}
\caption{\label{fig:2subs}(Left) The normalized plasmon propagation length ($q_r/q_i$) as a function of frequency for graphene with two different carrier densities [$n=3\times 10^{12} \mathrm{cm}^{-2}$ (solid) and $n=7\times 10^{12} \mathrm{cm}^{-2}$ (dashed)] and on two different substrates: SiO$_2$ (red) and hBN (blue). The impurity density in each sample is chosen in a way that yields the measured electron mobility of 2500 $\frac{\text{cm}^2}{\text{Vs}}$ for SiO$_2$ and 9900 $\frac{\text{cm}^2}{\text{Vs}}$ for hBN at a carrier density of $n=3\times 10^{12} \mathrm{cm}^{-2}$ \cite{gannett2011boron}. Inset: Plasmon dispersions. (Right) Plasmon propagation length as a function of frequency. }
\end{figure}

\begin{figure}
\includegraphics[width=\columnwidth]{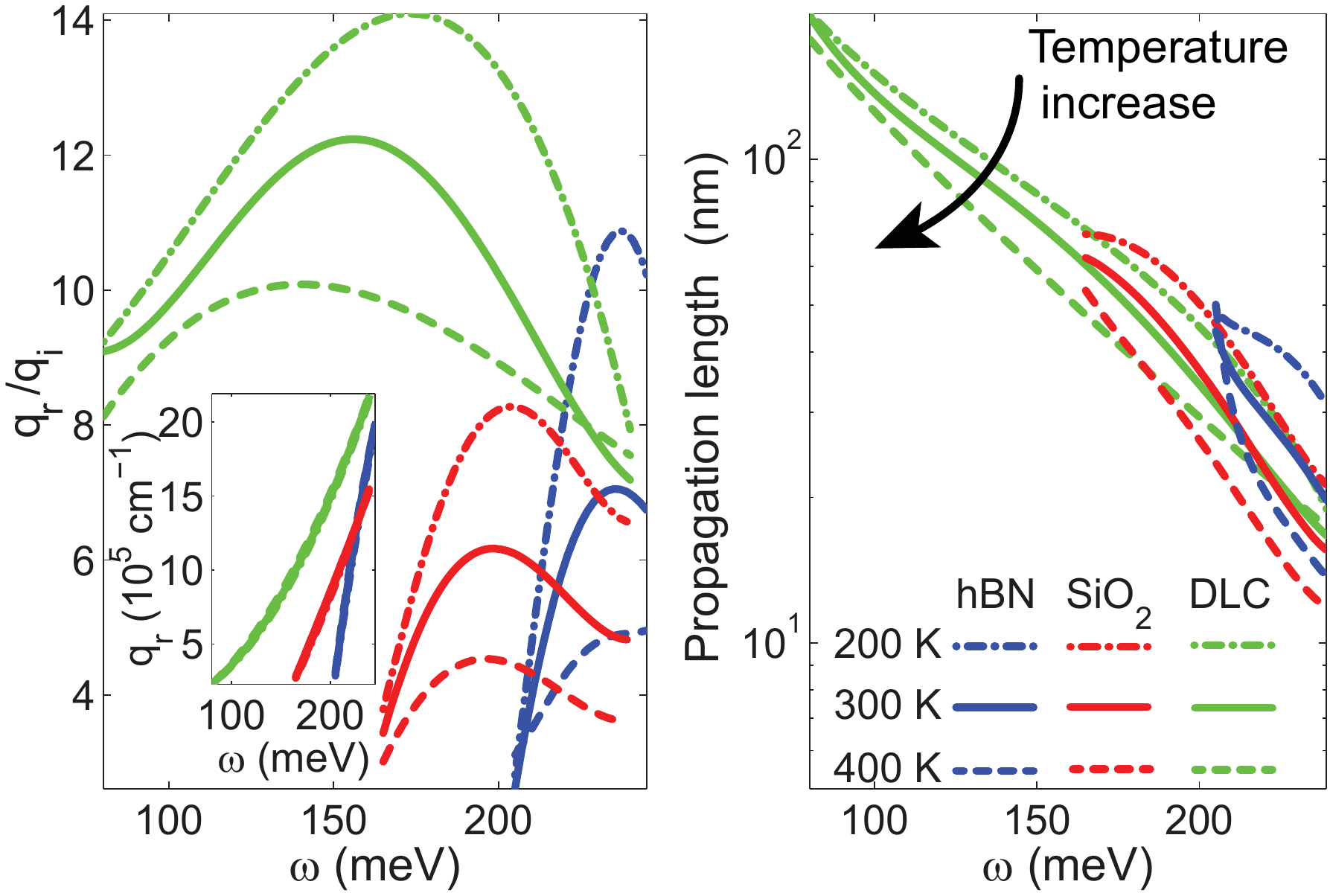}
\caption{\label{fig:3subs} (Left) The normalized plasmon propagation length ($q_r/q_i$) as a function of frequency for graphene with a carrier density of $n=3\times 10^{12} \mathrm{cm}^{-2}$ at three different temperatures (200, 300, and 400 K) and on three different substrates: DLC, SiO$_2$, and hBN. The impurity densities are chosen to yield the room-temperature electron mobility of 3000 $\frac{\text{cm}^2}{\text{Vs}}$ (DLC) \cite{wu2012state}, 2500 $\frac{\text{cm}^2}{\text{Vs}}$ (SiO$_2$) \cite{gannett2011boron}, and 9900 $\frac{\text{cm}^2}{\text{Vs}}$ (hBN) \cite{gannett2011boron}. Inset: Plasmon dispersions. (Right) Plasmon propagation length as a function of frequency. }
\end{figure}

\textcolor{black}{What our work captures is the effect of scattering [specifically, of the dominant ionized-impurity  scattering (see Appendix \ref{app3c})] on the plasmon propagation length (Figs. \ref{fig:2subs} and \ref{fig:3subs}), which the dissipation-free RPA calculations cannot do.} Because of dissipation, the plasmon wave vector is complex and the plasmon propagation length is limited. We can write $q=q_r+iq_i$, and the propagation length will be $\frac{1}{2 q_i}$. The plasmon dispersion $\omega_p(q_r)$ is obtained from the loss-function maximum \cite{brar2014hybrid,woessner2014highly,yan2013damping}. We write a Taylor expansion of $\varepsilon (\mathbf{q},\omega)$ in the vicinity of $\left(q_r,\omega_p(q_r)\right)$ in terms of $q-q_r=iq_i$ up to the fourth order. We then re-calculate the loss function based on the expansion; its maximum gives us the imaginary part of the complex wave vector of a plasmon. Figure \ref{fig:3loss_func}(d) shows the loss function of the star-marked point on Fig. \ref{fig:3loss_func}(c) as a function of ${q_r}/{q_i}$, the normalized propagation length. ${q_r}/{q_i}$ is the number of wavelengthes a plasmon propagates before dying out, and quantifies plasmon damping; higher normalized propagation length corresponds to lower damping.

In Fig. \ref{fig:2subs}, it can be seen that increasing the carrier density not only increases the normalized propagation length (equivalently decreases the plasmon damping), but also moves the minimum damping to higher frequencies. Changing carrier density, though, pushes the plasmon dispersion to higher wave vectors. But, in general, increasing the carrier density enhances the plasmon propagation length.

Figure \ref{fig:3subs} shows the plasmon propagation length versus frequency for the carrier density of $3\times 10^{13} $ cm$^{-2}$ on three different substrates and at three different temperatures. The complex dielectric function of SiO$_2$ and hBN is provided in Appendix \ref{app2}. As a nonpolar material with a relative permittivity of 3 \cite{grill1999electrical}, DLC provides a longer plasmon propagation length at low frequencies than the polar substrates. In contrast, plasmons are highly suppressed for frequencies below the highest surface modes of SiO$_2$ and hBN. Graphene on DLC shows a much greater  normalized propagation length $\sim q_r/q_i$, i.e., less plasmon damping (Fig.~\ref{fig:3subs},  left). However, because the plasmon wave vector ($\sim q_r$) in graphene on DLC is considerably longer [inset to left panel of Fig.~\ref{fig:3subs}], the absolute value of the plasmon propagation length ($\sim q_i^{-1}$) in graphene-on-DLC is not considerably improved over the two polar substrates.

Finally, plasmon dispersion remains almost independent of temperature (inset to Fig.~\ref{fig:3subs}, left), while decreasing temperature reduces plasmon attenuation and the shifts the damping minimum to higher frequencies (Fig.~\ref{fig:3subs}, left).

\section{Conclusion}\label{sect4}

In summary, we presented a method to calculate the linear-response dielectric function of a dissipative electronic system with an arbitrary band structure and Bloch wave functions. We calculated the induced charge density as a function of the self-consistent field by solving a generalized Markovian master equation of motion for coherences. The SCF-MMEF preserves the positivity of the density matrix and, consequently, conserves the number of electrons. The technique can be readily generalized to low-dimensional structures on other materials.

We employed the SCF-MMEF to study the electrical and optical properties of graphene. In our calculation, we considered intrinsic phonon scattering, ionized-impurity scattering, and SO-phonon scattering.  We calculated the complex conductivity of graphene and showed that ionized impurities improve screening when the carrier density is comparable to or lower than the impurity density, a phenomenon that the Mermin-Lindhard approach cannot capture.

We calculated the dielectric function and loss function for graphene on three substrates (the nonpolar DLC, and the polar SiO$_2$ and hBN) and for different values of the impurity density, carrier density, and temperature. From the loss-function maximum, the plasmon dispersions and propagation length were computed. On polar substrates (SiO$_2$ and hBN), plasmons are strongly suppressed and have short propagation lengths at frequencies below the highest SO phonon mode. DLC, being nonpolar, provides a broader spectrum for graphene plasmons.

We also investigated the effect of impurity density, carrier density, and temperature on plasmon dispersion and propagation length. Plasmon dispersion is fairly insensitive to varying impurity density or temperature, but is pushed towards shorter wave vectors with increasing carrier density. Plasmon damping -- inversely proportional to the number of wave lengths that a plasmon propagates before dying out -- worsens with more pronounced scattering, such as when the impurity density or temperature is increased. However, damping drops with increasing carrier density, benefiting plasmon propagation. Overall, the plasmon propagation length in absolute units is better on substrates with fewer impurities, and improves at lower temperatures and at higher carrier densities. We note that the calculated propagation lengths, comparable on polar and nonpolar substrates and roughly tens of nanometers, are an order of magnitude shorter than previously reported based on the Mermin-Lindhard approach \cite{jablan2009plasmonics}.

This work underscores the importance of treating the dissipative mechanisms accurately. The SCF-MMEF may lead to improved understanding of the dielectric function and collective electronic excitations in graphene and related nanomaterials, such as nanoribbons or van der Waals structures \cite{geim2013van}.

\begin{acknowledgments}
The authors gratefully acknowledge support by the U.S. Department of Energy, Office of Basic Energy Sciences, Division of Materials Sciences and Engineering, Physical Behavior of Materials Program, under Award DE-SC0008712. This work was performed using the compute resources and assistance of the UW-Madison Center for High Throughput Computing (CHTC) in the Department of Computer Sciences.
\end{acknowledgments}
%

\appendix

\section{\label{app} The SCF-MMEF for a quasi-one-dimensional material}

For quasi-one-dimensional materials such as graphene nanoribbons (GNRs), Eqs. (\ref{eps}) and  (\ref{sigma}) are modified. Assuming a GNR is extended in the $x$-direction and has a width of $W$ in the $y$-direction, the induced charge density can be written as
\begin{equation}
    n(x,y,z,t)=n_l(x,t)\left[\frac{1}{W}\Pi\left(\frac{y}{W}\right)\right]\delta(z).
\end{equation}
$\delta(\cdot)$ denotes the Dirac delta function and $\Pi(.)$ denotes the rectangular function [1 for its argument being within (0,1), 1/2 for argument equal to 0 or 1, and zero elsewhere]. With an approach similar to the derivation of \eq{Vind}, the induced potential reads
\begin{equation}\label{Vindl}
\begin{split}
 V_{\text{ind}}(q,y=0,z=0,\omega)&=\frac{-ie^2}{4\varepsilon_r\varepsilon_0 }n_l(q,\omega)\\
 & \times \int_{\frac{-1}{2}}^{\frac{1}{2}} d\eta ~H_0^{(1)}(iQW|\eta|),
\end{split}
\end{equation}
where $H_0^{(1)}(\cdot)$ is the zeroth-order Hankel function of the first kind. It has been assumed that the potential does not vary significantly across the GNR. To simplify the notation, henceforth we drop the $y$ and $z$ arguments. The dielectric function for a quasi-one-dimensional system may be written as
\begin{equation}\label{epsl}
\begin{split}
\varepsilon (q,\omega)=& 1+\frac{ie}{4\varepsilon_r\varepsilon_0 }P_l(q,\omega) \int_{\frac{-1}{2}}^{\frac{1}{2}} d\eta ~H_0^{(1)}(iQW|\eta|) ,
\end{split}
\end{equation}
where here $P_l(q,\omega)\equiv \frac{ e n_l(q,\omega)}{V_{\text{SCF}}(q,\omega)}$ is the linear polarization. Analogously, the conductivity of GNRs is
\begin{equation}\label{sigmal}
\sigma_l(q,\omega) = \frac{ -ie\omega}{q^2 }P_l(q,\omega).
\end{equation}
In solving the SCF-MMEF for a quasi-one-dimensional material, Eqs.  (\ref{matrix_eq}) and (\ref{matrices}) remain unchanged, but the Brillouin zone is one-dimensional. Also, \eq{polarization} should be modified to
\begin{equation}
P_l(\mathbf{q},\omega)= \frac{-e}{L} \mathcal{C}^T\mathcal{X}.
\end{equation}

\section{\label{app2} The complex dielectric function of SiO$_2$ and hBN}

We use the Lorentz oscillator model for the frequency-dependent dielectric function of SiO$_2$ and hBN.
\begin{equation}
\varepsilon=\varepsilon_\infty +\sum_j \frac{s_j^2}{\omega_j^2-\omega^2-i\Gamma_j \omega_j}
\end{equation}

\begin{figure}
\includegraphics[width=\columnwidth]{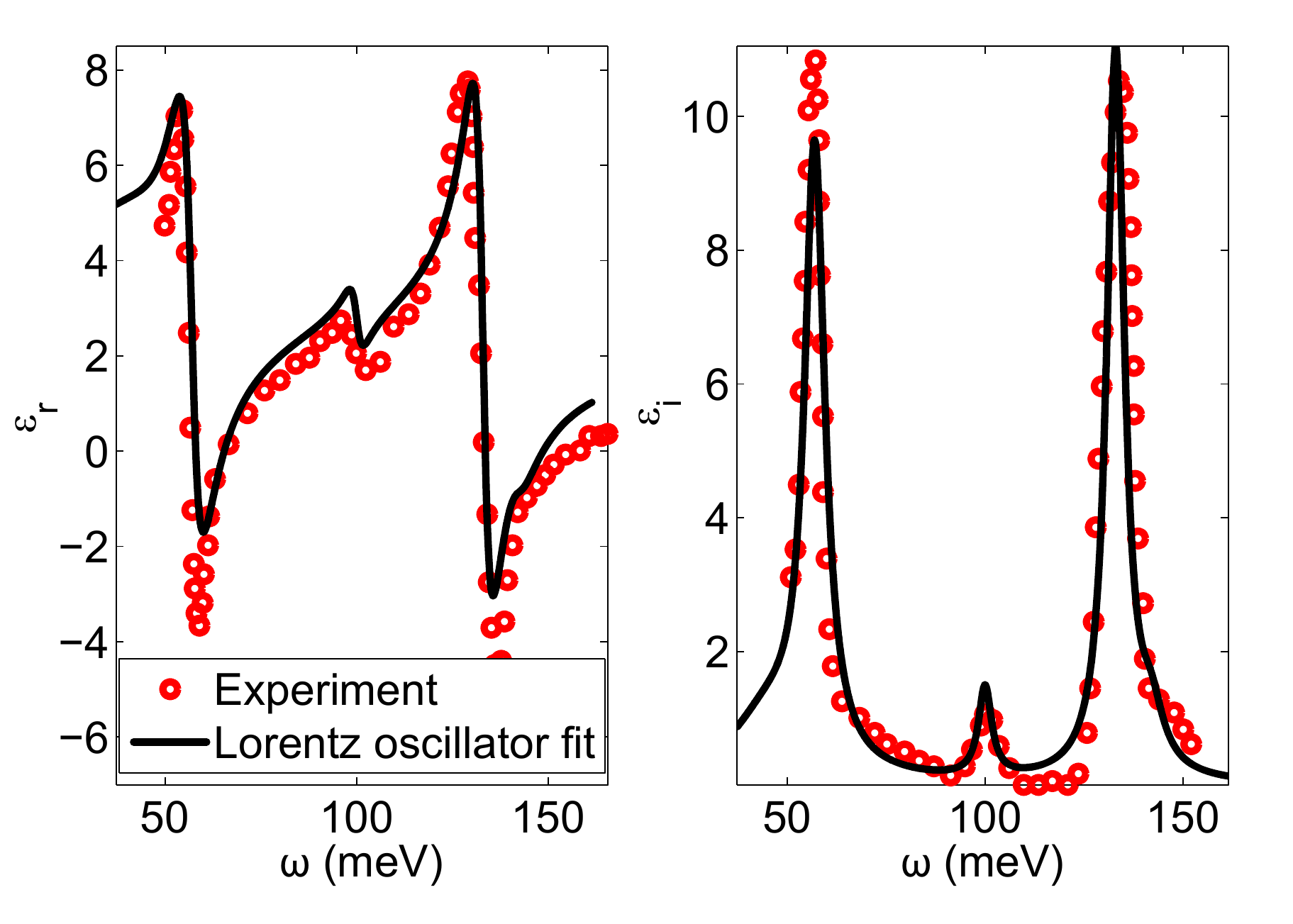}
\begin{center}
\caption{\label{fig:SiO2} Real part (left) and imaginary part (right) of the dielectric function of 300-nm-thick SiO$_2$ from experimental results \cite{kuvcirkova1994interpretation,philipp1979infrared} (circles) and the Lorentz oscillator model (solid line).}
\end{center}
\end{figure}

\noindent In Fig. \ref{fig:SiO2}, the real part and imaginary parts of the complex dielectric function of a roughly 300-nm-thick slab of SiO$_{2}$ adopted from measurements \cite{kuvcirkova1994interpretation,philipp1979infrared} and the Lorentz oscillator fit are shown. In the Lorentz oscillator model, $\varepsilon_\infty=2.3$, and the other parameters are
\begin{equation}
\begin{tabular}{c|c|c}
$\omega (\text{meV})$   & $s^2(\text{meV}^2)$ & $\Gamma(\text{meV})$\\ \hline
$142$  & $812$ & $7.4$\\
$133$  & $7832$  & $5.4$\\
$100$  & $537$ & $4$\\
$57$   & $3226$  & $6.2$\\
$47$   & $1069$  & $24.5$\\
\end{tabular}
\end{equation}

\noindent The above choice of parameters results in $\varepsilon_0=4.4$. The surface phonon modes can be obtained by solving $\varepsilon+1=0$ \cite{konar2010effect}, which for SiO$_2$ results in three dominant modes: 65 meV, 102 meV, and 147 meV.

The Lorentz oscillator fit of the complex dielectric function of hBN is adopted from  Ref. \cite{geick1966normal}. For hBN, $\varepsilon_\infty=4.95$, and the other parameters are
\begin{equation}
\begin{tabular}{c|c|c}
$\omega (\text{meV})$   & $s^2(\text{meV}^2)$ & $\Gamma(\text{meV})$\\ \hline
$169$  & $5364$ & $3.6$\\
$95$  & $1895$  & $4.34$\\
\end{tabular}
\end{equation}

\begin{figure}
\includegraphics[width=\columnwidth]{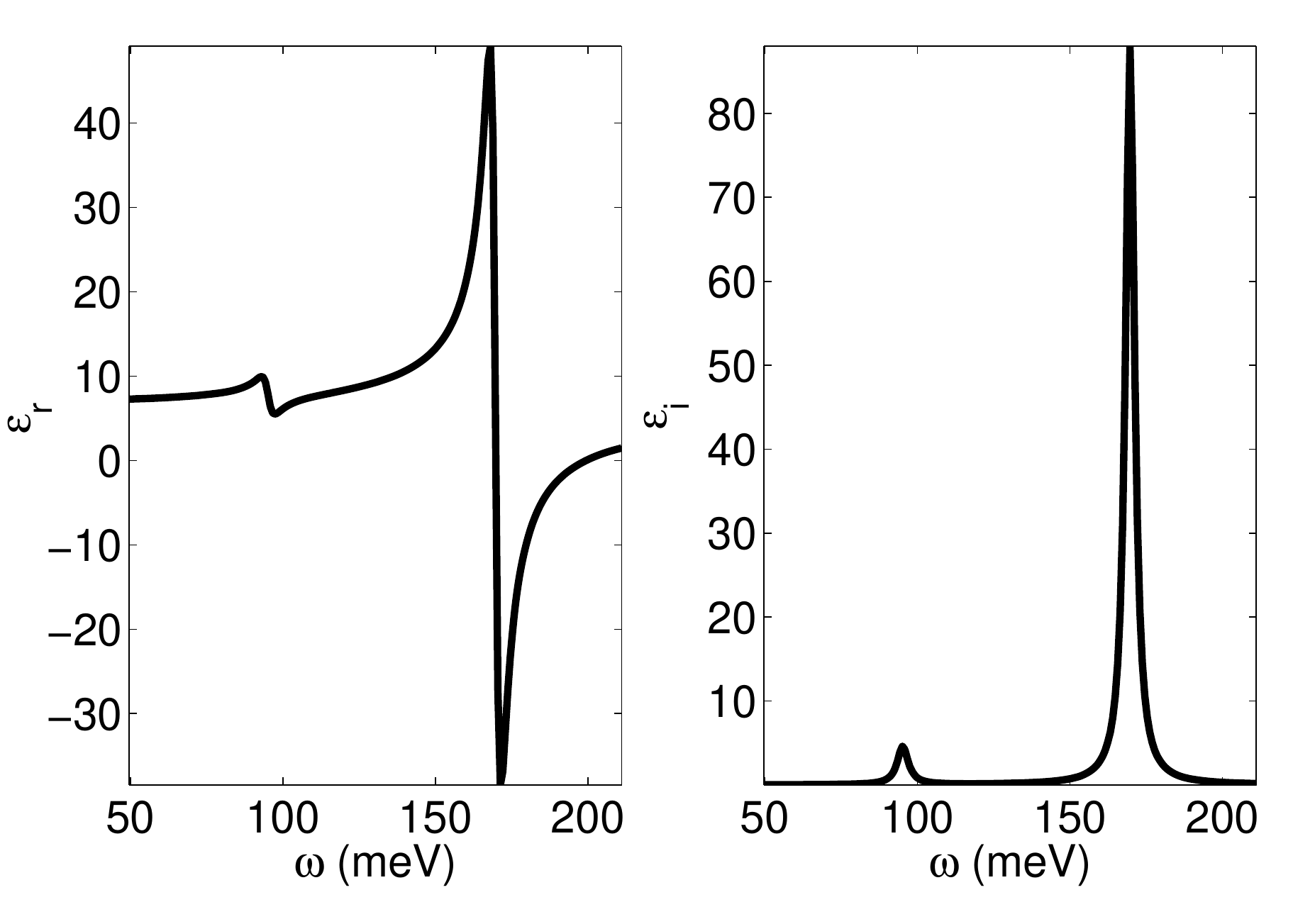}
\begin{center}
\caption{\label{fig:hBN} Real part (left) and imaginary part (right) of the hBN dielectric function}
\end{center}
\end{figure}
\noindent The above choice of parameters results in $\varepsilon_0=7.03$ for hBN. Similarly, we obtain the surface phonon modes by solving $\varepsilon+1=0$,  which results in two dominant modes for hBN: 98 meV and 195 meV.

\section{\label{app3} Scattering Mechanisms}
\subsection{Phonon scattering}

The interaction Hamiltonian of electrons and phonons reads
\begin{equation}
 \begin{split}
 &{\mathbb {H}_{\text{e-ph}}}=\sum_{\mathbf{kq},l'l} \mathcal{M}_\text{ph}(\mathbf{q})(\mathbf{k+q}l'|\mathbf{k}l)c^\dagger_{\mathbf{k+q}l'} c_{\mathbf{k}l} (b_{\mathbf{q}}+b^\dagger_{-\mathbf{q}}),\\
\end{split}
\end{equation}
where $b^\dagger$ and $b$ are the phonon creation and destruction operators, respectively. The scattering weights are defined as
\begin{equation}
\begin{split}
&\mathfrak{W}^{+}_{\mathbf{k-k'},\text{ph}}=N_{\mathbf{k-k'},\text{ph}} |\mathcal{M}_{\text{ph}}(\mathbf{k-k'})|^2,\\
&\mathfrak{W}^{-}_{\mathbf{k-k'},\text{ph}}=(N_{\mathbf{k-k'},\text{ph}}+1) |\mathcal{M}_{\text{ph}}(\mathbf{k-k'})|^2.
\end{split}
\end{equation}
For  longitudinal acoustic (LA) phonons
\begin{equation}
\mathcal{M}_{\text{LA}}(\mathbf{q})=D_{\text{ac}}\left(\frac{1}{2 m \omega_{\mathbf{q},\text{LA}}}\right)^{\frac{1}{2}} (i\mathbf{q \cdot e_q}).
\end{equation}
$m$ is the mass of the graphene sheet (with mass density of $7.6\times10^{-7}~{\text{kg}}/{\text{m}^2}$), $\omega_{\mathbf{q},\text{LA}}$ is the frequency of phonons in the acoustic branch, $D_{\text{ac}}=12$ eV  is the deformation potential for acoustic phonons, and $\mathbf{e_q}$ is the unit vector along the displacement direction.
The acoustic phonon scattering may be approximated as an elastic scattering mechanism. We assume linear dispersion for acoustic phonons ($\omega_\mathbf{q}=v_s|\mathbf{q}|$), where $v_s=2\times 10^4  ~\frac{\text{m}}{\text{s}}$ is the sound velocity in graphene. By employing the equipartition approximation at room temperature, it can be shown that
\begin{equation}
\begin{split}
\mathfrak{W}^{+}_{\mathbf{k,k'},\text{LA}}\approx \mathfrak{W}^{-}_{\mathbf{k,k'},\text{LA}} \approx \frac{D_{\mathrm{ac}}^2 k_B T}{2 m v_s^2}.
\end{split}
\end{equation}
For longitudinal optical (LO) phonons in a nonpolar materail
\begin{equation}
\mathcal{M}_{\text{LO}}(\mathbf{q})=D_{\text{op}}\left(\frac{1}{2 m \omega_{\mathbf{q},\text{LO}}}\right)^{\frac{1}{2}},
\end{equation}
where $\omega_{\mathbf{q},\text{LO}}$ represents the frequency of LO phonons, and $D_{\text{op}}=10^{11}\frac{\text{eV}}{\text{m}}$ is the deformation potential for nonpolar electron-LO phonon scattering.
LO phonons may be assumed dispersionless, $\omega_\mathbf{q,\text{LO}}=\omega_\text{LO}=195$~meV. It can be shown that
\begin{equation}
\begin{split}
&\mathfrak{W}^{+}_{\mathbf{k,k'},\text{LO}}\approx N_{\text{LO}} \frac{D_{\text{op}}^2 }{2 m \omega_0},\\
&\mathfrak{W}^{-}_{\mathbf{k,k'},\text{LO}}\approx(N_{\text{LO}}+1) \frac{D_{\text{op}}^2 }{2 m \omega_0}.
\end{split}
\end{equation}
For the electron-SO phonon interaction we have
\begin{equation}
  \mathcal{M}_{\text{SO}}(\mathbf{q})=\left[\frac{ e^2  \omega_{\text{SO}}}{2 A \varepsilon_0} \frac{\tilde{\varepsilon}}{\varepsilon^\infty_s} \left(\frac{e^{-2qd}}{q \varepsilon^2(\mathbf{q},\omega=0)}\right)  \right] ^{\frac{1}{2}},
\end{equation}\\
where $\tilde{\varepsilon}=\varepsilon^\infty_s\left(\frac{1}{\varepsilon^\infty_s+1}-\frac{1}{\varepsilon^0_s+1}\right)$, $\varepsilon^\infty_s$ and $\varepsilon^0_s$ are the high-frequency and low-frequency relative permittivities of the substrate, respectively. $\varepsilon(\mathbf{q},\omega=0)$, calculated self-consistently according to \eq{eps}, denotes the static dielectric function of the graphene sheet. The SO phonons may also be assumed dispersionless.\\

\subsection{Ionized-impurity scattering}

In graphene, the electron-ion interaction is related to the Coulomb potential in a two-dimensional electron gas. The screened potential of a buried ionized impurity at $z=-d$ in the substrate observed at the graphene sheet is
\begin{equation}
V_{\text{eff}}(\mathbf{q})= \frac{e^2}{2 A \varepsilon^0_b \varepsilon(\mathbf{q},\omega=0) } \frac{e^{-qd}}{q},
\end{equation}
where $\varepsilon^0_b\equiv\frac{1+\varepsilon_s(\omega=0)}{2}$. Thus, the interaction Hamiltonian of electrons and  ionized impurities may be written as
\begin{equation}
 \begin{split}
 &{\mathbb {H}_{\text{e-ii}}}=\sum_{\mathbf{kq},l'l} \frac{e^2}{A^2 \varepsilon^0_b \varepsilon(\mathbf{q},\omega=0) } \frac{e^{-qd}}{2q}(\mathbf{k+q}l'|\mathbf{k}l)c^\dagger_{\mathbf{k+q}l'} c_{\mathbf{k}l}.\\
\end{split}
\end{equation}
The corresponding scattering weight for a sheet of impurities  distributed uniformly on a plane located at a distance $d$ beneath the graphene sheet with the density of $N_{\mathrm{i}}$ reads
\begin{equation}
\mathfrak{W}_{\mathbf{k,k'},\text{ii}}=\frac{N_{\mathrm{i}}}{A}\left(\frac{e^2}{ \varepsilon_b^0 \varepsilon(\mathbf{k'-k},\omega=0) } \frac{e^{-|\mathbf{k'-k}|d}}{2|\mathbf{k'-k}|}\right)^2
\end{equation}

To have a unified notation, we define $\mathfrak{W}^+_{\mathbf{k,k'},\text{ii}}=\mathfrak{W}^-_{\mathbf{k,k'},\text{ii}}=\frac{1}{2}\mathfrak{W}_{\mathbf{k,k'},\text{ii}}.$
\subsection{Comparison between SO-phonon scattering and ionized-impurity scattering}\label{app3c}
\begin{figure}
\includegraphics[width=\columnwidth]{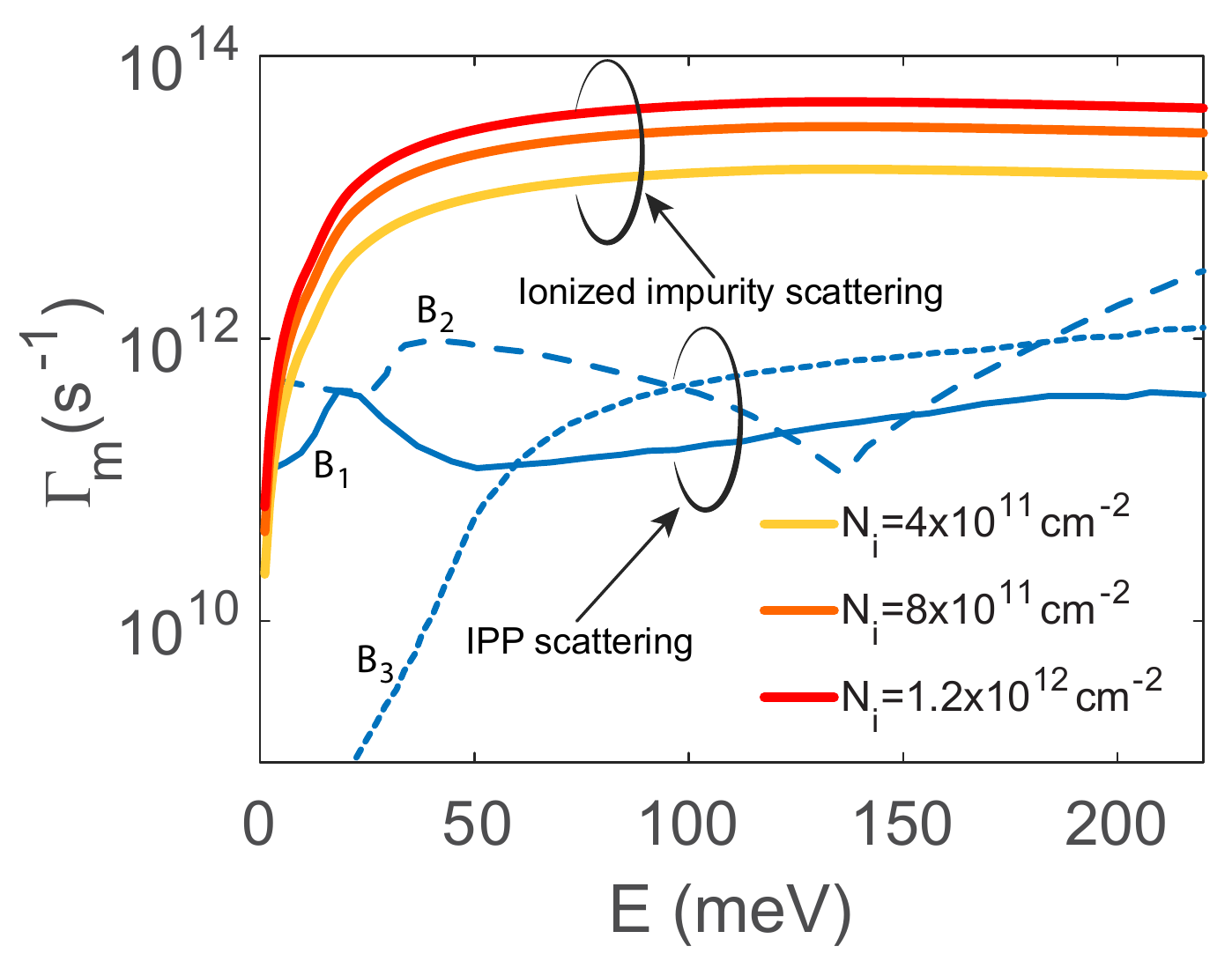}
\begin{center}
\caption{\label{fig:IPP} \color{black} The momentum relaxation rate, $\Gamma_m$,  of electrons due to ionized-impurity scattering at the impurity sheet densities of $4\times 10^{11}\,\mathrm{cm}^{-2}$ (yellow), $8\times 10^{11}\,\mathrm{cm}^{-2}$ (orange), and  $1.2\times 10^{12}\,\mathrm{cm}^{-2}$ (red) and due to scattering from three IPP branches (blue; data from \cite{ong2012theory}). Carrier density is $n=10^{12}\,\mathrm{cm}^{-2}$.}
\end{center}
\end{figure}

\color{black}
Surface plasmons and SO phonons couple with each other and form interfacial plasmon-phonon (IPP) modes \cite{ong2012theory}. The IPP modes are different from both pure plasmon and pure SO-phonon modes. Therefore, the electron--IPP scattering rates are different from the electron--SO phonon scattering rates. However, even in high-quality samples of supported graphene, ionized-impurity scattering is the dominant mechanism. In Fig. \ref{fig:IPP}, we compare the momentum relaxation rates for electron scattering with ionized impurities at three different impurity sheet densities against the rates for scattering with IPPs (data from \cite{ong2012theory}) for graphene on SiO$_2$. Given that the ionized-impurity scattering dominates by over an order of magnitude, we consider the approximation of uncoupled SO phonons and plasmons (instead of IPPs) and the static screening of SO phonons \cite{WangMahan_PRB_1972,konar2010effect,hwang2010plasmon} to be acceptable approximations, balancing simplicity with accuracy. The loss function does not get  considerably altered in shape by the inclusion of IPP versus pure SO-phonon and plasmon modes.
\color{black}


\begin{thebibliography}{97}%
\makeatletter
\providecommand \@ifxundefined [1]{%
 \@ifx{#1\undefined}
}%
\providecommand \@ifnum [1]{%
 \ifnum #1\expandafter \@firstoftwo
 \else \expandafter \@secondoftwo
 \fi
}%
\providecommand \@ifx [1]{%
 \ifx #1\expandafter \@firstoftwo
 \else \expandafter \@secondoftwo
 \fi
}%
\providecommand \natexlab [1]{#1}%
\providecommand \enquote  [1]{``#1''}%
\providecommand \bibnamefont  [1]{#1}%
\providecommand \bibfnamefont [1]{#1}%
\providecommand \citenamefont [1]{#1}%
\providecommand \href@noop [0]{\@secondoftwo}%
\providecommand \href [0]{\begingroup \@sanitize@url \@href}%
\providecommand \@href[1]{\@@startlink{#1}\@@href}%
\providecommand \@@href[1]{\endgroup#1\@@endlink}%
\providecommand \@sanitize@url [0]{\catcode `\\12\catcode `\$12\catcode
  `\&12\catcode `\#12\catcode `\^12\catcode `\_12\catcode `\%12\relax}%
\providecommand \@@startlink[1]{}%
\providecommand \@@endlink[0]{}%
\providecommand \url  [0]{\begingroup\@sanitize@url \@url }%
\providecommand \@url [1]{\endgroup\@href {#1}{\urlprefix }}%
\providecommand \urlprefix  [0]{URL }%
\providecommand \Eprint [0]{\href }%
\providecommand \doibase [0]{http://dx.doi.org/}%
\providecommand \selectlanguage [0]{\@gobble}%
\providecommand \bibinfo  [0]{\@secondoftwo}%
\providecommand \bibfield  [0]{\@secondoftwo}%
\providecommand \translation [1]{[#1]}%
\providecommand \BibitemOpen [0]{}%
\providecommand \bibitemStop [0]{}%
\providecommand \bibitemNoStop [0]{.\EOS\space}%
\providecommand \EOS [0]{\spacefactor3000\relax}%
\providecommand \BibitemShut  [1]{\csname bibitem#1\endcsname}%
\let\auto@bib@innerbib\@empty
\bibitem [{\citenamefont {Maier}\ \emph {et~al.}(2001)\citenamefont {Maier},
  \citenamefont {Brongersma}, \citenamefont {Kik}, \citenamefont {Meltzer},
  \citenamefont {Requicha},\ and\ \citenamefont
  {Atwater}}]{maier2001plasmonics}%
  \BibitemOpen
  \bibfield  {author} {\bibinfo {author} {\bibfnamefont {S.~A.}\ \bibnamefont
  {Maier}}, \bibinfo {author} {\bibfnamefont {M.~L.}\ \bibnamefont
  {Brongersma}}, \bibinfo {author} {\bibfnamefont {P.~G.}\ \bibnamefont {Kik}},
  \bibinfo {author} {\bibfnamefont {S.}~\bibnamefont {Meltzer}}, \bibinfo
  {author} {\bibfnamefont {A.~A.}\ \bibnamefont {Requicha}}, \ and\ \bibinfo
  {author} {\bibfnamefont {H.~A.}\ \bibnamefont {Atwater}},\ }\href@noop {}
  {\bibfield  {journal} {\bibinfo  {journal} {Adv. Mater.}\ }\textbf {\bibinfo
  {volume} {13}},\ \bibinfo {pages} {1501} (\bibinfo {year}
  {2001})}\BibitemShut {NoStop}%
\bibitem [{\citenamefont {Schuller}\ \emph {et~al.}(2010)\citenamefont
  {Schuller}, \citenamefont {Barnard}, \citenamefont {Cai}, \citenamefont
  {Jun}, \citenamefont {White},\ and\ \citenamefont
  {Brongersma}}]{schuller2010plasmonics}%
  \BibitemOpen
  \bibfield  {author} {\bibinfo {author} {\bibfnamefont {J.~A.}\ \bibnamefont
  {Schuller}}, \bibinfo {author} {\bibfnamefont {E.~S.}\ \bibnamefont
  {Barnard}}, \bibinfo {author} {\bibfnamefont {W.}~\bibnamefont {Cai}},
  \bibinfo {author} {\bibfnamefont {Y.~C.}\ \bibnamefont {Jun}}, \bibinfo
  {author} {\bibfnamefont {J.~S.}\ \bibnamefont {White}}, \ and\ \bibinfo
  {author} {\bibfnamefont {M.~L.}\ \bibnamefont {Brongersma}},\ }\href@noop {}
  {\bibfield  {journal} {\bibinfo  {journal} {Nat. Mater.}\ }\textbf {\bibinfo
  {volume} {9}},\ \bibinfo {pages} {193} (\bibinfo {year} {2010})}\BibitemShut
  {NoStop}%
\bibitem [{\citenamefont {Maier}\ and\ \citenamefont
  {Atwater}(2005)}]{maier2005plasmonics}%
  \BibitemOpen
  \bibfield  {author} {\bibinfo {author} {\bibfnamefont {S.~A.}\ \bibnamefont
  {Maier}}\ and\ \bibinfo {author} {\bibfnamefont {H.~A.}\ \bibnamefont
  {Atwater}},\ }\href@noop {} {\bibfield  {journal} {\bibinfo  {journal} {J.
  Appl. Phys.}\ }\textbf {\bibinfo {volume} {98}},\ \bibinfo {pages} {011101}
  (\bibinfo {year} {2005})}\BibitemShut {NoStop}%
\bibitem [{\citenamefont {Karalis}\ \emph {et~al.}(2005)\citenamefont
  {Karalis}, \citenamefont {Lidorikis}, \citenamefont {Ibanescu}, \citenamefont
  {Joannopoulos},\ and\ \citenamefont
  {Solja{\v{c}}i{\'c}}}]{karalis2005surface}%
  \BibitemOpen
  \bibfield  {author} {\bibinfo {author} {\bibfnamefont {A.}~\bibnamefont
  {Karalis}}, \bibinfo {author} {\bibfnamefont {E.}~\bibnamefont {Lidorikis}},
  \bibinfo {author} {\bibfnamefont {M.}~\bibnamefont {Ibanescu}}, \bibinfo
  {author} {\bibfnamefont {J.}~\bibnamefont {Joannopoulos}}, \ and\ \bibinfo
  {author} {\bibfnamefont {M.}~\bibnamefont {Solja{\v{c}}i{\'c}}},\ }\href@noop
  {} {\bibfield  {journal} {\bibinfo  {journal} {Phys. Rev. Lett.}\ }\textbf
  {\bibinfo {volume} {95}},\ \bibinfo {pages} {063901} (\bibinfo {year}
  {2005})}\BibitemShut {NoStop}%
\bibitem [{\citenamefont {Gramotnev}\ and\ \citenamefont
  {Bozhevolnyi}(2010)}]{gramotnev2010plasmonics}%
  \BibitemOpen
  \bibfield  {author} {\bibinfo {author} {\bibfnamefont {D.~K.}\ \bibnamefont
  {Gramotnev}}\ and\ \bibinfo {author} {\bibfnamefont {S.~I.}\ \bibnamefont
  {Bozhevolnyi}},\ }\href@noop {} {\bibfield  {journal} {\bibinfo  {journal}
  {Nat. Photonics}\ }\textbf {\bibinfo {volume} {4}},\ \bibinfo {pages} {83}
  (\bibinfo {year} {2010})}\BibitemShut {NoStop}%
\bibitem [{\citenamefont {Novotny}\ and\ \citenamefont
  {Van~Hulst}(2011)}]{novotny2011antennas}%
  \BibitemOpen
  \bibfield  {author} {\bibinfo {author} {\bibfnamefont {L.}~\bibnamefont
  {Novotny}}\ and\ \bibinfo {author} {\bibfnamefont {N.}~\bibnamefont
  {Van~Hulst}},\ }\href@noop {} {\bibfield  {journal} {\bibinfo  {journal}
  {Nat. Photonics}\ }\textbf {\bibinfo {volume} {5}},\ \bibinfo {pages} {83}
  (\bibinfo {year} {2011})}\BibitemShut {NoStop}%
\bibitem [{\citenamefont {Yu}\ \emph {et~al.}(2008)\citenamefont {Yu},
  \citenamefont {Veronis}, \citenamefont {Wang},\ and\ \citenamefont
  {Fan}}]{Yu2008One}%
  \BibitemOpen
  \bibfield  {author} {\bibinfo {author} {\bibfnamefont {Z.}~\bibnamefont
  {Yu}}, \bibinfo {author} {\bibfnamefont {G.}~\bibnamefont {Veronis}},
  \bibinfo {author} {\bibfnamefont {Z.}~\bibnamefont {Wang}}, \ and\ \bibinfo
  {author} {\bibfnamefont {S.}~\bibnamefont {Fan}},\ }\href {\doibase
  10.1103/PhysRevLett.100.023902} {\bibfield  {journal} {\bibinfo  {journal}
  {Phys. Rev. Lett.}\ }\textbf {\bibinfo {volume} {100}},\ \bibinfo {pages}
  {023902} (\bibinfo {year} {2008})}\BibitemShut {NoStop}%
\bibitem [{\citenamefont {Khanikaev}\ \emph {et~al.}(2010)\citenamefont
  {Khanikaev}, \citenamefont {Mousavi}, \citenamefont {Shvets},\ and\
  \citenamefont {Kivshar}}]{Khanikav2008One}%
  \BibitemOpen
  \bibfield  {author} {\bibinfo {author} {\bibfnamefont {A.~B.}\ \bibnamefont
  {Khanikaev}}, \bibinfo {author} {\bibfnamefont {S.~H.}\ \bibnamefont
  {Mousavi}}, \bibinfo {author} {\bibfnamefont {G.}~\bibnamefont {Shvets}}, \
  and\ \bibinfo {author} {\bibfnamefont {Y.~S.}\ \bibnamefont {Kivshar}},\
  }\href {\doibase 10.1103/PhysRevLett.105.126804} {\bibfield  {journal}
  {\bibinfo  {journal} {Phys. Rev. Lett.}\ }\textbf {\bibinfo {volume} {105}},\
  \bibinfo {pages} {126804} (\bibinfo {year} {2010})}\BibitemShut {NoStop}%
\bibitem [{\citenamefont {Abbasi}\ \emph {et~al.}(2015)\citenamefont {Abbasi},
  \citenamefont {Davoyan},\ and\ \citenamefont {Engheta}}]{abbasi2015one}%
  \BibitemOpen
  \bibfield  {author} {\bibinfo {author} {\bibfnamefont {F.}~\bibnamefont
  {Abbasi}}, \bibinfo {author} {\bibfnamefont {A.~R.}\ \bibnamefont {Davoyan}},
  \ and\ \bibinfo {author} {\bibfnamefont {N.}~\bibnamefont {Engheta}},\
  }\href@noop {} {\bibfield  {journal} {\bibinfo  {journal} {New J. Phys.}\
  }\textbf {\bibinfo {volume} {17}},\ \bibinfo {pages} {063014} (\bibinfo
  {year} {2015})}\BibitemShut {NoStop}%
\bibitem [{\citenamefont {Shalaev}(2007)}]{shalaev2007optical}%
  \BibitemOpen
  \bibfield  {author} {\bibinfo {author} {\bibfnamefont {V.~M.}\ \bibnamefont
  {Shalaev}},\ }\href@noop {} {\bibfield  {journal} {\bibinfo  {journal} {Nat.
  Photonics}\ }\textbf {\bibinfo {volume} {1}},\ \bibinfo {pages} {41}
  (\bibinfo {year} {2007})}\BibitemShut {NoStop}%
\bibitem [{\citenamefont {Luk'yanchuk}\ \emph {et~al.}(2010)\citenamefont
  {Luk'yanchuk}, \citenamefont {Zheludev}, \citenamefont {Maier}, \citenamefont
  {Halas}, \citenamefont {Nordlander}, \citenamefont {Giessen},\ and\
  \citenamefont {Chong}}]{luk2010fano}%
  \BibitemOpen
  \bibfield  {author} {\bibinfo {author} {\bibfnamefont {B.}~\bibnamefont
  {Luk'yanchuk}}, \bibinfo {author} {\bibfnamefont {N.~I.}\ \bibnamefont
  {Zheludev}}, \bibinfo {author} {\bibfnamefont {S.~A.}\ \bibnamefont {Maier}},
  \bibinfo {author} {\bibfnamefont {N.~J.}\ \bibnamefont {Halas}}, \bibinfo
  {author} {\bibfnamefont {P.}~\bibnamefont {Nordlander}}, \bibinfo {author}
  {\bibfnamefont {H.}~\bibnamefont {Giessen}}, \ and\ \bibinfo {author}
  {\bibfnamefont {C.~T.}\ \bibnamefont {Chong}},\ }\href@noop {} {\bibfield
  {journal} {\bibinfo  {journal} {Nat. Mater.}\ }\textbf {\bibinfo {volume}
  {9}},\ \bibinfo {pages} {707} (\bibinfo {year} {2010})}\BibitemShut {NoStop}%
\bibitem [{\citenamefont {Kawata}\ \emph {et~al.}(2009)\citenamefont {Kawata},
  \citenamefont {Inouye},\ and\ \citenamefont {Verma}}]{kawata2009plasmonics}%
  \BibitemOpen
  \bibfield  {author} {\bibinfo {author} {\bibfnamefont {S.}~\bibnamefont
  {Kawata}}, \bibinfo {author} {\bibfnamefont {Y.}~\bibnamefont {Inouye}}, \
  and\ \bibinfo {author} {\bibfnamefont {P.}~\bibnamefont {Verma}},\
  }\href@noop {} {\bibfield  {journal} {\bibinfo  {journal} {Nat. Photonics}\
  }\textbf {\bibinfo {volume} {3}},\ \bibinfo {pages} {388} (\bibinfo {year}
  {2009})}\BibitemShut {NoStop}%
\bibitem [{\citenamefont {Al{\`u}}\ and\ \citenamefont
  {Engheta}(2005)}]{alu2005achieving}%
  \BibitemOpen
  \bibfield  {author} {\bibinfo {author} {\bibfnamefont {A.}~\bibnamefont
  {Al{\`u}}}\ and\ \bibinfo {author} {\bibfnamefont {N.}~\bibnamefont
  {Engheta}},\ }\href@noop {} {\bibfield  {journal} {\bibinfo  {journal} {Phys.
  Rev. E}\ }\textbf {\bibinfo {volume} {72}},\ \bibinfo {pages} {016623}
  (\bibinfo {year} {2005})}\BibitemShut {NoStop}%
\bibitem [{\citenamefont {Anker}\ \emph {et~al.}(2008)\citenamefont {Anker},
  \citenamefont {Hall}, \citenamefont {Lyandres}, \citenamefont {Shah},
  \citenamefont {Zhao},\ and\ \citenamefont {Van~Duyne}}]{anker2008biosensing}%
  \BibitemOpen
  \bibfield  {author} {\bibinfo {author} {\bibfnamefont {J.~N.}\ \bibnamefont
  {Anker}}, \bibinfo {author} {\bibfnamefont {W.~P.}\ \bibnamefont {Hall}},
  \bibinfo {author} {\bibfnamefont {O.}~\bibnamefont {Lyandres}}, \bibinfo
  {author} {\bibfnamefont {N.~C.}\ \bibnamefont {Shah}}, \bibinfo {author}
  {\bibfnamefont {J.}~\bibnamefont {Zhao}}, \ and\ \bibinfo {author}
  {\bibfnamefont {R.~P.}\ \bibnamefont {Van~Duyne}},\ }\href@noop {} {\bibfield
   {journal} {\bibinfo  {journal} {Nat. Mater.}\ }\textbf {\bibinfo {volume}
  {7}},\ \bibinfo {pages} {442} (\bibinfo {year} {2008})}\BibitemShut {NoStop}%
\bibitem [{\citenamefont {Kabashin}\ \emph {et~al.}(2009)\citenamefont
  {Kabashin}, \citenamefont {Evans}, \citenamefont {Pastkovsky}, \citenamefont
  {Hendren}, \citenamefont {Wurtz}, \citenamefont {Atkinson}, \citenamefont
  {Pollard}, \citenamefont {Podolskiy},\ and\ \citenamefont
  {Zayats}}]{kabashin2009plasmonic}%
  \BibitemOpen
  \bibfield  {author} {\bibinfo {author} {\bibfnamefont {A.}~\bibnamefont
  {Kabashin}}, \bibinfo {author} {\bibfnamefont {P.}~\bibnamefont {Evans}},
  \bibinfo {author} {\bibfnamefont {S.}~\bibnamefont {Pastkovsky}}, \bibinfo
  {author} {\bibfnamefont {W.}~\bibnamefont {Hendren}}, \bibinfo {author}
  {\bibfnamefont {G.}~\bibnamefont {Wurtz}}, \bibinfo {author} {\bibfnamefont
  {R.}~\bibnamefont {Atkinson}}, \bibinfo {author} {\bibfnamefont
  {R.}~\bibnamefont {Pollard}}, \bibinfo {author} {\bibfnamefont
  {V.}~\bibnamefont {Podolskiy}}, \ and\ \bibinfo {author} {\bibfnamefont
  {A.}~\bibnamefont {Zayats}},\ }\href@noop {} {\bibfield  {journal} {\bibinfo
  {journal} {Nat. Mater.}\ }\textbf {\bibinfo {volume} {8}},\ \bibinfo {pages}
  {867} (\bibinfo {year} {2009})}\BibitemShut {NoStop}%
\bibitem [{\citenamefont {Atwater}\ and\ \citenamefont
  {Polman}(2010)}]{atwater2010plasmonics}%
  \BibitemOpen
  \bibfield  {author} {\bibinfo {author} {\bibfnamefont {H.~A.}\ \bibnamefont
  {Atwater}}\ and\ \bibinfo {author} {\bibfnamefont {A.}~\bibnamefont
  {Polman}},\ }\href@noop {} {\bibfield  {journal} {\bibinfo  {journal} {Nat.
  Mater.}\ }\textbf {\bibinfo {volume} {9}},\ \bibinfo {pages} {205} (\bibinfo
  {year} {2010})}\BibitemShut {NoStop}%
\bibitem [{\citenamefont {Low}\ and\ \citenamefont
  {Avouris}(2014)}]{low2014graphene}%
  \BibitemOpen
  \bibfield  {author} {\bibinfo {author} {\bibfnamefont {T.}~\bibnamefont
  {Low}}\ and\ \bibinfo {author} {\bibfnamefont {P.}~\bibnamefont {Avouris}},\
  }\href@noop {} {\bibfield  {journal} {\bibinfo  {journal} {Acs Nano}\
  }\textbf {\bibinfo {volume} {8}},\ \bibinfo {pages} {1086} (\bibinfo {year}
  {2014})}\BibitemShut {NoStop}%
\bibitem [{\citenamefont {Tonouchi}(2007)}]{tonouchi2007cutting}%
  \BibitemOpen
  \bibfield  {author} {\bibinfo {author} {\bibfnamefont {M.}~\bibnamefont
  {Tonouchi}},\ }\href@noop {} {\bibfield  {journal} {\bibinfo  {journal} {Nat.
  Photonics}\ }\textbf {\bibinfo {volume} {1}},\ \bibinfo {pages} {97}
  (\bibinfo {year} {2007})}\BibitemShut {NoStop}%
\bibitem [{\citenamefont {Ferguson}\ and\ \citenamefont
  {Zhang}(2002)}]{ferguson2002materials}%
  \BibitemOpen
  \bibfield  {author} {\bibinfo {author} {\bibfnamefont {B.}~\bibnamefont
  {Ferguson}}\ and\ \bibinfo {author} {\bibfnamefont {X.-C.}\ \bibnamefont
  {Zhang}},\ }\href@noop {} {\bibfield  {journal} {\bibinfo  {journal} {Nat.
  Mater.}\ }\textbf {\bibinfo {volume} {1}},\ \bibinfo {pages} {26} (\bibinfo
  {year} {2002})}\BibitemShut {NoStop}%
\bibitem [{\citenamefont {Soref}(2010)}]{soref2010mid}%
  \BibitemOpen
  \bibfield  {author} {\bibinfo {author} {\bibfnamefont {R.}~\bibnamefont
  {Soref}},\ }\href@noop {} {\bibfield  {journal} {\bibinfo  {journal} {Nat.
  Photonics}\ }\textbf {\bibinfo {volume} {4}},\ \bibinfo {pages} {495}
  (\bibinfo {year} {2010})}\BibitemShut {NoStop}%
\bibitem [{\citenamefont {West}\ \emph {et~al.}(2010)\citenamefont {West},
  \citenamefont {Ishii}, \citenamefont {Naik}, \citenamefont {Emani},
  \citenamefont {Shalaev},\ and\ \citenamefont
  {Boltasseva}}]{west2010searching}%
  \BibitemOpen
  \bibfield  {author} {\bibinfo {author} {\bibfnamefont {P.~R.}\ \bibnamefont
  {West}}, \bibinfo {author} {\bibfnamefont {S.}~\bibnamefont {Ishii}},
  \bibinfo {author} {\bibfnamefont {G.~V.}\ \bibnamefont {Naik}}, \bibinfo
  {author} {\bibfnamefont {N.~K.}\ \bibnamefont {Emani}}, \bibinfo {author}
  {\bibfnamefont {V.~M.}\ \bibnamefont {Shalaev}}, \ and\ \bibinfo {author}
  {\bibfnamefont {A.}~\bibnamefont {Boltasseva}},\ }\href@noop {} {\bibfield
  {journal} {\bibinfo  {journal} {Laser Photon. Rev.}\ }\textbf {\bibinfo
  {volume} {4}},\ \bibinfo {pages} {795} (\bibinfo {year} {2010})}\BibitemShut
  {NoStop}%
\bibitem [{\citenamefont {Novoselov}\ \emph {et~al.}(2004)\citenamefont
  {Novoselov}, \citenamefont {Geim}, \citenamefont {Morozov}, \citenamefont
  {Jiang}, \citenamefont {Zhang}, \citenamefont {Dubonos}, , \citenamefont
  {Grigorieva},\ and\ \citenamefont {Firsov}}]{novoselov2004electric}%
  \BibitemOpen
  \bibfield  {author} {\bibinfo {author} {\bibfnamefont {K.~S.}\ \bibnamefont
  {Novoselov}}, \bibinfo {author} {\bibfnamefont {A.~K.}\ \bibnamefont {Geim}},
  \bibinfo {author} {\bibfnamefont {S.}~\bibnamefont {Morozov}}, \bibinfo
  {author} {\bibfnamefont {D.}~\bibnamefont {Jiang}}, \bibinfo {author}
  {\bibfnamefont {Y.}~\bibnamefont {Zhang}}, \bibinfo {author} {\bibfnamefont
  {S.}~\bibnamefont {Dubonos}}, , \bibinfo {author} {\bibfnamefont
  {I.}~\bibnamefont {Grigorieva}}, \ and\ \bibinfo {author} {\bibfnamefont
  {A.}~\bibnamefont {Firsov}},\ }\href@noop {} {\bibfield  {journal} {\bibinfo
  {journal} {Science}\ }\textbf {\bibinfo {volume} {306}},\ \bibinfo {pages}
  {666} (\bibinfo {year} {2004})}\BibitemShut {NoStop}%
\bibitem [{\citenamefont {Geim}\ and\ \citenamefont
  {Novoselov}(2007)}]{geim2007rise}%
  \BibitemOpen
  \bibfield  {author} {\bibinfo {author} {\bibfnamefont {A.~K.}\ \bibnamefont
  {Geim}}\ and\ \bibinfo {author} {\bibfnamefont {K.~S.}\ \bibnamefont
  {Novoselov}},\ }\href@noop {} {\bibfield  {journal} {\bibinfo  {journal}
  {Nat. Mater.}\ }\textbf {\bibinfo {volume} {6}},\ \bibinfo {pages} {183}
  (\bibinfo {year} {2007})}\BibitemShut {NoStop}%
\bibitem [{\citenamefont {Neto}\ \emph {et~al.}(2009)\citenamefont {Neto},
  \citenamefont {Guinea}, \citenamefont {Peres}, \citenamefont {Novoselov},\
  and\ \citenamefont {Geim}}]{neto2009electronic}%
  \BibitemOpen
  \bibfield  {author} {\bibinfo {author} {\bibfnamefont {A.~C.}\ \bibnamefont
  {Neto}}, \bibinfo {author} {\bibfnamefont {F.}~\bibnamefont {Guinea}},
  \bibinfo {author} {\bibfnamefont {N.}~\bibnamefont {Peres}}, \bibinfo
  {author} {\bibfnamefont {K.~S.}\ \bibnamefont {Novoselov}}, \ and\ \bibinfo
  {author} {\bibfnamefont {A.~K.}\ \bibnamefont {Geim}},\ }\href@noop {}
  {\bibfield  {journal} {\bibinfo  {journal} {Rev. Mod. Phys.}\ }\textbf
  {\bibinfo {volume} {81}},\ \bibinfo {pages} {109} (\bibinfo {year}
  {2009})}\BibitemShut {NoStop}%
\bibitem [{\citenamefont {Sarma}\ \emph {et~al.}(2011)\citenamefont {Sarma},
  \citenamefont {Adam}, \citenamefont {Hwang},\ and\ \citenamefont
  {Rossi}}]{sarma2011electronic}%
  \BibitemOpen
  \bibfield  {author} {\bibinfo {author} {\bibfnamefont {S.~D.}\ \bibnamefont
  {Sarma}}, \bibinfo {author} {\bibfnamefont {S.}~\bibnamefont {Adam}},
  \bibinfo {author} {\bibfnamefont {E.}~\bibnamefont {Hwang}}, \ and\ \bibinfo
  {author} {\bibfnamefont {E.}~\bibnamefont {Rossi}},\ }\href@noop {}
  {\bibfield  {journal} {\bibinfo  {journal} {Rev. Mod. Phys.}\ }\textbf
  {\bibinfo {volume} {83}},\ \bibinfo {pages} {407} (\bibinfo {year}
  {2011})}\BibitemShut {NoStop}%
\bibitem [{\citenamefont {Mayorov}\ \emph {et~al.}(2011)\citenamefont
  {Mayorov}, \citenamefont {Gorbachev}, \citenamefont {Morozov}, \citenamefont
  {Britnell}, \citenamefont {Jalil}, \citenamefont {Ponomarenko}, \citenamefont
  {Blake}, \citenamefont {Novoselov}, \citenamefont {Watanabe}, \citenamefont
  {Taniguchi} \emph {et~al.}}]{mayorov2011micrometer}%
  \BibitemOpen
  \bibfield  {author} {\bibinfo {author} {\bibfnamefont {A.~S.}\ \bibnamefont
  {Mayorov}}, \bibinfo {author} {\bibfnamefont {R.~V.}\ \bibnamefont
  {Gorbachev}}, \bibinfo {author} {\bibfnamefont {S.~V.}\ \bibnamefont
  {Morozov}}, \bibinfo {author} {\bibfnamefont {L.}~\bibnamefont {Britnell}},
  \bibinfo {author} {\bibfnamefont {R.}~\bibnamefont {Jalil}}, \bibinfo
  {author} {\bibfnamefont {L.~A.}\ \bibnamefont {Ponomarenko}}, \bibinfo
  {author} {\bibfnamefont {P.}~\bibnamefont {Blake}}, \bibinfo {author}
  {\bibfnamefont {K.~S.}\ \bibnamefont {Novoselov}}, \bibinfo {author}
  {\bibfnamefont {K.}~\bibnamefont {Watanabe}}, \bibinfo {author}
  {\bibfnamefont {T.}~\bibnamefont {Taniguchi}},  \emph {et~al.},\ }\href@noop
  {} {\bibfield  {journal} {\bibinfo  {journal} {Nano Lett.}\ }\textbf
  {\bibinfo {volume} {11}},\ \bibinfo {pages} {2396} (\bibinfo {year}
  {2011})}\BibitemShut {NoStop}%
\bibitem [{\citenamefont {Stauber}(2014)}]{stauber2014plasmonics}%
  \BibitemOpen
  \bibfield  {author} {\bibinfo {author} {\bibfnamefont {T.}~\bibnamefont
  {Stauber}},\ }\href@noop {} {\bibfield  {journal} {\bibinfo  {journal} {J.
  Phys. Condens. Matter}\ }\textbf {\bibinfo {volume} {26}},\ \bibinfo {pages}
  {123201} (\bibinfo {year} {2014})}\BibitemShut {NoStop}%
\bibitem [{\citenamefont {Christensen}\ \emph {et~al.}(2014)\citenamefont
  {Christensen}, \citenamefont {Wang}, \citenamefont {Jauho}, \citenamefont
  {Wubs},\ and\ \citenamefont {Mortensen}}]{christensen2014classical}%
  \BibitemOpen
  \bibfield  {author} {\bibinfo {author} {\bibfnamefont {T.}~\bibnamefont
  {Christensen}}, \bibinfo {author} {\bibfnamefont {W.}~\bibnamefont {Wang}},
  \bibinfo {author} {\bibfnamefont {A.-P.}\ \bibnamefont {Jauho}}, \bibinfo
  {author} {\bibfnamefont {M.}~\bibnamefont {Wubs}}, \ and\ \bibinfo {author}
  {\bibfnamefont {N.~A.}\ \bibnamefont {Mortensen}},\ }\href@noop {} {\bibfield
   {journal} {\bibinfo  {journal} {Phys. Rev. B}\ }\textbf {\bibinfo {volume}
  {90}},\ \bibinfo {pages} {241414} (\bibinfo {year} {2014})}\BibitemShut
  {NoStop}%
\bibitem [{\citenamefont {Kravets}\ \emph {et~al.}(2014)\citenamefont
  {Kravets}, \citenamefont {Jalil}, \citenamefont {Kim}, \citenamefont
  {Ansell}, \citenamefont {Aznakayeva}, \citenamefont {Thackray}, \citenamefont
  {Britnell}, \citenamefont {Belle}, \citenamefont {Withers}, \citenamefont
  {Radko} \emph {et~al.}}]{kravets2014graphene}%
  \BibitemOpen
  \bibfield  {author} {\bibinfo {author} {\bibfnamefont {V.}~\bibnamefont
  {Kravets}}, \bibinfo {author} {\bibfnamefont {R.}~\bibnamefont {Jalil}},
  \bibinfo {author} {\bibfnamefont {Y.-J.}\ \bibnamefont {Kim}}, \bibinfo
  {author} {\bibfnamefont {D.}~\bibnamefont {Ansell}}, \bibinfo {author}
  {\bibfnamefont {D.}~\bibnamefont {Aznakayeva}}, \bibinfo {author}
  {\bibfnamefont {B.}~\bibnamefont {Thackray}}, \bibinfo {author}
  {\bibfnamefont {L.}~\bibnamefont {Britnell}}, \bibinfo {author}
  {\bibfnamefont {B.}~\bibnamefont {Belle}}, \bibinfo {author} {\bibfnamefont
  {F.}~\bibnamefont {Withers}}, \bibinfo {author} {\bibfnamefont
  {I.}~\bibnamefont {Radko}},  \emph {et~al.},\ }\href@noop {} {\bibfield
  {journal} {\bibinfo  {journal} {Sci. Rep.}\ }\textbf {\bibinfo {volume} {4}}
  (\bibinfo {year} {2014})}\BibitemShut {NoStop}%
\bibitem [{\citenamefont {Grigorenko}\ \emph {et~al.}(2012)\citenamefont
  {Grigorenko}, \citenamefont {Polini},\ and\ \citenamefont
  {Novoselov}}]{grigorenko2012graphene}%
  \BibitemOpen
  \bibfield  {author} {\bibinfo {author} {\bibfnamefont {A.}~\bibnamefont
  {Grigorenko}}, \bibinfo {author} {\bibfnamefont {M.}~\bibnamefont {Polini}},
  \ and\ \bibinfo {author} {\bibfnamefont {K.}~\bibnamefont {Novoselov}},\
  }\href@noop {} {\bibfield  {journal} {\bibinfo  {journal} {Nat. Photonics}\
  }\textbf {\bibinfo {volume} {6}},\ \bibinfo {pages} {749} (\bibinfo {year}
  {2012})}\BibitemShut {NoStop}%
\bibitem [{\citenamefont {Koppens}\ \emph {et~al.}(2011)\citenamefont
  {Koppens}, \citenamefont {Chang},\ and\ \citenamefont {Garcia~de
  Abajo}}]{koppens2011graphene}%
  \BibitemOpen
  \bibfield  {author} {\bibinfo {author} {\bibfnamefont {F.~H.}\ \bibnamefont
  {Koppens}}, \bibinfo {author} {\bibfnamefont {D.~E.}\ \bibnamefont {Chang}},
  \ and\ \bibinfo {author} {\bibfnamefont {F.~J.}\ \bibnamefont {Garcia~de
  Abajo}},\ }\href@noop {} {\bibfield  {journal} {\bibinfo  {journal} {Nano
  Lett.}\ }\textbf {\bibinfo {volume} {11}},\ \bibinfo {pages} {3370} (\bibinfo
  {year} {2011})}\BibitemShut {NoStop}%
\bibitem [{\citenamefont {Hwang}\ and\ \citenamefont
  {Sarma}(2007)}]{hwang2007dielectric}%
  \BibitemOpen
  \bibfield  {author} {\bibinfo {author} {\bibfnamefont {E.}~\bibnamefont
  {Hwang}}\ and\ \bibinfo {author} {\bibfnamefont {S.~D.}\ \bibnamefont
  {Sarma}},\ }\href@noop {} {\bibfield  {journal} {\bibinfo  {journal} {Phys.
  Rev. B}\ }\textbf {\bibinfo {volume} {75}},\ \bibinfo {pages} {205418}
  (\bibinfo {year} {2007})}\BibitemShut {NoStop}%
\bibitem [{\citenamefont {Jablan}\ \emph {et~al.}(2009)\citenamefont {Jablan},
  \citenamefont {Buljan},\ and\ \citenamefont
  {Solja{\v{c}}i{\'c}}}]{jablan2009plasmonics}%
  \BibitemOpen
  \bibfield  {author} {\bibinfo {author} {\bibfnamefont {M.}~\bibnamefont
  {Jablan}}, \bibinfo {author} {\bibfnamefont {H.}~\bibnamefont {Buljan}}, \
  and\ \bibinfo {author} {\bibfnamefont {M.}~\bibnamefont
  {Solja{\v{c}}i{\'c}}},\ }\href@noop {} {\bibfield  {journal} {\bibinfo
  {journal} {Phys. Rev. B}\ }\textbf {\bibinfo {volume} {80}},\ \bibinfo
  {pages} {245435} (\bibinfo {year} {2009})}\BibitemShut {NoStop}%
\bibitem [{\citenamefont {Ando}\ \emph {et~al.}(1982)\citenamefont {Ando},
  \citenamefont {Fowler},\ and\ \citenamefont {Stern}}]{ando1982electronic}%
  \BibitemOpen
  \bibfield  {author} {\bibinfo {author} {\bibfnamefont {T.}~\bibnamefont
  {Ando}}, \bibinfo {author} {\bibfnamefont {A.~B.}\ \bibnamefont {Fowler}}, \
  and\ \bibinfo {author} {\bibfnamefont {F.}~\bibnamefont {Stern}},\
  }\href@noop {} {\bibfield  {journal} {\bibinfo  {journal} {Rev. Mod. Phys.}\
  }\textbf {\bibinfo {volume} {54}},\ \bibinfo {pages} {437} (\bibinfo {year}
  {1982})}\BibitemShut {NoStop}%
\bibitem [{\citenamefont {Liou}\ \emph {et~al.}(2015)\citenamefont {Liou},
  \citenamefont {Shie}, \citenamefont {Chen}, \citenamefont {Breitwieser},
  \citenamefont {Pai}, \citenamefont {Guo},\ and\ \citenamefont
  {Chu}}]{liou2015pi}%
  \BibitemOpen
  \bibfield  {author} {\bibinfo {author} {\bibfnamefont {S.}~\bibnamefont
  {Liou}}, \bibinfo {author} {\bibfnamefont {C.-S.}\ \bibnamefont {Shie}},
  \bibinfo {author} {\bibfnamefont {C.}~\bibnamefont {Chen}}, \bibinfo {author}
  {\bibfnamefont {R.}~\bibnamefont {Breitwieser}}, \bibinfo {author}
  {\bibfnamefont {W.}~\bibnamefont {Pai}}, \bibinfo {author} {\bibfnamefont
  {G.}~\bibnamefont {Guo}}, \ and\ \bibinfo {author} {\bibfnamefont {M.-W.}\
  \bibnamefont {Chu}},\ }\href@noop {} {\bibfield  {journal} {\bibinfo
  {journal} {Phys. Rev. B}\ }\textbf {\bibinfo {volume} {91}},\ \bibinfo
  {pages} {045418} (\bibinfo {year} {2015})}\BibitemShut {NoStop}%
\bibitem [{\citenamefont {Eberlein}\ \emph {et~al.}(2008)\citenamefont
  {Eberlein}, \citenamefont {Bangert}, \citenamefont {Nair}, \citenamefont
  {Jones}, \citenamefont {Gass}, \citenamefont {Bleloch}, \citenamefont
  {Novoselov}, \citenamefont {Geim},\ and\ \citenamefont
  {Briddon}}]{eberlein2008plasmon}%
  \BibitemOpen
  \bibfield  {author} {\bibinfo {author} {\bibfnamefont {T.}~\bibnamefont
  {Eberlein}}, \bibinfo {author} {\bibfnamefont {U.}~\bibnamefont {Bangert}},
  \bibinfo {author} {\bibfnamefont {R.}~\bibnamefont {Nair}}, \bibinfo {author}
  {\bibfnamefont {R.}~\bibnamefont {Jones}}, \bibinfo {author} {\bibfnamefont
  {M.}~\bibnamefont {Gass}}, \bibinfo {author} {\bibfnamefont {A.}~\bibnamefont
  {Bleloch}}, \bibinfo {author} {\bibfnamefont {K.}~\bibnamefont {Novoselov}},
  \bibinfo {author} {\bibfnamefont {A.}~\bibnamefont {Geim}}, \ and\ \bibinfo
  {author} {\bibfnamefont {P.}~\bibnamefont {Briddon}},\ }\href@noop {}
  {\bibfield  {journal} {\bibinfo  {journal} {Phys. Rev. B}\ }\textbf {\bibinfo
  {volume} {77}},\ \bibinfo {pages} {233406} (\bibinfo {year}
  {2008})}\BibitemShut {NoStop}%
\bibitem [{\citenamefont {Liu}\ and\ \citenamefont
  {Willis}(2010)}]{liu2010plasmon}%
  \BibitemOpen
  \bibfield  {author} {\bibinfo {author} {\bibfnamefont {Y.}~\bibnamefont
  {Liu}}\ and\ \bibinfo {author} {\bibfnamefont {R.~F.}\ \bibnamefont
  {Willis}},\ }\href@noop {} {\bibfield  {journal} {\bibinfo  {journal} {Phys.
  Rev. B}\ }\textbf {\bibinfo {volume} {81}},\ \bibinfo {pages} {081406}
  (\bibinfo {year} {2010})}\BibitemShut {NoStop}%
\bibitem [{\citenamefont {Kramberger}\ \emph {et~al.}(2008)\citenamefont
  {Kramberger}, \citenamefont {Hambach}, \citenamefont {Giorgetti},
  \citenamefont {R{\"u}mmeli}, \citenamefont {Knupfer}, \citenamefont {Fink},
  \citenamefont {B{\"u}chner}, \citenamefont {Reining}, \citenamefont
  {Einarsson}, \citenamefont {Maruyama},\ and\ \citenamefont
  {Sottile}}]{kramberger2008linear}%
  \BibitemOpen
  \bibfield  {author} {\bibinfo {author} {\bibfnamefont {C.}~\bibnamefont
  {Kramberger}}, \bibinfo {author} {\bibfnamefont {C.}~\bibnamefont {Hambach}},
  \bibinfo {author} {\bibfnamefont {C.}~\bibnamefont {Giorgetti}}, \bibinfo
  {author} {\bibfnamefont {C.}~\bibnamefont {R{\"u}mmeli}}, \bibinfo {author}
  {\bibfnamefont {C.}~\bibnamefont {Knupfer}}, \bibinfo {author} {\bibfnamefont
  {C.}~\bibnamefont {Fink}}, \bibinfo {author} {\bibfnamefont {C.}~\bibnamefont
  {B{\"u}chner}}, \bibinfo {author} {\bibfnamefont {C.}~\bibnamefont
  {Reining}}, \bibinfo {author} {\bibfnamefont {C.}~\bibnamefont {Einarsson}},
  \bibinfo {author} {\bibfnamefont {C.}~\bibnamefont {Maruyama}}, \ and\
  \bibinfo {author} {\bibfnamefont {C.}~\bibnamefont {Sottile}},\ }\href@noop
  {} {\bibfield  {journal} {\bibinfo  {journal} {Phys. Rev. Lett.}\ }\textbf
  {\bibinfo {volume} {100}},\ \bibinfo {pages} {196803} (\bibinfo {year}
  {2008})}\BibitemShut {NoStop}%
\bibitem [{\citenamefont {Brar}\ \emph {et~al.}(2014)\citenamefont {Brar},
  \citenamefont {Jang}, \citenamefont {Sherrott}, \citenamefont {Kim},
  \citenamefont {Lopez}, \citenamefont {Kim}, \citenamefont {Choi},\ and\
  \citenamefont {Atwater}}]{brar2014hybrid}%
  \BibitemOpen
  \bibfield  {author} {\bibinfo {author} {\bibfnamefont {V.~W.}\ \bibnamefont
  {Brar}}, \bibinfo {author} {\bibfnamefont {M.~S.}\ \bibnamefont {Jang}},
  \bibinfo {author} {\bibfnamefont {M.}~\bibnamefont {Sherrott}}, \bibinfo
  {author} {\bibfnamefont {S.}~\bibnamefont {Kim}}, \bibinfo {author}
  {\bibfnamefont {J.~J.}\ \bibnamefont {Lopez}}, \bibinfo {author}
  {\bibfnamefont {L.~B.}\ \bibnamefont {Kim}}, \bibinfo {author} {\bibfnamefont
  {M.}~\bibnamefont {Choi}}, \ and\ \bibinfo {author} {\bibfnamefont
  {H.}~\bibnamefont {Atwater}},\ }\href@noop {} {\bibfield  {journal} {\bibinfo
   {journal} {Nano Lett.}\ }\textbf {\bibinfo {volume} {14}},\ \bibinfo {pages}
  {3876} (\bibinfo {year} {2014})}\BibitemShut {NoStop}%
\bibitem [{\citenamefont {Yan}\ \emph {et~al.}(2013)\citenamefont {Yan},
  \citenamefont {Low}, \citenamefont {Zhu}, \citenamefont {Wu}, \citenamefont
  {Freitag}, \citenamefont {Li}, \citenamefont {Guinea}, \citenamefont
  {Avouris},\ and\ \citenamefont {Xia}}]{yan2013damping}%
  \BibitemOpen
  \bibfield  {author} {\bibinfo {author} {\bibfnamefont {H.}~\bibnamefont
  {Yan}}, \bibinfo {author} {\bibfnamefont {T.}~\bibnamefont {Low}}, \bibinfo
  {author} {\bibfnamefont {W.}~\bibnamefont {Zhu}}, \bibinfo {author}
  {\bibfnamefont {Y.}~\bibnamefont {Wu}}, \bibinfo {author} {\bibfnamefont
  {M.}~\bibnamefont {Freitag}}, \bibinfo {author} {\bibfnamefont
  {X.}~\bibnamefont {Li}}, \bibinfo {author} {\bibfnamefont {F.}~\bibnamefont
  {Guinea}}, \bibinfo {author} {\bibfnamefont {P.}~\bibnamefont {Avouris}}, \
  and\ \bibinfo {author} {\bibfnamefont {F.}~\bibnamefont {Xia}},\ }\href@noop
  {} {\bibfield  {journal} {\bibinfo  {journal} {Nat. Photonics}\ }\textbf
  {\bibinfo {volume} {7}},\ \bibinfo {pages} {394} (\bibinfo {year}
  {2013})}\BibitemShut {NoStop}%
\bibitem [{\citenamefont {Woessner}\ \emph {et~al.}(2014)\citenamefont
  {Woessner}, \citenamefont {Lundeberg}, \citenamefont {Gao}, \citenamefont
  {Principi}, \citenamefont {Alonso-Gonz{\'a}lez}, \citenamefont {Carrega},
  \citenamefont {Watanabe}, \citenamefont {Taniguchi}, \citenamefont {Vignale},
  \citenamefont {Polini} \emph {et~al.}}]{woessner2014highly}%
  \BibitemOpen
  \bibfield  {author} {\bibinfo {author} {\bibfnamefont {A.}~\bibnamefont
  {Woessner}}, \bibinfo {author} {\bibfnamefont {M.~B.}\ \bibnamefont
  {Lundeberg}}, \bibinfo {author} {\bibfnamefont {Y.}~\bibnamefont {Gao}},
  \bibinfo {author} {\bibfnamefont {A.}~\bibnamefont {Principi}}, \bibinfo
  {author} {\bibfnamefont {P.}~\bibnamefont {Alonso-Gonz{\'a}lez}}, \bibinfo
  {author} {\bibfnamefont {M.}~\bibnamefont {Carrega}}, \bibinfo {author}
  {\bibfnamefont {K.}~\bibnamefont {Watanabe}}, \bibinfo {author}
  {\bibfnamefont {T.}~\bibnamefont {Taniguchi}}, \bibinfo {author}
  {\bibfnamefont {G.}~\bibnamefont {Vignale}}, \bibinfo {author} {\bibfnamefont
  {M.}~\bibnamefont {Polini}},  \emph {et~al.},\ }\href@noop {} {\bibfield
  {journal} {\bibinfo  {journal} {Nat. Mater.}\ } (\bibinfo {year}
  {2014})}\BibitemShut {NoStop}%
\bibitem [{\citenamefont {Fei}\ \emph {et~al.}(2012)\citenamefont {Fei},
  \citenamefont {Rodin}, \citenamefont {Andreev}, \citenamefont {Bao},
  \citenamefont {McLeod}, \citenamefont {Wagner}, \citenamefont {Zhang},
  \citenamefont {Zhao}, \citenamefont {Thiemens}, \citenamefont {Dominguez}
  \emph {et~al.}}]{fei2012gate}%
  \BibitemOpen
  \bibfield  {author} {\bibinfo {author} {\bibfnamefont {Z.}~\bibnamefont
  {Fei}}, \bibinfo {author} {\bibfnamefont {A.}~\bibnamefont {Rodin}}, \bibinfo
  {author} {\bibfnamefont {G.}~\bibnamefont {Andreev}}, \bibinfo {author}
  {\bibfnamefont {W.}~\bibnamefont {Bao}}, \bibinfo {author} {\bibfnamefont
  {A.}~\bibnamefont {McLeod}}, \bibinfo {author} {\bibfnamefont
  {M.}~\bibnamefont {Wagner}}, \bibinfo {author} {\bibfnamefont
  {L.}~\bibnamefont {Zhang}}, \bibinfo {author} {\bibfnamefont
  {Z.}~\bibnamefont {Zhao}}, \bibinfo {author} {\bibfnamefont {M.}~\bibnamefont
  {Thiemens}}, \bibinfo {author} {\bibfnamefont {G.}~\bibnamefont {Dominguez}},
   \emph {et~al.},\ }\href@noop {} {\bibfield  {journal} {\bibinfo  {journal}
  {Nature}\ }\textbf {\bibinfo {volume} {487}},\ \bibinfo {pages} {82}
  (\bibinfo {year} {2012})}\BibitemShut {NoStop}%
\bibitem [{\citenamefont {Lu}\ \emph {et~al.}(2009)\citenamefont {Lu},
  \citenamefont {Loh}, \citenamefont {Huang}, \citenamefont {Chen},\ and\
  \citenamefont {Wee}}]{lu2009plasmon}%
  \BibitemOpen
  \bibfield  {author} {\bibinfo {author} {\bibfnamefont {J.}~\bibnamefont
  {Lu}}, \bibinfo {author} {\bibfnamefont {K.~P.}\ \bibnamefont {Loh}},
  \bibinfo {author} {\bibfnamefont {H.}~\bibnamefont {Huang}}, \bibinfo
  {author} {\bibfnamefont {W.}~\bibnamefont {Chen}}, \ and\ \bibinfo {author}
  {\bibfnamefont {A.~T.}\ \bibnamefont {Wee}},\ }\href@noop {} {\bibfield
  {journal} {\bibinfo  {journal} {Phys. Rev. B}\ }\textbf {\bibinfo {volume}
  {80}},\ \bibinfo {pages} {113410} (\bibinfo {year} {2009})}\BibitemShut
  {NoStop}%
\bibitem [{\citenamefont {Hwang}\ \emph {et~al.}(2010)\citenamefont {Hwang},
  \citenamefont {Sensarma},\ and\ \citenamefont {Sarma}}]{hwang2010plasmon}%
  \BibitemOpen
  \bibfield  {author} {\bibinfo {author} {\bibfnamefont {E.}~\bibnamefont
  {Hwang}}, \bibinfo {author} {\bibfnamefont {R.}~\bibnamefont {Sensarma}}, \
  and\ \bibinfo {author} {\bibfnamefont {S.~D.}\ \bibnamefont {Sarma}},\
  }\href@noop {} {\bibfield  {journal} {\bibinfo  {journal} {Phys. Rev. B}\
  }\textbf {\bibinfo {volume} {82}},\ \bibinfo {pages} {195406} (\bibinfo
  {year} {2010})}\BibitemShut {NoStop}%
\bibitem [{\citenamefont {Hwang}\ and\ \citenamefont
  {Sarma}(2013)}]{hwang2013surface}%
  \BibitemOpen
  \bibfield  {author} {\bibinfo {author} {\bibfnamefont {E.}~\bibnamefont
  {Hwang}}\ and\ \bibinfo {author} {\bibfnamefont {S.~D.}\ \bibnamefont
  {Sarma}},\ }\href@noop {} {\bibfield  {journal} {\bibinfo  {journal} {Phys.
  Rev. B}\ }\textbf {\bibinfo {volume} {87}},\ \bibinfo {pages} {115432}
  (\bibinfo {year} {2013})}\BibitemShut {NoStop}%
\bibitem [{\citenamefont {Ahn}\ \emph {et~al.}(2014)\citenamefont {Ahn},
  \citenamefont {Hwang},\ and\ \citenamefont {Min}}]{ahn2014inelastic}%
  \BibitemOpen
  \bibfield  {author} {\bibinfo {author} {\bibfnamefont {S.}~\bibnamefont
  {Ahn}}, \bibinfo {author} {\bibfnamefont {E.}~\bibnamefont {Hwang}}, \ and\
  \bibinfo {author} {\bibfnamefont {H.}~\bibnamefont {Min}},\ }\href@noop {}
  {\bibfield  {journal} {\bibinfo  {journal} {Phys. Rev. B}\ }\textbf {\bibinfo
  {volume} {90}},\ \bibinfo {pages} {245436} (\bibinfo {year}
  {2014})}\BibitemShut {NoStop}%
\bibitem [{\citenamefont {Jablan}\ \emph {et~al.}(2011)\citenamefont {Jablan},
  \citenamefont {Solja{\v{c}}i{\'c}},\ and\ \citenamefont
  {Buljan}}]{jablan2011unconventional}%
  \BibitemOpen
  \bibfield  {author} {\bibinfo {author} {\bibfnamefont {M.}~\bibnamefont
  {Jablan}}, \bibinfo {author} {\bibfnamefont {M.}~\bibnamefont
  {Solja{\v{c}}i{\'c}}}, \ and\ \bibinfo {author} {\bibfnamefont
  {H.}~\bibnamefont {Buljan}},\ }\href@noop {} {\bibfield  {journal} {\bibinfo
  {journal} {Phys. Rev. B}\ }\textbf {\bibinfo {volume} {83}},\ \bibinfo
  {pages} {161409} (\bibinfo {year} {2011})}\BibitemShut {NoStop}%
\bibitem [{\citenamefont {Ong}\ and\ \citenamefont
  {Fischetti}(2012)}]{ong2012theory}%
  \BibitemOpen
  \bibfield  {author} {\bibinfo {author} {\bibfnamefont {Z.-Y.}\ \bibnamefont
  {Ong}}\ and\ \bibinfo {author} {\bibfnamefont {M.~V.}\ \bibnamefont
  {Fischetti}},\ }\href@noop {} {\bibfield  {journal} {\bibinfo  {journal}
  {Phys. Rev. B}\ }\textbf {\bibinfo {volume} {86}},\ \bibinfo {pages} {165422}
  (\bibinfo {year} {2012})}\BibitemShut {NoStop}%
\bibitem [{\citenamefont {Perebeinos}\ and\ \citenamefont
  {Avouris}(2010)}]{perebeinos2010inelastic}%
  \BibitemOpen
  \bibfield  {author} {\bibinfo {author} {\bibfnamefont {V.}~\bibnamefont
  {Perebeinos}}\ and\ \bibinfo {author} {\bibfnamefont {P.}~\bibnamefont
  {Avouris}},\ }\href@noop {} {\bibfield  {journal} {\bibinfo  {journal} {Phys.
  Rev. B}\ }\textbf {\bibinfo {volume} {81}},\ \bibinfo {pages} {195442}
  (\bibinfo {year} {2010})}\BibitemShut {NoStop}%
\bibitem [{\citenamefont {Sule}\ and\ \citenamefont
  {Knezevic}(2012)}]{sule2012phonon}%
  \BibitemOpen
  \bibfield  {author} {\bibinfo {author} {\bibfnamefont {N.}~\bibnamefont
  {Sule}}\ and\ \bibinfo {author} {\bibfnamefont {I.}~\bibnamefont
  {Knezevic}},\ }\href@noop {} {\bibfield  {journal} {\bibinfo  {journal} {J.
  Appl. Phys.}\ }\textbf {\bibinfo {volume} {112}},\ \bibinfo {pages} {053702}
  (\bibinfo {year} {2012})}\BibitemShut {NoStop}%
\bibitem [{\citenamefont {Li}\ \emph {et~al.}(2010)\citenamefont {Li},
  \citenamefont {Barry}, \citenamefont {Zavada}, \citenamefont {Nardelli},\
  and\ \citenamefont {Kim}}]{li2010surface}%
  \BibitemOpen
  \bibfield  {author} {\bibinfo {author} {\bibfnamefont {X.}~\bibnamefont
  {Li}}, \bibinfo {author} {\bibfnamefont {E.}~\bibnamefont {Barry}}, \bibinfo
  {author} {\bibfnamefont {J.}~\bibnamefont {Zavada}}, \bibinfo {author}
  {\bibfnamefont {M.~B.}\ \bibnamefont {Nardelli}}, \ and\ \bibinfo {author}
  {\bibfnamefont {K.}~\bibnamefont {Kim}},\ }\href@noop {} {\bibfield
  {journal} {\bibinfo  {journal} {Appl. Phys. Lett.}\ }\textbf {\bibinfo
  {volume} {97}},\ \bibinfo {pages} {232105} (\bibinfo {year}
  {2010})}\BibitemShut {NoStop}%
\bibitem [{\citenamefont {Stauber}\ \emph {et~al.}(2007)\citenamefont
  {Stauber}, \citenamefont {Peres},\ and\ \citenamefont
  {Guinea}}]{stauber2007electronic}%
  \BibitemOpen
  \bibfield  {author} {\bibinfo {author} {\bibfnamefont {T.}~\bibnamefont
  {Stauber}}, \bibinfo {author} {\bibfnamefont {N.}~\bibnamefont {Peres}}, \
  and\ \bibinfo {author} {\bibfnamefont {F.}~\bibnamefont {Guinea}},\
  }\href@noop {} {\bibfield  {journal} {\bibinfo  {journal} {Phys. Rev. B}\
  }\textbf {\bibinfo {volume} {76}},\ \bibinfo {pages} {205423} (\bibinfo
  {year} {2007})}\BibitemShut {NoStop}%
\bibitem [{\citenamefont {Hwang}\ \emph {et~al.}(2007)\citenamefont {Hwang},
  \citenamefont {Adam},\ and\ \citenamefont {Sarma}}]{hwang2007carrier}%
  \BibitemOpen
  \bibfield  {author} {\bibinfo {author} {\bibfnamefont {E.}~\bibnamefont
  {Hwang}}, \bibinfo {author} {\bibfnamefont {S.}~\bibnamefont {Adam}}, \ and\
  \bibinfo {author} {\bibfnamefont {S.~D.}\ \bibnamefont {Sarma}},\ }\href@noop
  {} {\bibfield  {journal} {\bibinfo  {journal} {Phys. Rev. Lett.}\ }\textbf
  {\bibinfo {volume} {98}},\ \bibinfo {pages} {186806} (\bibinfo {year}
  {2007})}\BibitemShut {NoStop}%
\bibitem [{\citenamefont {Sule}\ \emph
  {et~al.}(2014{\natexlab{a}})\citenamefont {Sule}, \citenamefont {Hagness},\
  and\ \citenamefont {Knezevic}}]{sule2014clustered}%
  \BibitemOpen
  \bibfield  {author} {\bibinfo {author} {\bibfnamefont {N.}~\bibnamefont
  {Sule}}, \bibinfo {author} {\bibfnamefont {S.}~\bibnamefont {Hagness}}, \
  and\ \bibinfo {author} {\bibfnamefont {I.}~\bibnamefont {Knezevic}},\
  }\href@noop {} {\bibfield  {journal} {\bibinfo  {journal} {Phys. Rev. B}\
  }\textbf {\bibinfo {volume} {89}},\ \bibinfo {pages} {165402} (\bibinfo
  {year} {2014}{\natexlab{a}})}\BibitemShut {NoStop}%
\bibitem [{\citenamefont {Fratini}\ and\ \citenamefont
  {Guinea}(2008)}]{fratini2008substrate}%
  \BibitemOpen
  \bibfield  {author} {\bibinfo {author} {\bibfnamefont {S.}~\bibnamefont
  {Fratini}}\ and\ \bibinfo {author} {\bibfnamefont {F.}~\bibnamefont
  {Guinea}},\ }\href@noop {} {\bibfield  {journal} {\bibinfo  {journal} {Phys.
  Rev. B}\ }\textbf {\bibinfo {volume} {77}},\ \bibinfo {pages} {195415}
  (\bibinfo {year} {2008})}\BibitemShut {NoStop}%
\bibitem [{\citenamefont {Schiefele}\ \emph {et~al.}(2012)\citenamefont
  {Schiefele}, \citenamefont {Sols},\ and\ \citenamefont
  {Guinea}}]{schiefele2012temperature}%
  \BibitemOpen
  \bibfield  {author} {\bibinfo {author} {\bibfnamefont {J.}~\bibnamefont
  {Schiefele}}, \bibinfo {author} {\bibfnamefont {F.}~\bibnamefont {Sols}}, \
  and\ \bibinfo {author} {\bibfnamefont {F.}~\bibnamefont {Guinea}},\
  }\href@noop {} {\bibfield  {journal} {\bibinfo  {journal} {Phys. Rev. B}\
  }\textbf {\bibinfo {volume} {85}},\ \bibinfo {pages} {195420} (\bibinfo
  {year} {2012})}\BibitemShut {NoStop}%
\bibitem [{\citenamefont {Konar}\ \emph {et~al.}(2010)\citenamefont {Konar},
  \citenamefont {Fang},\ and\ \citenamefont {Jena}}]{konar2010effect}%
  \BibitemOpen
  \bibfield  {author} {\bibinfo {author} {\bibfnamefont {A.}~\bibnamefont
  {Konar}}, \bibinfo {author} {\bibfnamefont {T.}~\bibnamefont {Fang}}, \ and\
  \bibinfo {author} {\bibfnamefont {D.}~\bibnamefont {Jena}},\ }\href@noop {}
  {\bibfield  {journal} {\bibinfo  {journal} {Phys. Rev. B}\ }\textbf {\bibinfo
  {volume} {82}},\ \bibinfo {pages} {115452} (\bibinfo {year}
  {2010})}\BibitemShut {NoStop}%
\bibitem [{\citenamefont {Mermin}(1970)}]{mermin1970lindhard}%
  \BibitemOpen
  \bibfield  {author} {\bibinfo {author} {\bibfnamefont {N.~D.}\ \bibnamefont
  {Mermin}},\ }\href@noop {} {\bibfield  {journal} {\bibinfo  {journal} {Phys.
  Rev. B}\ }\textbf {\bibinfo {volume} {1}},\ \bibinfo {pages} {2362} (\bibinfo
  {year} {1970})}\BibitemShut {NoStop}%
\bibitem [{\citenamefont {Lundstrom}(2009)}]{lundstrom2009fundamentals}%
  \BibitemOpen
  \bibfield  {author} {\bibinfo {author} {\bibfnamefont {M.}~\bibnamefont
  {Lundstrom}},\ }\href {https://books.google.com/books?id=Vrkd\_dSC3zwC}
  {\emph {\bibinfo {title} {Fundamentals of Carrier Transport}}}\ (\bibinfo
  {publisher} {Cambridge University Press},\ \bibinfo {year}
  {2009})\BibitemShut {NoStop}%
\bibitem [{\citenamefont {Yager}(1936)}]{YagerJAP36}%
  \BibitemOpen
  \bibfield  {author} {\bibinfo {author} {\bibfnamefont {W.~A.}\ \bibnamefont
  {Yager}},\ }\href@noop {} {\bibfield  {journal} {\bibinfo  {journal} {J.
  Appl. Phys.}\ }\textbf {\bibinfo {volume} {7}} (\bibinfo {year}
  {1936})}\BibitemShut {NoStop}%
\bibitem [{\citenamefont {Hill}\ and\ \citenamefont
  {Dissado}(1985)}]{HillDissadoJPC85}%
  \BibitemOpen
  \bibfield  {author} {\bibinfo {author} {\bibfnamefont {R.~M.}\ \bibnamefont
  {Hill}}\ and\ \bibinfo {author} {\bibfnamefont {L.~A.}\ \bibnamefont
  {Dissado}},\ }\href {http://stacks.iop.org/0022-3719/18/i=19/a=021}
  {\bibfield  {journal} {\bibinfo  {journal} {Journal of Physics C: Solid State
  Physics}\ }\textbf {\bibinfo {volume} {18}},\ \bibinfo {pages} {3829}
  (\bibinfo {year} {1985})}\BibitemShut {NoStop}%
\bibitem [{\citenamefont {Beard}\ \emph {et~al.}(2000)\citenamefont {Beard},
  \citenamefont {Turner},\ and\ \citenamefont {Schmuttenmaer}}]{BeardPRB2000}%
  \BibitemOpen
  \bibfield  {author} {\bibinfo {author} {\bibfnamefont {M.~C.}\ \bibnamefont
  {Beard}}, \bibinfo {author} {\bibfnamefont {G.~M.}\ \bibnamefont {Turner}}, \
  and\ \bibinfo {author} {\bibfnamefont {C.~A.}\ \bibnamefont
  {Schmuttenmaer}},\ }\href {\doibase 10.1103/PhysRevB.62.15764} {\bibfield
  {journal} {\bibinfo  {journal} {Phys. Rev. B}\ }\textbf {\bibinfo {volume}
  {62}},\ \bibinfo {pages} {15764} (\bibinfo {year} {2000})}\BibitemShut
  {NoStop}%
\bibitem [{\citenamefont {Willis}\ \emph {et~al.}(2013)\citenamefont {Willis},
  \citenamefont {Hagness},\ and\ \citenamefont
  {Knezevic}}]{willis2013generalized}%
  \BibitemOpen
  \bibfield  {author} {\bibinfo {author} {\bibfnamefont {K.}~\bibnamefont
  {Willis}}, \bibinfo {author} {\bibfnamefont {S.}~\bibnamefont {Hagness}}, \
  and\ \bibinfo {author} {\bibfnamefont {I.}~\bibnamefont {Knezevic}},\
  }\href@noop {} {\bibfield  {journal} {\bibinfo  {journal} {Applied Physics
  Letters}\ }\textbf {\bibinfo {volume} {102}},\ \bibinfo {pages} {122113}
  (\bibinfo {year} {2013})}\BibitemShut {NoStop}%
\bibitem [{\citenamefont {Sule}\ \emph
  {et~al.}(2014{\natexlab{b}})\citenamefont {Sule}, \citenamefont {Willis},
  \citenamefont {Hagness},\ and\ \citenamefont {Knezevic}}]{sule2014terahertz}%
  \BibitemOpen
  \bibfield  {author} {\bibinfo {author} {\bibfnamefont {N.}~\bibnamefont
  {Sule}}, \bibinfo {author} {\bibfnamefont {K.}~\bibnamefont {Willis}},
  \bibinfo {author} {\bibfnamefont {S.}~\bibnamefont {Hagness}}, \ and\
  \bibinfo {author} {\bibfnamefont {I.}~\bibnamefont {Knezevic}},\ }\href@noop
  {} {\bibfield  {journal} {\bibinfo  {journal} {Phys. Rev. B}\ }\textbf
  {\bibinfo {volume} {90}},\ \bibinfo {pages} {045431} (\bibinfo {year}
  {2014}{\natexlab{b}})}\BibitemShut {NoStop}%
\bibitem [{\citenamefont {Breuer}\ and\ \citenamefont
  {Petruccione}(2002)}]{breuer2002theory}%
  \BibitemOpen
  \bibfield  {author} {\bibinfo {author} {\bibfnamefont {H.-P.}\ \bibnamefont
  {Breuer}}\ and\ \bibinfo {author} {\bibfnamefont {F.}~\bibnamefont
  {Petruccione}},\ }\href@noop {} {\emph {\bibinfo {title} {The theory of open
  quantum systems}}}\ (\bibinfo  {publisher} {Oxford university press},\
  \bibinfo {year} {2002})\BibitemShut {NoStop}%
\bibitem [{\citenamefont {Knezevic}\ and\ \citenamefont
  {Novakovic}(2013)}]{knezevic2013time}%
  \BibitemOpen
  \bibfield  {author} {\bibinfo {author} {\bibfnamefont {I.}~\bibnamefont
  {Knezevic}}\ and\ \bibinfo {author} {\bibfnamefont {B.}~\bibnamefont
  {Novakovic}},\ }\href@noop {} {\bibfield  {journal} {\bibinfo  {journal} {J.
  Comput. Electron.}\ }\textbf {\bibinfo {volume} {12}},\ \bibinfo {pages}
  {363} (\bibinfo {year} {2013})}\BibitemShut {NoStop}%
\bibitem [{\citenamefont {Wu}\ \emph {et~al.}(2012)\citenamefont {Wu},
  \citenamefont {Jenkins}, \citenamefont {Valdes-Garcia}, \citenamefont
  {Farmer}, \citenamefont {Zhu}, \citenamefont {Bol}, \citenamefont
  {Dimitrakopoulos}, \citenamefont {Zhu}, \citenamefont {Xia}, \citenamefont
  {Avouris} \emph {et~al.}}]{wu2012state}%
  \BibitemOpen
  \bibfield  {author} {\bibinfo {author} {\bibfnamefont {Y.}~\bibnamefont
  {Wu}}, \bibinfo {author} {\bibfnamefont {K.~A.}\ \bibnamefont {Jenkins}},
  \bibinfo {author} {\bibfnamefont {A.}~\bibnamefont {Valdes-Garcia}}, \bibinfo
  {author} {\bibfnamefont {D.~B.}\ \bibnamefont {Farmer}}, \bibinfo {author}
  {\bibfnamefont {Y.}~\bibnamefont {Zhu}}, \bibinfo {author} {\bibfnamefont
  {A.~A.}\ \bibnamefont {Bol}}, \bibinfo {author} {\bibfnamefont
  {C.}~\bibnamefont {Dimitrakopoulos}}, \bibinfo {author} {\bibfnamefont
  {W.}~\bibnamefont {Zhu}}, \bibinfo {author} {\bibfnamefont {F.}~\bibnamefont
  {Xia}}, \bibinfo {author} {\bibfnamefont {P.}~\bibnamefont {Avouris}},  \emph
  {et~al.},\ }\href@noop {} {\bibfield  {journal} {\bibinfo  {journal} {Nano
  Lett.}\ }\textbf {\bibinfo {volume} {12}},\ \bibinfo {pages} {3062} (\bibinfo
  {year} {2012})}\BibitemShut {NoStop}%
\bibitem [{\citenamefont {Chen}\ \emph {et~al.}(2008)\citenamefont {Chen},
  \citenamefont {Jang}, \citenamefont {Xiao}, \citenamefont {Ishigami},\ and\
  \citenamefont {Fuhrer}}]{chen2008intrinsic}%
  \BibitemOpen
  \bibfield  {author} {\bibinfo {author} {\bibfnamefont {J.-H.}\ \bibnamefont
  {Chen}}, \bibinfo {author} {\bibfnamefont {C.}~\bibnamefont {Jang}}, \bibinfo
  {author} {\bibfnamefont {S.}~\bibnamefont {Xiao}}, \bibinfo {author}
  {\bibfnamefont {M.}~\bibnamefont {Ishigami}}, \ and\ \bibinfo {author}
  {\bibfnamefont {M.~S.}\ \bibnamefont {Fuhrer}},\ }\href@noop {} {\bibfield
  {journal} {\bibinfo  {journal} {Nat. Nanotechnol.}\ }\textbf {\bibinfo
  {volume} {3}},\ \bibinfo {pages} {206} (\bibinfo {year} {2008})}\BibitemShut
  {NoStop}%
\bibitem [{\citenamefont {Dorgan}\ \emph {et~al.}(2010)\citenamefont {Dorgan},
  \citenamefont {Bae},\ and\ \citenamefont {Pop}}]{dorgan2010mobility}%
  \BibitemOpen
  \bibfield  {author} {\bibinfo {author} {\bibfnamefont {V.~E.}\ \bibnamefont
  {Dorgan}}, \bibinfo {author} {\bibfnamefont {M.-H.}\ \bibnamefont {Bae}}, \
  and\ \bibinfo {author} {\bibfnamefont {E.}~\bibnamefont {Pop}},\ }\href@noop
  {} {\bibfield  {journal} {\bibinfo  {journal} {Appl. Phys. Lett.}\ }\textbf
  {\bibinfo {volume} {97}},\ \bibinfo {pages} {082112} (\bibinfo {year}
  {2010})}\BibitemShut {NoStop}%
\bibitem [{\citenamefont {Fei}\ \emph {et~al.}(2011)\citenamefont {Fei},
  \citenamefont {Andreev}, \citenamefont {Bao}, \citenamefont {Zhang},
  \citenamefont {S.~McLeod}, \citenamefont {Wang}, \citenamefont {Stewart},
  \citenamefont {Zhao}, \citenamefont {Dominguez}, \citenamefont {Thiemens}
  \emph {et~al.}}]{fei2011infrared}%
  \BibitemOpen
  \bibfield  {author} {\bibinfo {author} {\bibfnamefont {Z.}~\bibnamefont
  {Fei}}, \bibinfo {author} {\bibfnamefont {G.~O.}\ \bibnamefont {Andreev}},
  \bibinfo {author} {\bibfnamefont {W.}~\bibnamefont {Bao}}, \bibinfo {author}
  {\bibfnamefont {L.~M.}\ \bibnamefont {Zhang}}, \bibinfo {author}
  {\bibfnamefont {A.}~\bibnamefont {S.~McLeod}}, \bibinfo {author}
  {\bibfnamefont {C.}~\bibnamefont {Wang}}, \bibinfo {author} {\bibfnamefont
  {M.~K.}\ \bibnamefont {Stewart}}, \bibinfo {author} {\bibfnamefont
  {Z.}~\bibnamefont {Zhao}}, \bibinfo {author} {\bibfnamefont {G.}~\bibnamefont
  {Dominguez}}, \bibinfo {author} {\bibfnamefont {M.}~\bibnamefont {Thiemens}},
   \emph {et~al.},\ }\href@noop {} {\bibfield  {journal} {\bibinfo  {journal}
  {Nano Lett.}\ }\textbf {\bibinfo {volume} {11}},\ \bibinfo {pages} {4701}
  (\bibinfo {year} {2011})}\BibitemShut {NoStop}%
\bibitem [{\citenamefont {Principi}\ \emph {et~al.}(2014)\citenamefont
  {Principi}, \citenamefont {Carrega}, \citenamefont {Lundeberg}, \citenamefont
  {Woessner}, \citenamefont {Koppens}, \citenamefont {Vignale},\ and\
  \citenamefont {Polini}}]{principi2014plasmon}%
  \BibitemOpen
  \bibfield  {author} {\bibinfo {author} {\bibfnamefont {A.}~\bibnamefont
  {Principi}}, \bibinfo {author} {\bibfnamefont {M.}~\bibnamefont {Carrega}},
  \bibinfo {author} {\bibfnamefont {M.~B.}\ \bibnamefont {Lundeberg}}, \bibinfo
  {author} {\bibfnamefont {A.}~\bibnamefont {Woessner}}, \bibinfo {author}
  {\bibfnamefont {F.~H.}\ \bibnamefont {Koppens}}, \bibinfo {author}
  {\bibfnamefont {G.}~\bibnamefont {Vignale}}, \ and\ \bibinfo {author}
  {\bibfnamefont {M.}~\bibnamefont {Polini}},\ }\href@noop {} {\bibfield
  {journal} {\bibinfo  {journal} {Phys. Rev. B}\ }\textbf {\bibinfo {volume}
  {90}},\ \bibinfo {pages} {165408} (\bibinfo {year} {2014})}\BibitemShut
  {NoStop}%
\bibitem [{\citenamefont {Wang}\ \emph {et~al.}(2013)\citenamefont {Wang},
  \citenamefont {Meric}, \citenamefont {Huang}, \citenamefont {Gao},
  \citenamefont {Gao}, \citenamefont {Tran}, \citenamefont {Taniguchi},
  \citenamefont {Watanabe}, \citenamefont {Campos}, \citenamefont {Muller}
  \emph {et~al.}}]{wang2013one}%
  \BibitemOpen
  \bibfield  {author} {\bibinfo {author} {\bibfnamefont {L.}~\bibnamefont
  {Wang}}, \bibinfo {author} {\bibfnamefont {I.}~\bibnamefont {Meric}},
  \bibinfo {author} {\bibfnamefont {P.}~\bibnamefont {Huang}}, \bibinfo
  {author} {\bibfnamefont {Q.}~\bibnamefont {Gao}}, \bibinfo {author}
  {\bibfnamefont {Y.}~\bibnamefont {Gao}}, \bibinfo {author} {\bibfnamefont
  {H.}~\bibnamefont {Tran}}, \bibinfo {author} {\bibfnamefont {T.}~\bibnamefont
  {Taniguchi}}, \bibinfo {author} {\bibfnamefont {K.}~\bibnamefont {Watanabe}},
  \bibinfo {author} {\bibfnamefont {L.}~\bibnamefont {Campos}}, \bibinfo
  {author} {\bibfnamefont {D.}~\bibnamefont {Muller}},  \emph {et~al.},\
  }\href@noop {} {\bibfield  {journal} {\bibinfo  {journal} {Science}\ }\textbf
  {\bibinfo {volume} {342}},\ \bibinfo {pages} {614} (\bibinfo {year}
  {2013})}\BibitemShut {NoStop}%
\bibitem [{\citenamefont {Dean}\ \emph {et~al.}(2010)\citenamefont {Dean},
  \citenamefont {Young}, \citenamefont {Meric}, \citenamefont {Lee},
  \citenamefont {Wang}, \citenamefont {Sorgenfrei}, \citenamefont {Watanabe},
  \citenamefont {Taniguchi}, \citenamefont {Kim}, \citenamefont {Shepard} \emph
  {et~al.}}]{dean2010boron}%
  \BibitemOpen
  \bibfield  {author} {\bibinfo {author} {\bibfnamefont {C.}~\bibnamefont
  {Dean}}, \bibinfo {author} {\bibfnamefont {A.}~\bibnamefont {Young}},
  \bibinfo {author} {\bibfnamefont {I.}~\bibnamefont {Meric}}, \bibinfo
  {author} {\bibfnamefont {C.}~\bibnamefont {Lee}}, \bibinfo {author}
  {\bibfnamefont {L.}~\bibnamefont {Wang}}, \bibinfo {author} {\bibfnamefont
  {S.}~\bibnamefont {Sorgenfrei}}, \bibinfo {author} {\bibfnamefont
  {K.}~\bibnamefont {Watanabe}}, \bibinfo {author} {\bibfnamefont
  {T.}~\bibnamefont {Taniguchi}}, \bibinfo {author} {\bibfnamefont
  {P.}~\bibnamefont {Kim}}, \bibinfo {author} {\bibfnamefont {K.}~\bibnamefont
  {Shepard}},  \emph {et~al.},\ }\href@noop {} {\bibfield  {journal} {\bibinfo
  {journal} {Nat. Nanotechnol.}\ }\textbf {\bibinfo {volume} {5}},\ \bibinfo
  {pages} {722} (\bibinfo {year} {2010})}\BibitemShut {NoStop}%
\bibitem [{\citenamefont {Ren}\ \emph {et~al.}(2012{\natexlab{a}})\citenamefont
  {Ren}, \citenamefont {Zhang}, \citenamefont {Yao}, \citenamefont {Sun},
  \citenamefont {Kaneko}, \citenamefont {Yan}, \citenamefont {Nanot},
  \citenamefont {Jin}, \citenamefont {Kawayama}, \citenamefont {Tonouchi},
  \citenamefont {Tour},\ and\ \citenamefont {Kono}}]{Ren_NanoLett2012}%
  \BibitemOpen
  \bibfield  {author} {\bibinfo {author} {\bibfnamefont {L.}~\bibnamefont
  {Ren}}, \bibinfo {author} {\bibfnamefont {Q.}~\bibnamefont {Zhang}}, \bibinfo
  {author} {\bibfnamefont {J.}~\bibnamefont {Yao}}, \bibinfo {author}
  {\bibfnamefont {Z.}~\bibnamefont {Sun}}, \bibinfo {author} {\bibfnamefont
  {R.}~\bibnamefont {Kaneko}}, \bibinfo {author} {\bibfnamefont
  {Z.}~\bibnamefont {Yan}}, \bibinfo {author} {\bibfnamefont {S.}~\bibnamefont
  {Nanot}}, \bibinfo {author} {\bibfnamefont {Z.}~\bibnamefont {Jin}}, \bibinfo
  {author} {\bibfnamefont {I.}~\bibnamefont {Kawayama}}, \bibinfo {author}
  {\bibfnamefont {M.}~\bibnamefont {Tonouchi}}, \bibinfo {author}
  {\bibfnamefont {J.~M.}\ \bibnamefont {Tour}}, \ and\ \bibinfo {author}
  {\bibfnamefont {J.}~\bibnamefont {Kono}},\ }\href {\doibase
  10.1021/nl301496r} {\bibfield  {journal} {\bibinfo  {journal} {Nano Letters}\
  }\textbf {\bibinfo {volume} {12}},\ \bibinfo {pages} {3711} (\bibinfo {year}
  {2012}{\natexlab{a}})},\ \bibinfo {note} {pMID: 22663563},\ \Eprint
  {http://arxiv.org/abs/http://dx.doi.org/10.1021/nl301496r}
  {http://dx.doi.org/10.1021/nl301496r} \BibitemShut {NoStop}%
\bibitem [{\citenamefont {Rouhi}\ \emph {et~al.}(2012)\citenamefont {Rouhi},
  \citenamefont {Capdevila}, \citenamefont {Jain}, \citenamefont {Zand},
  \citenamefont {Wang}, \citenamefont {Brown}, \citenamefont {Jofre},\ and\
  \citenamefont {Burke}}]{Rouhi2012}%
  \BibitemOpen
  \bibfield  {author} {\bibinfo {author} {\bibfnamefont {N.}~\bibnamefont
  {Rouhi}}, \bibinfo {author} {\bibfnamefont {S.}~\bibnamefont {Capdevila}},
  \bibinfo {author} {\bibfnamefont {D.}~\bibnamefont {Jain}}, \bibinfo {author}
  {\bibfnamefont {K.}~\bibnamefont {Zand}}, \bibinfo {author} {\bibfnamefont
  {Y.~Y.}\ \bibnamefont {Wang}}, \bibinfo {author} {\bibfnamefont
  {E.}~\bibnamefont {Brown}}, \bibinfo {author} {\bibfnamefont
  {L.}~\bibnamefont {Jofre}}, \ and\ \bibinfo {author} {\bibfnamefont
  {P.}~\bibnamefont {Burke}},\ }\href {\doibase 10.1007/s12274-012-0251-0}
  {\bibfield  {journal} {\bibinfo  {journal} {Nano Research}\ }\textbf
  {\bibinfo {volume} {5}},\ \bibinfo {pages} {667} (\bibinfo {year}
  {2012})}\BibitemShut {NoStop}%
\bibitem [{\citenamefont {Ju}\ \emph {et~al.}(2011)\citenamefont {Ju},
  \citenamefont {Geng}, \citenamefont {Horng}, \citenamefont {Girit},
  \citenamefont {Martin}, \citenamefont {Hao}, \citenamefont {Bechtel},
  \citenamefont {Liang}, \citenamefont {Zettl}, \citenamefont {Shen} \emph
  {et~al.}}]{ju2011graphene}%
  \BibitemOpen
  \bibfield  {author} {\bibinfo {author} {\bibfnamefont {L.}~\bibnamefont
  {Ju}}, \bibinfo {author} {\bibfnamefont {B.}~\bibnamefont {Geng}}, \bibinfo
  {author} {\bibfnamefont {J.}~\bibnamefont {Horng}}, \bibinfo {author}
  {\bibfnamefont {C.}~\bibnamefont {Girit}}, \bibinfo {author} {\bibfnamefont
  {M.}~\bibnamefont {Martin}}, \bibinfo {author} {\bibfnamefont
  {Z.}~\bibnamefont {Hao}}, \bibinfo {author} {\bibfnamefont {H.~A.}\
  \bibnamefont {Bechtel}}, \bibinfo {author} {\bibfnamefont {X.}~\bibnamefont
  {Liang}}, \bibinfo {author} {\bibfnamefont {A.}~\bibnamefont {Zettl}},
  \bibinfo {author} {\bibfnamefont {Y.~R.}\ \bibnamefont {Shen}},  \emph
  {et~al.},\ }\href@noop {} {\bibfield  {journal} {\bibinfo  {journal} {Nature
  nanotechnology}\ }\textbf {\bibinfo {volume} {6}},\ \bibinfo {pages} {630}
  (\bibinfo {year} {2011})}\BibitemShut {NoStop}%
\bibitem [{\citenamefont {Rossi}\ and\ \citenamefont
  {Das~Sarma}(2008)}]{Rossi_PRL2008}%
  \BibitemOpen
  \bibfield  {author} {\bibinfo {author} {\bibfnamefont {E.}~\bibnamefont
  {Rossi}}\ and\ \bibinfo {author} {\bibfnamefont {S.}~\bibnamefont
  {Das~Sarma}},\ }\href {\doibase 10.1103/PhysRevLett.101.166803} {\bibfield
  {journal} {\bibinfo  {journal} {Phys. Rev. Lett.}\ }\textbf {\bibinfo
  {volume} {101}},\ \bibinfo {pages} {166803} (\bibinfo {year}
  {2008})}\BibitemShut {NoStop}%
\bibitem [{\citenamefont {Abergel}\ \emph {et~al.}(2013)\citenamefont
  {Abergel}, \citenamefont {Rodriguez-Vega}, \citenamefont {Rossi},\ and\
  \citenamefont {Das~Sarma}}]{Abergel_PRB2013}%
  \BibitemOpen
  \bibfield  {author} {\bibinfo {author} {\bibfnamefont {D.~S.~L.}\
  \bibnamefont {Abergel}}, \bibinfo {author} {\bibfnamefont {M.}~\bibnamefont
  {Rodriguez-Vega}}, \bibinfo {author} {\bibfnamefont {E.}~\bibnamefont
  {Rossi}}, \ and\ \bibinfo {author} {\bibfnamefont {S.}~\bibnamefont
  {Das~Sarma}},\ }\href {\doibase 10.1103/PhysRevB.88.235402} {\bibfield
  {journal} {\bibinfo  {journal} {Phys. Rev. B}\ }\textbf {\bibinfo {volume}
  {88}},\ \bibinfo {pages} {235402} (\bibinfo {year} {2013})}\BibitemShut
  {NoStop}%
\bibitem [{\citenamefont {Kharitonov}\ and\ \citenamefont
  {Efetov}(2008)}]{Kharitonov_PRB2008}%
  \BibitemOpen
  \bibfield  {author} {\bibinfo {author} {\bibfnamefont {M.~Y.}\ \bibnamefont
  {Kharitonov}}\ and\ \bibinfo {author} {\bibfnamefont {K.~B.}\ \bibnamefont
  {Efetov}},\ }\href {\doibase 10.1103/PhysRevB.78.241401} {\bibfield
  {journal} {\bibinfo  {journal} {Phys. Rev. B}\ }\textbf {\bibinfo {volume}
  {78}},\ \bibinfo {pages} {241401} (\bibinfo {year} {2008})}\BibitemShut
  {NoStop}%
\bibitem [{\citenamefont {Fetter}\ and\ \citenamefont
  {Walecka}(2003)}]{fetter2003quantum}%
  \BibitemOpen
  \bibfield  {author} {\bibinfo {author} {\bibfnamefont {A.~L.}\ \bibnamefont
  {Fetter}}\ and\ \bibinfo {author} {\bibfnamefont {J.~D.}\ \bibnamefont
  {Walecka}},\ }\href@noop {} {\emph {\bibinfo {title} {Quantum theory of
  many-particle systems}}}\ (\bibinfo  {publisher} {Courier Corporation},\
  \bibinfo {year} {2003})\BibitemShut {NoStop}%
\bibitem [{\citenamefont {Fischetti}(1999)}]{fischetti1999master}%
  \BibitemOpen
  \bibfield  {author} {\bibinfo {author} {\bibfnamefont {M.}~\bibnamefont
  {Fischetti}},\ }\href@noop {} {\bibfield  {journal} {\bibinfo  {journal}
  {Phys. Rev. B}\ }\textbf {\bibinfo {volume} {59}},\ \bibinfo {pages} {4901}
  (\bibinfo {year} {1999})}\BibitemShut {NoStop}%
\bibitem [{\citenamefont {Kohn}\ and\ \citenamefont
  {Luttinger}(1957)}]{kohn1957quantum}%
  \BibitemOpen
  \bibfield  {author} {\bibinfo {author} {\bibfnamefont {W.}~\bibnamefont
  {Kohn}}\ and\ \bibinfo {author} {\bibfnamefont {J.}~\bibnamefont
  {Luttinger}},\ }\href@noop {} {\bibfield  {journal} {\bibinfo  {journal}
  {Physical Review}\ }\textbf {\bibinfo {volume} {108}},\ \bibinfo {pages}
  {590} (\bibinfo {year} {1957})}\BibitemShut {NoStop}%
\bibitem [{\citenamefont {Buecking}\ \emph {et~al.}(2007)\citenamefont
  {Buecking}, \citenamefont {Scheffler}, \citenamefont {Kratzer},\ and\
  \citenamefont {Knorr}}]{buecking2007theory}%
  \BibitemOpen
  \bibfield  {author} {\bibinfo {author} {\bibfnamefont {N.}~\bibnamefont
  {Buecking}}, \bibinfo {author} {\bibfnamefont {M.}~\bibnamefont {Scheffler}},
  \bibinfo {author} {\bibfnamefont {P.}~\bibnamefont {Kratzer}}, \ and\
  \bibinfo {author} {\bibfnamefont {A.}~\bibnamefont {Knorr}},\ }\href@noop {}
  {\bibfield  {journal} {\bibinfo  {journal} {Appl. Phys. A}\ }\textbf
  {\bibinfo {volume} {88}},\ \bibinfo {pages} {505} (\bibinfo {year}
  {2007})}\BibitemShut {NoStop}%
\bibitem [{\citenamefont {Ren}\ \emph {et~al.}(2012{\natexlab{b}})\citenamefont
  {Ren}, \citenamefont {Zhang}, \citenamefont {Yao}, \citenamefont {Sun},
  \citenamefont {Kaneko}, \citenamefont {Yan}, \citenamefont {Nanot},
  \citenamefont {Jin}, \citenamefont {Kawayama}, \citenamefont {Tonouchi} \emph
  {et~al.}}]{ren2012terahertz}%
  \BibitemOpen
  \bibfield  {author} {\bibinfo {author} {\bibfnamefont {L.}~\bibnamefont
  {Ren}}, \bibinfo {author} {\bibfnamefont {Q.}~\bibnamefont {Zhang}}, \bibinfo
  {author} {\bibfnamefont {J.}~\bibnamefont {Yao}}, \bibinfo {author}
  {\bibfnamefont {Z.}~\bibnamefont {Sun}}, \bibinfo {author} {\bibfnamefont
  {R.}~\bibnamefont {Kaneko}}, \bibinfo {author} {\bibfnamefont
  {Z.}~\bibnamefont {Yan}}, \bibinfo {author} {\bibfnamefont {S.}~\bibnamefont
  {Nanot}}, \bibinfo {author} {\bibfnamefont {Z.}~\bibnamefont {Jin}}, \bibinfo
  {author} {\bibfnamefont {I.}~\bibnamefont {Kawayama}}, \bibinfo {author}
  {\bibfnamefont {M.}~\bibnamefont {Tonouchi}},  \emph {et~al.},\ }\href@noop
  {} {\bibfield  {journal} {\bibinfo  {journal} {Nano Lett.}\ }\textbf
  {\bibinfo {volume} {12}},\ \bibinfo {pages} {3711} (\bibinfo {year}
  {2012}{\natexlab{b}})}\BibitemShut {NoStop}%
\bibitem [{\citenamefont {Gannett}\ \emph {et~al.}(2011)\citenamefont
  {Gannett}, \citenamefont {Regan}, \citenamefont {Watanabe}, \citenamefont
  {Taniguchi}, \citenamefont {Crommie},\ and\ \citenamefont
  {Zettl}}]{gannett2011boron}%
  \BibitemOpen
  \bibfield  {author} {\bibinfo {author} {\bibfnamefont {W.}~\bibnamefont
  {Gannett}}, \bibinfo {author} {\bibfnamefont {W.}~\bibnamefont {Regan}},
  \bibinfo {author} {\bibfnamefont {K.}~\bibnamefont {Watanabe}}, \bibinfo
  {author} {\bibfnamefont {T.}~\bibnamefont {Taniguchi}}, \bibinfo {author}
  {\bibfnamefont {M.}~\bibnamefont {Crommie}}, \ and\ \bibinfo {author}
  {\bibfnamefont {A.}~\bibnamefont {Zettl}},\ }\href@noop {} {\bibfield
  {journal} {\bibinfo  {journal} {Appl. Phys. Lett.}\ }\textbf {\bibinfo
  {volume} {98}},\ \bibinfo {pages} {242105} (\bibinfo {year}
  {2011})}\BibitemShut {NoStop}%
\bibitem [{\citenamefont {Polini}\ \emph {et~al.}(2008)\citenamefont {Polini},
  \citenamefont {Asgari}, \citenamefont {Borghi}, \citenamefont {Barlas},
  \citenamefont {Pereg-Barnea},\ and\ \citenamefont {MacDonald}}]{barlas1}%
  \BibitemOpen
  \bibfield  {author} {\bibinfo {author} {\bibfnamefont {M.}~\bibnamefont
  {Polini}}, \bibinfo {author} {\bibfnamefont {R.}~\bibnamefont {Asgari}},
  \bibinfo {author} {\bibfnamefont {G.}~\bibnamefont {Borghi}}, \bibinfo
  {author} {\bibfnamefont {Y.}~\bibnamefont {Barlas}}, \bibinfo {author}
  {\bibfnamefont {T.}~\bibnamefont {Pereg-Barnea}}, \ and\ \bibinfo {author}
  {\bibfnamefont {A.}~\bibnamefont {MacDonald}},\ }\href@noop {} {\bibfield
  {journal} {\bibinfo  {journal} {Physical Review B}\ }\textbf {\bibinfo
  {volume} {77}},\ \bibinfo {pages} {081411} (\bibinfo {year}
  {2008})}\BibitemShut {NoStop}%
\bibitem [{\citenamefont {Barlas}\ \emph {et~al.}(2007)\citenamefont {Barlas},
  \citenamefont {Pereg-Barnea}, \citenamefont {Polini}, \citenamefont
  {Asgari},\ and\ \citenamefont {MacDonald}}]{barlas2}%
  \BibitemOpen
  \bibfield  {author} {\bibinfo {author} {\bibfnamefont {Y.}~\bibnamefont
  {Barlas}}, \bibinfo {author} {\bibfnamefont {T.}~\bibnamefont
  {Pereg-Barnea}}, \bibinfo {author} {\bibfnamefont {M.}~\bibnamefont
  {Polini}}, \bibinfo {author} {\bibfnamefont {R.}~\bibnamefont {Asgari}}, \
  and\ \bibinfo {author} {\bibfnamefont {A.}~\bibnamefont {MacDonald}},\
  }\href@noop {} {\bibfield  {journal} {\bibinfo  {journal} {Physical review
  letters}\ }\textbf {\bibinfo {volume} {98}},\ \bibinfo {pages} {236601}
  (\bibinfo {year} {2007})}\BibitemShut {NoStop}%
\bibitem [{\citenamefont {Barlas}\ and\ \citenamefont {Yang}(2009)}]{barlas3}%
  \BibitemOpen
  \bibfield  {author} {\bibinfo {author} {\bibfnamefont {Y.}~\bibnamefont
  {Barlas}}\ and\ \bibinfo {author} {\bibfnamefont {K.}~\bibnamefont {Yang}},\
  }\href@noop {} {\bibfield  {journal} {\bibinfo  {journal} {Physical Review
  B}\ }\textbf {\bibinfo {volume} {80}},\ \bibinfo {pages} {161408} (\bibinfo
  {year} {2009})}\BibitemShut {NoStop}%
\bibitem [{\citenamefont {Polini}\ \emph {et~al.}(2007)\citenamefont {Polini},
  \citenamefont {Asgari}, \citenamefont {Barlas}, \citenamefont
  {Pereg-Barnea},\ and\ \citenamefont {MacDonald}}]{barlas4}%
  \BibitemOpen
  \bibfield  {author} {\bibinfo {author} {\bibfnamefont {M.}~\bibnamefont
  {Polini}}, \bibinfo {author} {\bibfnamefont {R.}~\bibnamefont {Asgari}},
  \bibinfo {author} {\bibfnamefont {Y.}~\bibnamefont {Barlas}}, \bibinfo
  {author} {\bibfnamefont {T.}~\bibnamefont {Pereg-Barnea}}, \ and\ \bibinfo
  {author} {\bibfnamefont {A.~H.}\ \bibnamefont {MacDonald}},\ }\href@noop {}
  {\bibfield  {journal} {\bibinfo  {journal} {Solid State Communications}\
  }\textbf {\bibinfo {volume} {143}},\ \bibinfo {pages} {58} (\bibinfo {year}
  {2007})}\BibitemShut {NoStop}%
\bibitem [{\citenamefont {Barlas}(2008)}]{barlasPhD}%
  \BibitemOpen
  \bibfield  {author} {\bibinfo {author} {\bibfnamefont {Y.}~\bibnamefont
  {Barlas}},\ }\href@noop {} {\emph {\bibinfo {title} {Role of
  electron-electron interactions in chiral 2DEGs}}}\ (\bibinfo  {publisher}
  {University of Texas at Austin},\ \bibinfo {year} {2008})\ \bibinfo {note}
  {{PhD} Dissertation}\BibitemShut {NoStop}%
\bibitem [{\citenamefont {Wunsch}\ \emph {et~al.}(2006)\citenamefont {Wunsch},
  \citenamefont {Stauber}, \citenamefont {Sols},\ and\ \citenamefont
  {Guinea}}]{Wunsch_NJP06}%
  \BibitemOpen
  \bibfield  {author} {\bibinfo {author} {\bibfnamefont {B.}~\bibnamefont
  {Wunsch}}, \bibinfo {author} {\bibfnamefont {T.}~\bibnamefont {Stauber}},
  \bibinfo {author} {\bibfnamefont {F.}~\bibnamefont {Sols}}, \ and\ \bibinfo
  {author} {\bibfnamefont {F.}~\bibnamefont {Guinea}},\ }\href
  {http://stacks.iop.org/1367-2630/8/i=12/a=318} {\bibfield  {journal}
  {\bibinfo  {journal} {New Journal of Physics}\ }\textbf {\bibinfo {volume}
  {8}},\ \bibinfo {pages} {318} (\bibinfo {year} {2006})}\BibitemShut {NoStop}%
\bibitem [{\citenamefont {Grill}(1999)}]{grill1999electrical}%
  \BibitemOpen
  \bibfield  {author} {\bibinfo {author} {\bibfnamefont {A.}~\bibnamefont
  {Grill}},\ }\href@noop {} {\bibfield  {journal} {\bibinfo  {journal} {Thin
  Solid Films}\ }\textbf {\bibinfo {volume} {355}},\ \bibinfo {pages} {189}
  (\bibinfo {year} {1999})}\BibitemShut {NoStop}%
\bibitem [{\citenamefont {Geim}\ and\ \citenamefont
  {Grigorieva}(2013)}]{geim2013van}%
  \BibitemOpen
  \bibfield  {author} {\bibinfo {author} {\bibfnamefont {A.}~\bibnamefont
  {Geim}}\ and\ \bibinfo {author} {\bibfnamefont {I.}~\bibnamefont
  {Grigorieva}},\ }\href@noop {} {\bibfield  {journal} {\bibinfo  {journal}
  {Nature}\ }\textbf {\bibinfo {volume} {499}},\ \bibinfo {pages} {419}
  (\bibinfo {year} {2013})}\BibitemShut {NoStop}%
\bibitem [{\citenamefont {Ku{\v{c}}{\'\i}rkov{\'a}}\ and\ \citenamefont
  {Navr{\'a}til}(1994)}]{kuvcirkova1994interpretation}%
  \BibitemOpen
  \bibfield  {author} {\bibinfo {author} {\bibfnamefont {A.}~\bibnamefont
  {Ku{\v{c}}{\'\i}rkov{\'a}}}\ and\ \bibinfo {author} {\bibfnamefont
  {K.}~\bibnamefont {Navr{\'a}til}},\ }\href@noop {} {\bibfield  {journal}
  {\bibinfo  {journal} {Appl. Spectrosc.}\ }\textbf {\bibinfo {volume} {48}},\
  \bibinfo {pages} {113} (\bibinfo {year} {1994})}\BibitemShut {NoStop}%
\bibitem [{\citenamefont {Philipp}(1979)}]{philipp1979infrared}%
  \BibitemOpen
  \bibfield  {author} {\bibinfo {author} {\bibfnamefont {H.~R.}\ \bibnamefont
  {Philipp}},\ }\href@noop {} {\bibfield  {journal} {\bibinfo  {journal} {J.
  Appl. Phys.}\ }\textbf {\bibinfo {volume} {50}},\ \bibinfo {pages} {1053}
  (\bibinfo {year} {1979})}\BibitemShut {NoStop}%
\bibitem [{\citenamefont {Geick}\ \emph {et~al.}(1966)\citenamefont {Geick},
  \citenamefont {Perry},\ and\ \citenamefont {Rupprecht}}]{geick1966normal}%
  \BibitemOpen
  \bibfield  {author} {\bibinfo {author} {\bibfnamefont {R.}~\bibnamefont
  {Geick}}, \bibinfo {author} {\bibfnamefont {C.}~\bibnamefont {Perry}}, \ and\
  \bibinfo {author} {\bibfnamefont {G.}~\bibnamefont {Rupprecht}},\ }\href@noop
  {} {\bibfield  {journal} {\bibinfo  {journal} {Phys. Rev.}\ }\textbf
  {\bibinfo {volume} {146}},\ \bibinfo {pages} {543} (\bibinfo {year}
  {1966})}\BibitemShut {NoStop}%
\bibitem [{\citenamefont {Wang}\ and\ \citenamefont
  {Mahan}(1972)}]{WangMahan_PRB_1972}%
  \BibitemOpen
  \bibfield  {author} {\bibinfo {author} {\bibfnamefont {S.~Q.}\ \bibnamefont
  {Wang}}\ and\ \bibinfo {author} {\bibfnamefont {G.~D.}\ \bibnamefont
  {Mahan}},\ }\href {\doibase 10.1103/PhysRevB.6.4517} {\bibfield  {journal}
  {\bibinfo  {journal} {Phys. Rev. B}\ }\textbf {\bibinfo {volume} {6}},\
  \bibinfo {pages} {4517} (\bibinfo {year} {1972})}\BibitemShut {NoStop}%
\end{thebibliography}
%

\end{document}